\RequirePackage[l2tabu,orthodox]{nag}
\documentclass[preprint,number,a4paper,11pt]{elsarticle}
\usepackage[utf8]{inputenc}
\usepackage[T1]{fontenc}
\usepackage[english]{babel}
\usepackage[margin=1.5cm,footskip=1cm]{geometry}
\usepackage{amsmath}
\usepackage{amssymb}
\usepackage{braket}
\usepackage{bm}
\usepackage{hyperref}
\usepackage{latexsym}
\usepackage{graphicx,caption}
\usepackage{rotating}
\usepackage{grffile}
\usepackage{color}
\usepackage{braket}
\usepackage{amssymb}
\usepackage{listings}
\usepackage{placeins}
\usepackage{multirow}
\usepackage[normalem]{ulem}
\newcommand{\Imag}{\mathop{\text{Im}}}
\newcommand{\Real}{\mathop{\text{Re}}}
\newcommand{\FFT}{\mathop{\text{FFT}}}
\newcommand{\IFFT}{\mathop{\text{IFFT}}}
\renewcommand{\i}{\mathop{\text{i}}}
\newcommand{\ig}[1]{\includegraphics[width=.49\textwidth]{#1}}
\newcommand{\igg}[1]{\includegraphics[width=.99\textwidth]{#1}}
\newcommand{\igt}[1]{\includegraphics[width=.32\textwidth]{#1}}
\usepackage{mdwlist}
\renewenvironment{description}%
{
\begin{basedescript}{
\desclabelstyle{\nextlinelabel}
\desclabelwidth{2em}}}
{
\end{basedescript}
}
\usepackage{sidecap}
\sidecaptionvpos{figure}{c}
\usepackage{chngcntr}
\counterwithin{figure}{subsection}

\newcounter{foo}

\journal{Computer Physics Communications}
\begin{document}

\begin{frontmatter}

\title{The Relativistic Schrödinger Equation through \texttt{FFTW~3}: An Extension of \texorpdfstring{\texttt{quantumfdtd}}{quantumfdtd}}

\author[1,2,3]{Rafael L.~Delgado\corref{cor1}}
\ead{rafael.delgado@upm.es}
\author[3]{Sebastian Steinbeißer}
\ead{sebastian.steinbeisser@tum.de}
\author[4]{Michael Strickland}
\ead{mstrick6@kent.edu}
\author[3,5,6]{Johannes Heinrich Weber}
\ead{dr.rer.nat.weber@gmail.com}

\cortext[cor1]{Corresponding author}
\address[1]{ETSIS de Telecomunicación (UPM), Campus Sur, C/Nikola Tesla, s/n 28031 Madrid, Spain}
\address[2]{INFN-Firenze, Via Giovanni Sansone, 1, 50019 Sesto Fiorentino FI, Italy (moved to UPM)}
\address[3]{Physik-Department, Technische Universität München, James-Franck-Straße~1, D-85748 Garching b. München, Germany}
\address[4]{Department of Physics, Kent State University, Kent, OH 44242 USA}
\address[5]{Department of Computational Mathematics, Science and Engineering \& Department of Physics and Astronomy, Michigan State University, East Lansing, MI 48824, USA}
\address[6]{Institut für Physik \& IRIS Adlershof, Humboldt-Universität zu Berlin, Zum Großen Windkanal~6, D-12489 Berlin, Germany}

\begin{abstract}
In order to solve the time-independent three-dimensional Schrödinger equation, one can transform the time-dependent Schrödinger equation to imaginary time and use a parallelized iterative method to obtain the full three-dimensional eigen-states and eigen-values on very large lattices.
In the case of the non-relativistic Schrödinger equation, there exists a publicly available code called \texttt{quantumfdtd} which implements this algorithm.
In this paper, we (a) extend the \texttt{quantumfdtd} code to include the case of the relativistic Schrödinger equation and (b) add two optimized Fast Fourier Transform (FFT) based kinetic energy terms for non-relativistic cases.
The new kinetic energy terms (two non-relativistic and one relativistic) are computed using the parallelized FFT-algorithm provided by the \texttt{FFTW~3}~library.
The resulting \texttt{quantumfdtd~v3} code, which is publicly released with this paper, is backwards compatible with version~2, supporting explicit finite-differences schemes in addition to the new FFT-based schemes.
Finally, we (c) extend the original code so that it supports arbitrary external file-based potentials and the option to project out distinct parity eigen-states from the solutions.
Herein, we provide details of the \texttt{quantumfdtd~v3} implementation, comparisons and tests of the three new kinetic energy terms, and code documentation.
\end{abstract}


\end{frontmatter}

\clearpage


\textbf{NEW VERSION PROGRAM SUMMARY}\\\\

\begin{small}
\noindent
{\em Program Title:} \texttt{quantumfdtd~v3}\\\\
{\em CPC Library link to program files:} (to be added by Technical Editor)\\\\
{\em Developer's repository link:} \url{https://github.com/quantumfdtd/quantumfdtd_v3}\\\\
{\em Code Ocean capsule:} (to be added by Technical Editor)\\\\
{\em Licensing provisions:} GPLv3\\\\
{\em Programming language:} \texttt{C++}, \texttt{Python}, \texttt{Shell}\\\\
{\em Journal reference of previous version:} \cite{Dumitru:2009fy_summary, Dumitru:2009ni_summary, Strickland:2009ft_summary, Margotta:2011ta_summary}\\\\
{\em Does the new version supersede the previous version?:} Yes\\\\
{\em Reasons for the new version:} We extended the previous version of \texttt{quantumfdtd}, which solves the time-independent non-relativistic three-dimensional Schrödinger equation, to the case of the relativistic kinetic energy.
Additionally, we have added the capability of using arbitrary three-dimensional potentials from external files.
This functionality is required, for instance, in order to interface with lattice QCD code.\\\\
{\em Summary of revisions:} (a) Include the case of the relativistic Schrödinger equation, (b) add two optimized Fast Fourier Transform (FFT) based kinetic energy terms for non-relativistic cases, and (c) support arbitrary external file-based potentials and the option to project out distinct parity eigen-states from the solutions.
The new kinetic energy terms (two non-relativistic and one relativistic) are computed using the parallelized FFT-algorithm provided by the \texttt{FFTW~3}~library.\\\\
{\em Nature of problem:} We compute the ground, first, and second excited states of the time-independent three-dimensional Schrödinger equation.
As input, we accept a number of hard-coded, analytical, three-dimensional potentials, as well as arbitrary potentials via external files.
The outputs of the program are the corresponding eigen-values and eigen-vectors (energies and wave-functions).
We can also project out distinct parity eigen-states of the solutions.\\\\
{\em Solution method:} The time-dependent Schrödinger equation is transformed to imaginary time.
We use a parallelized iterative method to obtain the full three-dimensional time-independent eigen-states and eigen-values on large lattices.
The new kinetic energy terms (two non-relativistic and one relativistic) are computed using the parallelized FFT-algorithm provided by the \texttt{FFTW~3}~library.
The old non-relativistic kinetic term is computed using the finite-difference time-domain (FDTD) algorithm.
Finally, we provide several \texttt{Python} and \texttt{Shell} scripts for the analysis steps.
This includes a \texttt{Python} program for projecting out distinct parity eigen-states from the solutions.\\\\
{\em Additional comments including restrictions and unusual features:}\\
We require the number of \texttt{MPI} processes being a divisor of the number $N$ of spatial grid points.
The following external programs/libraries are used and thus required:
\begin{itemize}
\item \texttt{MPI}~library
\item \texttt{FFTW\_MPI}, version~3
\item GNU Scientific Library (\texttt{GSL}), linked to \texttt{CBLAS}
\item For some of the post-processing scripts, \texttt{Python~3} is required
\end{itemize}

\end{small}


\clearpage
\tableofcontents
\clearpage

\section{Introduction}
\label{sec:introduction}

\subsection{Motivation}
\label{subsec:motivation}

The three-dimensional Schrödinger equation has a closed-form analytical solution for only a very small class of systems (i.e., a free particle in a box, harmonic oscillator, or some central potentials).
In all other cases, one has to rely on approximations, perturbation theory, or numerical solutions.
In particular, most quark models used for phenomenological descriptions of QCD bound states are described by the three-dimensional Schrödinger equation with a wide variety of potentials.
Potential-based quark models have enjoyed a long history of success in describing below-threshold charmonium production and bottomonium spectra and have helped to establish confidence in QCD as the first-principles description of hadronic matter, see, e.g., Refs.~\cite{Ebert:2002pp, Eichten:2007qx, Segovia:2016xqb}.
In recent years, non-relativistic effective field theory methods have allowed for a first-principles approach to potential-based non-relativistic QCD (pNRQCD)~\cite{Brambilla:1999xf, Brambilla:1999qa}.
Additionally, in order to describe quarkonium evolution in the quark-gluon plasma, such potential models have been extended to finite temperature~\cite{Laine:2006ns, Brambilla:2008cx, Brambilla:2010xn} and non-equilibrium~\cite{Dumitru:2007hy, Burnier:2009yu, Dumitru:2009fy, Dumitru:2009ni, Guo:2018vwy}, in which case, the potentials are no longer real-valued or spherically symmetric.
In the full non-equilibrium case, it is necessary to use a full three-dimensional Schrödinger equation solver that allows for complex potentials.

In the past, a parallelized solver called \texttt{quantumfdtd} has been made available for real-valued~\cite{Strickland:2009ft} and complex-valued~\cite{Margotta:2011ta} potentials.
The two previous versions of \texttt{quantumfdtd} use local operators in a finite-difference time-domain (FDTD) scheme for solving the non-relativistic three-dimensional Schrödinger~equation.
However, if light quarks are involved, such as in heavy-light mesons, e.g., in B mesons, these have to be described as relativistic degrees of freedom.
The light quark's relativistic dispersion results in a much larger average radius of the state, namely with the wave-function's falloff in the case of a linearly rising Cornell potential $V(r) = -A/r + \sigma \cdot r$, like $\Psi(r) \sim \exp(-M r)$ instead of $\Psi(r) \sim \exp(-M r^{3/2})$ ($M$ being the reduced mass of the system)~\cite{Duncan:1994uq}.
While spin-orbit and spin-spin corrections can be accounted for in terms of angular-momentum-dependent contributions beyond a simple central potential, at a fundamental level, the change to relativistic dispersion requires an implementation of a relativistic kinetic term in the Schrödinger equation.
For this reason we extend the existing \texttt{quantumfdtd} parallelized 3D Schrödinger solver~\cite{Strickland:2009ft, Margotta:2011ta} to permit a relativistic kinetic term.

These relativistic quark model calculations have a purpose beyond purely phenomenological calculations.
The role of heavy-light mesons or heavy-heavy mesons (quarkonia) as an excellent laboratory of QCD (or the whole Standard Model) is well established.
The confinement property of QCD, which implies that individual gluons or quarks cannot be isolated as asymptotic states, precludes their direct observation.
Nevertheless, heavy-light mesons are bound states of QCD, where the dominant contribution to the meson mass can be understood as being due to its heavy-quark constituent, while any other contribution to its mass -- whether from the mass of its light-quark constituent or from the dynamical gluon fields that mediate the binding -- is small in comparison.
These extra contributions can be understood quantitatively using an effective field theory approach that permits an extraordinarily precise extraction of quark masses when applied to sufficiently accurate lattice QCD results; see Ref.~\cite{Komijani:2020kst} for a review.
Together with experimental data for their weak decay rates, the decay constants of heavy-light mesons provide powerful constraints to many elements of the CKM matrix.
Violation of CKM unitarity would represent evidence of new physics.
Heavy-light mesons are important probes in other, more direct searches for new physics, too.
In particular, the large heavy-quark mass implies that heavy-light mesons may couple rather strongly to currents that mediate interactions with new physics at some high energy scale.
The suppression of any new physics contributions is less pronounced for heavy-light systems, since the heavy-quark mass is much larger than typical hadronic scales.
The matrix elements of such currents interacting with heavy-light mesons can be computed using lattice QCD.
Thus, accurate lattice QCD results for heavy-light mesons can provide strong constraints on new physics.

\subsection{Interface of relativistic quark models with lattice QCD}
\label{subsec:lattice}

In order to extract the aforementioned information from lattice QCD simulations, hadron spectroscopy calculations are required; see Ref.~\cite{Wagner:2013tiz} for a pedagogical introduction to lattice hadron spectroscopy aimed at non-expert readers\footnote{%
This subsection can be skipped by readers not interested into the interface with lattice gauge theory.}.
Lattice spectroscopy with heavy quarks is challenging for many reasons.
The most obvious one is that the heavy-quark mass $M_h$ is quite large compared to the inverse of the lattice spacing $A$.
Hence, discretization artifacts in the heavy-quark sector are usually substantial.
If these artifacts are parameterized in the most general form as a series in $[\log(A M_h)]^i (A M_h)^j$, rather high orders may be needed due to poor convergence caused by the largeness of the expansion parameter $A M_h$.
Fixing many coefficients then necessitates the availability of a wide range of heavy-quark masses and lattice spacings.
Fine lattice spacings that would sustain a moderate size of $A M_h$ imply that a vast number of lattice sites in all directions are required to sustain a sufficient physical volume that prevents major finite size corrections due to correlations stretching across the lattice's boundaries.
Due to the computational cost associated with such a large lattice, one may then have to deal with further issues that are not directly related to the heavy quarks, such as limited statistics, unphysical sea quark masses, or similar drawbacks that may be even more severe than in simulations that do not aim at incorporating heavy quark physics.
There is a workaround for these heavy-quark mass discretization artifacts, which is in spirit close to the quark-model ideas.
Namely, one may treat a heavy quark in the static limit.
While the mass of such a static quark, which is due to the self-energy of the gluon field, is finite at finite lattice spacing, it diverges in the continuum limit.
Then one may study the operators appearing in an expansion in the inverse heavy-quark mass, i.e., in the heavy-quark effective theory (HQET)~\cite{Caswell:1985ui}, or determine mass differences between states involving static as well as non-static quarks on the lattice (see Ref.~\cite{Wagner:2013tiz}).

To make matters worse, heavy quarks in the valence sector give rise to further complications that are quite independent of the underlying ensemble.
On the one hand, the large mass (or energy) $E$ of a corresponding (ground state) meson implies a very rapid decay of any correlation functions $C(t_1-t_0) \sim e^{-E (t_1-t_0)}$ involving heavy quarks, such that numerical errors due to limited statistics, limited precision in matrix inversions, or simply due to machine precision may become quite significant at large times.
On the other hand, the splittings between different energy levels of heavy-quark bound states, e.g., the splitting $E_1-E_0$ between the first excited state and the ground state, are much smaller than the mass of the ground state, $E_0$.
Multi-particle thresholds are usually not far away either.
Taken in combination, these circumstances imply that one all too often has to analyze lattice correlation functions in a time window, where the excited-state contamination is non-negligible.
In the direct searches for new physics on the lattice, one typically studies three-point functions and extracts QCD matrix elements for external currents.
Excited-state contamination is particularly severe for these, as the current couples to the heavy quark at an intermediate time $t$, where $t_0 < t < t_1$.
In practice, it is usually not possible to simultaneously achieve $(t_1-t),(t-t_0) \gg 1/(E_1-E_0)$, which would be necessary to suppress the excited states to a satisfying degree.

Ground or excited state contributions are usually separated in the analysis of lattice correlators by utilizing a wide range of different hadron interpolating operators, which have different overlap factors for the various low-lying hadron (or multi-hadron) states, and then solving a generalized eigen-value problem by diagonalizing a matrix of correlation functions.
This is necessary, since typical hadron interpolating operators in lattice QCD encode three-dimensional shapes that are very different from those typical for the low-lying hadrons, and thereby include major contributions from many excited states or even from a wide range of scattering states.
The most common shapes of hadron interpolating operators -- \emph{Point}-type or \emph{Wall}-type -- are discussed later on.
It has been clearly demonstrated in the past~\cite{Duncan:1994uq} using valence light quarks in the Wilson fermion formulation~\cite{Wilson:1974sk}, heavy quarks in the static limit, and gauge ensembles with Wilson plaquette action~\cite{Wilson:1974sk} in the quenched approximation, i.e., without sea quarks, that heavy-light meson wave-functions from relativistic quark models provide interpolating operators (hereafter: \emph{RQM}-type) for heavy-light mesons with excellent excited-state suppression in first-principle lattice QCD calculations.
A more recent application uses gauge ensembles in (2+1+1)-flavor QCD with physical strange and charm quarks, and two degenerate light quarks at their physical or two unphysical values of the masses (corresponding to pion masses of $m_\pi \approx 140,~220,$ or $300$ MeV in the continuum limit)~\cite{Bazavov:2017lyh}.
They have been generated by the MILC collaboration with one-loop Symanzik improved gauge action~\cite{Luscher:1984xn,Luscher:1985zq} and highly improved staggered quark action (HISQ)~\cite{Follana:2006rc} for the sea.
The valence heavy-quark is implemented in the static limit such that it is propagating forward in time in the first half of the lattice and backward in the second half of the lattice, while the valence "light quark" is implemented using HISQ action and the sea strange quark mass.
This peculiar combination permits the usual symmetrization procedure for meson correlation functions (folding the backward propagating meson state on top of the forward propagating one at the midpoint in time).
The static quark propagator is built up from bare gauge links, or from links after a single iteration of four-dimensional hypercubic (HYP) smearing with standard coefficients~\cite{Hasenfratz:2001hp}.
The latter reduces the gauge noise and the mass of the static quark in the lattice simulation.
Due to the presence of the staggered antiquark field, the meson correlation function receives contributions from states of opposite parity.
For the pseudoscalar interpolating operator used in the correlation function that is being considered, the parity partner has scalar quantum numbers.
The correlation function takes the form of
\begin{equation}
C(t) = \sum\limits_{i=0}^{\infty} A_{i}^{\text{PS}} B_{i}^{\text{PS}\ast} \left[ e^{-E_{i}^{\text{PS}}\tau} + e^{-E_{i}^{\text{PS}}(aN_{\tau}-\tau)} \right] - (-1)^{\tau/a} \sum\limits_{j=0}^{\infty} A_{j}^{\text{S}}B_{j}^{\text{S}\ast} \left[ e^{-E_{j}^{\text{S}}\tau} + e^{-E_{j}^{\text{S}}(aN_{\tau}-\tau)} \right] \,,
\end{equation}
where PS stands for pseudoscalar and S stands for scalar.
Both towers of pseudoscalar or scalar states involve a significant, but finite number of bound states, and an infinite number of scattering states in an infinite volume; in a finite volume, there is only a huge, but finite number of scattering states.
In a lattice correlation function with finite amount of data in the time direction and a non-negligible statistical error, only a limited subset of these states can be resolved at all, and both towers get truncated in a lattice hadron spectroscopy analysis.
While the alternating factor in the parity partner's contribution (and the fact of its admixture) is an artifact of the staggered quark discretization, the parity partner state itself is a genuine state of QCD.
The masses of any individual states may be affected by different discretization artifacts.
The amplitudes $A_{i}^{\text{X}}$ or $B_{i}^{\text{X}}$ ($\text{X} = \{$PS,S$\}$) are associated with the details of the hadron interpolating operators at source or sink, respectively, and their products, e.g., $A_{i}^{\text{X}} B_{i}^{\text{X}\ast}$ are the overlap factors of the correlation function with specific hadrons.
In each square bracket, the second term is due to the backward propagating contribution, and can be safely neglected for $\tau \ll aN_{\tau}/2$.
After neglecting this backward propagating contribution, the two different parities can be separated using ratios of the correlation function spread over a 4-point stencil in time (see Ref.~\cite{Bazavov:2014cta} for details), and reduced correlators of the two individual parities can be constructed.
Obviously, for quark discretizations without doubling such procedures are not necessary.

Both the staggered quark field as well as the static quark field are single-component (or rather single spin component) fields, i.e., they cannot encode non-trivial spin-taste\footnote{%
\emph{Taste} is lattice QCD jargon and denotes the artificial internal $U(4)$ flavor-symmetry group of staggered quarks.
The taste degree of freedom is intertwined with the spin degree of freedom through a combination of phase factors and translations within a single unit hypercube.
Different taste combinations differ in terms of the size of discretization artifacts affecting the hadron masses.} structure in hadron interpolating operators that have the quark and antiquark at the same lattice site.
Other spin-taste combinations could be accessed using point-split hadron interpolating operators with quark and antiquark at different sites.
Those would give rise to unphysical mass differences caused by discretization artifacts associated with different staggered tastes, while physical mass differences associated with different orbital angular momenta and spins could be realized using specific (anti-)symmetrized geometries of the hadron interpolating operators, see Ref.~\cite{Wagner:2013tiz}.

\begin{figure}[t]
\centering
\igg{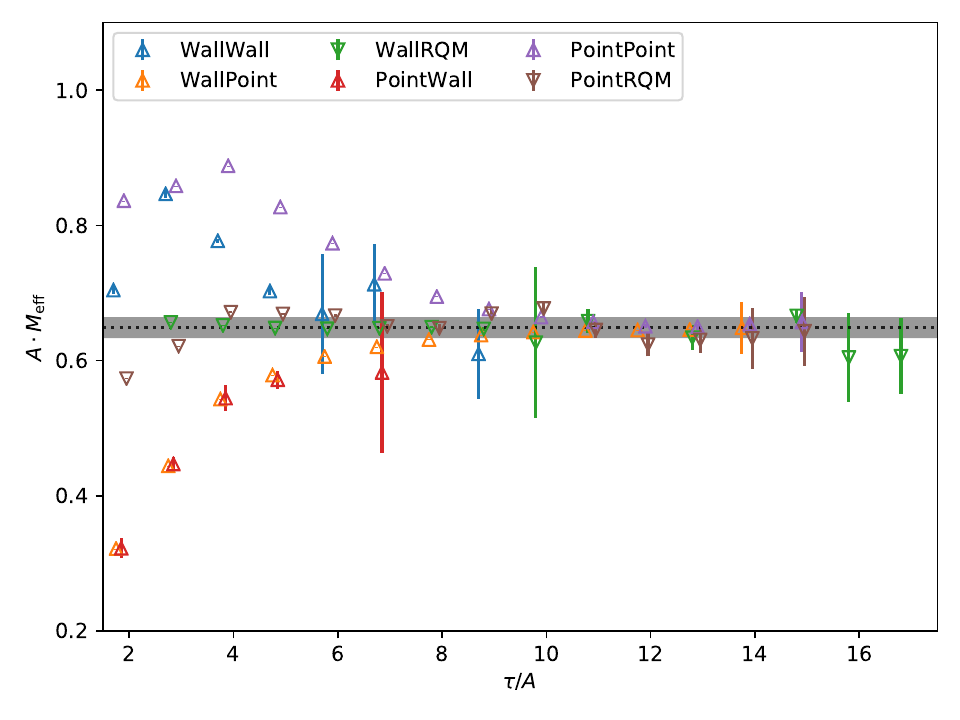}
\caption{\label{fig:effMass}
Effective mass plot for a heavy-strange meson correlation function (static limit for the heavy quark with 1 step of 4D hypercubic smearing~\cite{Hasenfratz:2001hp}, and the highly improved staggered quark discretization~\cite{Follana:2006rc} for the strange quark) with different source and sink operators.
The source operators are either \emph{Point}-type (namely, random Z(2) noise sources with no momentum projection) or \emph{Wall}-type (namely, corner-wall sources, with momenta $p_{i} = 0$~or~$\pi/A$).
The sinks are either \emph{Point} sinks, \emph{Wall} sinks or relativistic quark model inspired wave-functions obtained with the code discussed here.
Simulations were conducted in (2+1+1)-flavor QCD with physical sea and lattice spacing $A \approx 0.15$~fm using MILC ensembles~\cite{Bazavov:2017lyh} and the publicly available MILC code~\cite{MILC}.\\
The dotted line shows the average of the data obtained using the wave-function sink and the gray band shows the respective standard deviation.
It is clearly visible how much excited state suppression can be gained by using wave-functions at the sink.}
\end{figure}

In Fig.~\ref{fig:effMass}, we show the effective mass of the reduced negative parity correlator with a pseudoscalar static-strange hadron interpolating operator computed on the coarsest lattice of the MILC data set, i.e., $32^{3} \times 48$ with a lattice spacing of $A \approx 0.15$~fm and all sea quark masses at their physical values, after one step of HYP smearing is applied to the links making up the static quark propagator.
The lattice simulations were performed using the MILC code~\cite{MILC}.
In the following, we discuss the combinations of interpolating operators shown in the figure.

On the quark level, a \emph{Point}-type operator (at source or sink) has non-zero overlap factors with quark-antiquark states for all allowed momenta at uniform weights.
However, since low-lying, single hadron states have the largest overlap with quark-antiquark states, where both quark and antiquark have relatively small momenta, the \emph{Point} operators exhibit large overlap with high energy states and comparatively poor overlap with the lowest hadron states of interest.
Instead of using a \emph{Point} source utilizing only a single site, random-color-wall sources realize statistically independent gaussian $Z(2)$-noise (in all three color components) on all sites at the same time.
Hence, with (sufficiently large number of simultaneous) random color-wall sources for quark and antiquark, any combinations from different sites for quark and antiquark cancel eventually, and one projects the hadron state to vanishing spatial momentum, while the individual quark (or antiquark) may still cover the full range of allowed momenta.
The \emph{Point} sink is realized by a simple summation over all spatial sites at the same time (propagator endpoints), and thereby achieves the same projection at the sink.
In Fig.~\ref{fig:effMass}, such source and sink operators are denoted as \texttt{Point}.
On the quark level, a \emph{Wall}-type operator (at source or sink) has non-zero overlap factors with quark-antiquark states only if their momenta are at a few specific values (the Fermi points of the reciprocal lattice).
Since the quark and antiquark fields at different sites transform differently under local gauge transforms, one either has to guarantee gauge invariance by connecting them with Wilson lines or by fixing an appropriate gauge such that the Wilson lines can be neglected.
Since Wilson lines would introduce path dependence, it is common to simply fix a Coulomb gauge before constructing \emph{Wall} operators on the lattice.
When implementing \emph{Wall} operators for staggered quarks it is necessary to pay attention to the staggered taste structure.
In particular, non-trivial separations between quark and antiquark must be decomposed into one part that is an integer multiple of a unit hypercube and another part within a unit hypercube.
Non-trivial separations of the latter type (within a unit hypercube) would generate hadrons with non-trivial spin-taste structure.
If a variety of such non-trivial separations is included at both source and sink, this would lead to a correlation function, whose spectrum contains many states with similar masses differing only by discretization artifacts of order $A^2$, which are very challenging to disentangle numerically.
In particular, a common \emph{Wall}-type operator that does not pay heed to the taste structure is called an even-and-odd wall operator.
If an even-and-odd wall operator is combined with any taste-filtering operator at the other end (such as \emph{Point}-type) of the correlator, the latter filters out the unwanted tastes, whose remnants only show up as contributions to the gauge noise.
However, in a symmetric correlator with even-and-odd wall at both source and sink, the additional states due to the unwanted staggered tastes would spoil any attempts to efficiently solve the generalized eigen-value problem.
Thus, it is common practice to use so-called corner-wall sources that utilize only the lower corner of each spatial unit hypercube, i.e., $(0,0,0)$, which project to spatial momenta where all three components are $0$ or $\pi/A$.
These corner-wall operators eliminate the contributions from the other staggered tastes in the valence sector.
In this recent application corner-wall sinks -- the naive analog of this prescription applied at the sink -- have been used for the first time (to the authors' knowledge).
It is irrelevant whether the \emph{Wall} operator is applied at the sink to the quark or to the antiquark or to both, since it is followed by a summation over all spatial sites at the same time (propagator endpoints).
On the contrary, an application of a \emph{Wall} operator at the source yields different results whether it is applied either to the quark or to the antiquark or to both, since it is not followed by a summation over all spatial sites.
The latter is the "correct", symmetric choice for the generalized eigen-value problem.
In Fig.~\ref{fig:effMass}, such source and sink operators are denoted as \texttt{Wall}.
Finally, the last type of hadron interpolating operators considered here is the \emph{RQM}-type.
In this case, a real-valued bound state wave-function of the positive parity ground state in a relativistic quark model is directly read into a complex field from an ASCII table (in the format: $x_1 \text{\textvisiblespace} x_2 \text{\textvisiblespace} x_3 \text{\textvisiblespace} \Re(\Psi) \text{\textvisiblespace} \Im(\Psi)$) and then convoluted with the antiquark propagator, before it is contracted with the quark propagator at the sink, and all spatial sites at any given sink time are summed over.
For this reason, it does not matter, whether the wave-function is convoluted at the sink with the quark or with the antiquark.
The requirement of ensuring gauge-invariance with Wilson lines or through gauge fixing carries over from the discussion of \emph{Wall} operators.
In principle, the same amount of attention with regard to staggered taste structure as in the case of \emph{Wall} operators is in order here.
As soon as the desired taste structure is filtered out at the source, the omission of isolating the desired taste structure at the sink does not lead to non-vanishing contributions from the other tastes (although they still contribute to the gauge noise).
In the presented correlators, taste-filtering on the level of \emph{RQM} sink operator has been ignored, which leads to the larger gauge noise in comparison to the correlators that are taste filtered at both source and sink.
While it is not implemented for this data set, the \emph{RQM} wave-functions could also be used directly as source operators.
However, similar to the case with \emph{Wall} operators at the source, one has to distinguish between applying the \emph{RQM} operator to the "light" antiquark or to the "heavy" static quark (the latter being computationally advantageous, see Ref.~\cite{Duncan:1994uq}), since there is no spatial summation involved.
Moreover, one has to take care of enforcing the appropriate taste filtering.
Hence, the "correct", symmetric choice for the generalized eigen-value problem, would see a taste-filtered \emph{RQM} operator at the source for either propagator contracted with a corner-wall operator at the source for the other propagator.

The \emph{RQM} wave-functions are obtained by first computing the static quark-antiquark correlator on the same gauge ensemble, then using its effective mass at time $t \approx 0.5$~fm as an external potential in the \texttt{quantumfdtd~v3} code, extending it for large distances with a Cornell-type fit and flattening it off as a constant at even larger distances to mimic the string breaking (see subsection~\ref{subsec:external_potentials}).
Using the same type of box geometry, i.e., $32^{3}$, a similar lattice spacing, i.e., $A = 0.15$~fm, and a "light" constituent quark mass of $M = 300$~MeV with the relativistic kinetic term (see subsection~\ref{subsec:kin_terms}), the ground state wave-function is obtained using the \texttt{quantumfdtd~v3} code.
While this wave-function decays rapidly, it is still large enough to remain sensitive to the (periodic) boundary conditions if a physical 3D volume similar to the one available in typical 4D QCD lattice calculations is employed.
In order to prevent the operator of the RQM wave-function adding with the contributions across the boundaries, as one would encounter in these usually rather small 3D volumes, the wave-function's amplitude at larger distances is artificially suppressed to reduce the impact of finite size effects, before the wave-function is converted into a MILC-code readable file format and fed back into the lattice QCD simulation for convolution at the sink.
An ad-hoc form-factor, $\text{FF}$, is applied to the wave-function for $r > \text{\texttt{POTFLATR}}$ (see the definition in subsection~\ref{flag:POTFLATR} on page~\pageref{flag:POTFLATR}), that makes the potential flat for $r > \text{\texttt{POTFLATR}}$.
The form-factor $\text{FF}$ is defined as
\begin{equation}
\label{eq:formfactor}
\text{FF} = 10^{-50 \cdot \left(\frac{r}{\text{\texttt{POTFLATR}}} - 1\right)} \,.
\end{equation}
Furthermore, if $\text{FF} < 10^{-10}$, the wave-function is set to $0$.

In Fig.~\ref{fig:effMass}, this type of sink operator is denoted as \texttt{RQM}.
It is apparent from the figure, that the combination of a \texttt{Wall} at the source with an \texttt{RQM} at the sink eliminates the excited state contamination to an \emph{extraordinary} extent.
Whereas the excited state suppression is not as thorough with a \texttt{Point} at the source, it can still be of a great benefit when it comes to projecting out the ground state.
It appears likely that the use of a combination of such \emph{Wall} and/or \emph{RQM} operators at the two endpoints of a three-point function with a \emph{Point}-like operator for the external current and the corresponding two-point function by which it is divided, could make a \emph{major difference} for the excited state suppression and the accuracy of the QCD matrix elements that could be achieved with the \emph{same} underlying gauge ensembles.

In this context, the question of heavy-light tetraquarks also looms large.
A three-dimensional relativistic Schrödinger equation solver may be able to deliver interpolating operators for first-principles lattice QCD studies of tetraquarks that are able to enhance overlap factors for certain geometries\footnote{%
The present version of the \texttt{quantumfdtd} code is not particularly well suited for an extension to systems with two light degrees of freedom interacting with each other in the background of another potential, which would be needed if the two light quarks were not represented as a light diquark in the model.
Nonetheless, the present code could be straightforwardly extended to enable computations with two light degrees of freedom that interact with each other and are symmetrized or anti-symmetrized in the overall wave-function according to their spin statistics.} (i.e., spherical symmetry in a light-diquark heavy-antidiquark system, or cylindrical symmetry in a light-diquark heavy-antiquark heavy-antiquark system) of tetraquarks, while effectively suppressing the excited states.
Such an approach might make the resolution of the geometric structure of tetraquarks feasible and distinguish between a wide range of possible scenarios.

\subsection{Overview and structure of the paper}
\label{subsec:overview}

In this work, we extend the original \texttt{quantumfdtd} code~\cite{Strickland:2009ft, Margotta:2011ta}.
The new version, like the old version, allows one to compute the ground-state and low-lying excited state three-dimensional (3D) wave-functions and their corresponding energies for both real- and complex-valued potentials.
We implement a new relativistic kinetic term and add two new non-relativistic ones, both of which are Fast Fourier Transform (FFT) based and implemented using the parallelized \texttt{FFTW~3}~library.
In addition, the original code has been extended so that it supports arbitrary external potentials and the option to project out distinct parity eigen-states from the solutions.
These options also work with the original implementation of the finite-difference time-domain (FDTD) solver for the non-relativistic Schrödinger equation.
Because of the original code design, we use a 3-D Cartesian, homogeneous and isotropic lattice with $N$ points in each dimension and lattice spacing $A$ (in units of $\text{GeV}^{-1}$).
No adaptive engine has been implemented.
The solver depends on the \texttt{MPI}~library to distribute the lattice between the \texttt{MPI}~processes.
Finally, we note that by construction, the original FDTD solver for the non-relativistic Schrödinger equation used Dirichlet boundary conditions.
For the new kinetic terms, based on the FFT, periodic boundary conditions are used.
The potential can be centered either in the lattice volume at the off-grid point $(N_H,N_H,N_H)$ (with $N_H = (N+1)/2$) or in the origin at the grid point $(0,0,0)$.
We use the original iterative solver (see Ref.~\cite{Strickland:2009ft} and section~\ref{subsec:iterative_procedure} where we review the method).

This paper is organized as follows:
In section~\ref{sec:rel_kin_term} we introduce the newly implemented kinetic terms (subsection~\ref{subsec:kin_terms}), the iterative procedure (subsection~\ref{subsec:iterative_procedure}), and the parity projection scripts (subsection~\ref{subsec:parity_projection}).
In section~\ref{sec:examples} we show and compare results obtained using the different kinetic terms.
In section~\ref{sec:compile_run}, the program requirements are listed and basic instructions for compilation and running are provided.
In section~\ref{sec:params}, we show the basic configuration parameters of the program (subsection~\ref{subsec:basic_params}) and list the different initial conditions (subsection~\ref{subsec:initial_conditions}) that are implemented.
We furthermore list the hard-coded potentials in subsection~\ref{subsec:hard_coded_potentials}, and present a newly implemented option that allows the user to load potentials from external files (subsection~\ref{subsec:external_potentials}).
The new post-processing scripts, which include parity projection scripts for wave-functions, are explained in the subsection~\ref{subsec:post_processing_scripts} and~\ref{subsec:post_processing_python_module}.
We end with a conclusion in section~\ref{sec:conclusion}.

\section{The new relativistic kinetic term}
\label{sec:rel_kin_term}

The Schrödinger Hamiltonian
\begin{equation}
H = H_{K} + V(\vec{r}) \,,
\end{equation}
can be generally\footnote{%
In the absence of dissipation and other non-potential effects.} split into a kinetic piece, $H_{K}$, and into a potential $V(\vec{r})$.
The non-relativistic kinetic term is given by
\begin{equation}
\label{eq:HKnr}
H_{K}^{nr} = \sum\limits_{i=1}^{3} \frac{p_{i}^{2}}{2M} \,,
\end{equation}
where $\vec{p} = (p_{1},p_{2},p_{3})$ is the spatial three momentum.
On the other hand, the relativistic term is given by
\begin{equation}
\label{eq:HKrel}
H_{K}^{rel} = \sqrt{M^2 + \sum\limits_{i=1,2,3} p_{i}^{2}} \,,
\end{equation}
if neither spin nor antiparticle degrees of freedom are given consideration.

In the following subsections we describe the newly implemented (relativistic) kinetic terms.

\subsection{Kinetic terms}
\label{subsec:kin_terms}

There are three newly implemented kinetic terms, $H_{K}^{(i)}$, ($i = 1,2,3$), on top of the one based on the FDTD method, $H_{K}^{(0)}$, with Dirichlet boundary conditions ($\Psi(\text{boundary}) \equiv 0$).
Its action on a wave-function $\Psi = \Psi(\vec{r})$, ($\vec{r} = (Ax_{1},Ax_{2},Ax_{3})$ being spatial coordinates and $x_{1},x_{2},x_{3} \in \{0,1,\ldots, N-1\}$, and $\vec{\rho} = (x_{1},x_{2},x_{3})$ its analog in lattice units) is given by
\begin{equation}
\label{eq:HK0}
H_{K}^{(0)} \Psi = - \frac{1}{2M} \sum_{l = \pm 1}^{\pm 3} \frac{1}{2A}\frac{\Psi(\vec{r} + A\hat{e}_l) - \Psi(\vec{r})}{A} \,,
\end{equation}
where $\hat{e}_{l}$ denotes the (signed) unit vector in the $l$ direction ($l = \pm 1,\pm 2, \pm 3$), and we write $\Psi(\vec{r}) \equiv \Psi(Ax_{1}, Ax_{2}, Ax_{3}) \equiv \Psi(x_1,x_2,x_3)$ for convenience reasons.

In particular, the FDTD kinetic term $H_{K}^{(0)}$ is based on Taylor expansions on the lattice sites and assumes continuity and differentiability.
In momentum space this term corresponds to
\begin{equation}
\label{eq:HK0_momentum_space_1}
H_{K}^{(0)} \Psi(\vec{p}) = \frac{1}{2M} \sum\limits_{l=1}^{3} \frac{4}{A^2}\sin^{2}\left(\frac{A p_{l}}{2}\right) \Psi(\vec{p}) \,,
\end{equation}
where $A p_{l}$ is real-valued and satisfies $-\pi < A p_{l} \le \pi$.
These coefficients are not constrained to a finite set of momenta (due to the absence of periodic boundary conditions).
In the following we write $k_l \equiv A p_l$ for convenience.
Note that there are no fundamental limitations for a modified FDTD kinetic term with periodic boundary conditions, although it would require a substantial addition of structure encoding the communication across this boundary in the present version of the code.

The newly implemented kinetic terms depend on the FFT and, hence, use periodic boundary conditions, and have the momenta constrained to the eigenmodes of the finite box, i.e.,
\begin{equation}
\label{eq:eigenmodes}
k_{l} \equiv A p_{l} \in \{ -\pi+2\pi/N, \ldots, 0, 2\pi/N, 4\pi/N, \ldots \pi \} \,,
\end{equation}
for $l = 1,2,3$.
We note that the momentum space lattice is the reciprocal of the position space lattice.
The action of the new kinetic terms on a wave-function $\Psi = \Psi(\vec{r})$ ($\vec{r} = (Ax_{1}, Ax_{2}, Ax_{3})$ being spatial coordinates) is given in terms of the Fast Fourier Transform (FFT) and Inverse Fast Fourier Transform (IFFT) operators (realized through the \texttt{FFTW~3}~library) by
\begin{align}
\label{eq:HK1} & H_{K}^{(1)} \Psi = \frac{1}{2A^{2}M N^{3}} \cdot \IFFT\left[ \sum\limits_{l=1}^{3} \left(k_{l}\right)^{2} \cdot \FFT[\Psi] \right] \,, \\
\label{eq:HK2} & H_{K}^{(2)} \Psi = \frac{1}{2A^{2}M N^{3}} \cdot \IFFT\left[ 4\sum\limits_{l=1}^{3} \sin^{2}\!\left(\frac{k_{l}}{2}\right) \cdot \FFT[\Psi] \right] \,, \\
\label{eq:HK3} & H_{K}^{(3)} \Psi = \frac{1}{A N^{3}} \cdot \IFFT\left[ \sqrt{4\sum\limits_{l=1}^{3} \sin^{2}\!\left(\frac{k_{l}}{2}\right) + (AM)^{2}} \cdot \FFT[\Psi] \right] \,,
\end{align}
where the FFT and IFFT operations are defined via
\begin{align}
& \left(\FFT[\Psi]\right)(k_1,k_2,k_3) = \sum\limits_{x_1,x_2,x_3} \Psi \cdot \exp\left(-\i\sum\limits_{l=1}^{3} k_{l}\cdot x_l\right) \,, \\
& \left(\IFFT[\tilde{\Psi}]\right)(x_1,x_2,x_3) = \!\! \sum\limits_{k_{1},k_{2},k_{3}} \!\! \tilde{\Psi} \cdot \exp\left(\i\sum\limits_{l=1}^{3} k_{l}\cdot x_l\right) \,.
\end{align}

Due to the realization of the kinetic terms through the FFT, optimized communications across the inter-node boundaries, as done in the FDTD approach, are not required here.

The term $H_{K}^{(1)}$ is a discretized FFT implementation of the classical non-relativistic kinetic term $H_{K}^{nr}$, Eq.~\eqref{eq:HKnr}.
For this reason, the discretization errors of Eq.~\eqref{eq:HK0_momentum_space_1}, namely,
\begin{equation}
\label{eq:HK0_momentum_space}
\left[ H_{K}^{(0)} -H_{K}^{nr} \right] \Psi(\vec{p}) = \frac{1}{2M}\sum\limits_{l=1}^{3} \frac{2}{A^2}\sum\limits_{n=2}^{\infty} (-1)^{n} \frac{(A p_l)^{2n}}{(2n)!} \Psi(\vec{p}) \,,
\end{equation}
are completely absent in $H_{K}^{(1)}$.
Note that although $H_{K}^{(1)}$ formally seems to avoid the breaking of the $O(3)$ symmetry to the cubic group, this symmetry breaking is implicit in the FFT.
The $2\pi$ periodicity and restriction to the interval of Eq.~\eqref{eq:eigenmodes} is automatically realized in the FFT implementation.
Using the Symanzik effective field theory~\cite{Symanzik:1983dc, Symanzik:1983gh} the discretization errors of $H_{K}^{(0)}$ could be systematically reduced by including more extended stencils with adjusted coefficients, e.g.,
\begin{align}
\label{eq:HK0imp_momentum_space}
&\begin{aligned}
H_{K}^{(0), \text{imp}} \Psi(\vec{p}) &= \frac{1}{2M}\sum\limits_{l=1}^{3} \frac{4}{A^2} \left[ c_0\sin^2\left(\frac{A p_l}{2}\right)+ c_1\sin^2(A p_l) \right] \Psi(\vec{p}) \\
&= \frac{1}{2M}\sum\limits_{l=1}^{3} \frac{2}{A^2} \sum\limits_{n=1}^{\infty} \frac{c_0+2^{2n}c_1}{(2n)!} (Ap_l)^{2n} \Psi(\vec{p}) \,,
\end{aligned} \\
\label{eq:HK0imp_coordinate_space}
&\begin{aligned}
H_{K}^{(0), \text{imp}} \Psi(\vec{r}) &= \frac{1}{2M}\sum\limits_{l = \pm 1}^{\pm 3} \frac{1}{2A} \frac{c_0\left[\Psi(\vec{r}+A\hat{e}_l) - \Psi(\vec{r})\right] + c_1\left[\Psi(\vec{r}+2A\hat{e}_l) - \Psi(\vec{r})\right]}{A} \,.
\end{aligned}
\end{align}
The choice $c_0 = 4/3$, $c_1 = -1/12$ restores the leading term to its continuum coefficient and cancels the second term in the series.
Hence, there cannot be any local coordinate space representation of $H_{K}^{(1)}$, since infinitely many stencils up to infinite extent would be required to completely remove the discretization errors.

The term $H_{K}^{(2)}$ is an FFT implementation of the classical non-relativistic kinetic term $H_{K}^{(0)}$ for periodic boundary conditions.
Due to the different boundary conditions there are differences between those terms that vanish in the infinite volume limit.
However, the discretization errors of $H_{K}^{(0)}$ and $H_{K}^{(2)}$ are formally equivalent as they have the same coefficients in the Symanzik effective theory in the infinite volume limit.

The term $H_{K}^{(3)}$ is the newly implemented relativistic kinetic term, which realizes the most simple discretized version of Eq.~\eqref{eq:HKrel}.
It is inspired by a relativistic quark model (RQM)~\cite{Duncan:1994uq} and can be used by means of the \texttt{KINTERM} parameter (see the definition in subsection~\ref{flag:KINTERM} on page~\pageref{flag:KINTERM}).
There is no local or continuously differentiable coordinate space representation of the relativistic kinetic term, as is evident from the momentum dependent term under the square root.
Using Symanzik effective theory~\cite{Symanzik:1983dc, Symanzik:1983gh} the discretization errors of $H_{K}^{(3)}$ could be systematically reduced by including more extended stencils with adjusted coefficients just as in the non-relativistic case.
It is also straightforward to construct an relativistic analog to $H_{K}^{(1)}$.
None of these two obvious extensions have been pursued in this version of the program.

Note that the FDTD kinetic term, $H_{K}^{(0)}$, and each of the FFT kinetic terms, $H_{K}^{(i)}$, $i = 1,2,3$, are qualitatively different and, in practice, exhibit different numerical behavior.
On the one hand, comparing the results of $H_{K}^{(0)}$ and $H_{K}^{(2)}$ permits a quantitative study of finite volume errors while using a single volume, due to the different boundary conditions.
On the other hand, comparing the results of $H_{K}^{(1)}$ and $H_{K}^{(2)}$ permits a quantitative study of discretization errors while using a single lattice spacing, due to different coefficients in the Symanzik effective theory.
Nevertheless, the dependence of the low-energy spectrum on these differences is usually very small.
On the contrary, the relativistic kinetic term encodes qualitatively different physics already in the low-energy spectrum.

\subsection{The iterative procedure}
\label{subsec:iterative_procedure}

The iterative procedure is unmodified with respect to \texttt{quantumfdtd~v2}.
That is, it is the very same procedure as described in section~3 of Ref.~\cite{Strickland:2009ft},
\begin{equation}
\label{eq:iter_proced}
\Psi(\vec{r},\tau+\Delta\tau) = \mathcal{A} \Psi(\vec{r},\tau) - \mathcal{B} \Delta\tau H_{K} \Psi(\vec{r},\tau) \,.
\end{equation}
Equation~\eqref{eq:iter_proced} is the result of a Wick rotation of the Schrödinger Equation, $\i t \to \tau$.
The evolution with $\tau$ is given by Eq.~(2.5) of Ref.~\cite{Strickland:2009ft},
\begin{equation}
\label{eq:iter_decay}
\Psi(\vec{r},\tau) = \sum\limits_{i=0}^{\infty} a_{n}\Psi_{i}(\vec{r}) e^{-E_{i} \tau} \,,
\end{equation}
where $E_{i}$ is the $i$-th eigen-energy of the Hamiltonian and $\{a_0,a_1,\dots\}$ are the decomposition coefficients of the initial guess $\Psi(\vec{r},0)$ in the basis of eigen-vectors.

Each wave-function that is generated at each time step of the iterative procedure of Eq.~\eqref{eq:iter_proced} constitutes a \emph{snapshot}.
Before snapshots are computed, the wave-function is normalized in the full lattice volume, $V = (AN)^3$, as $\sum_{\vec{r} \in V} \lvert\Psi(\vec{r}\rvert^2 = 1$.
In order to compute excited states, overlaps between the last two wave-function snapshots are used.
These last two snapshots are temporarily kept in memory, and can also be saved to disk\footnote{%
See flag \texttt{SNAPUPDATE} in subsection~\ref{flag:SNAPUPDATE} on page~\pageref{flag:SNAPUPDATE}.}.
However, storing many of them consumes a lot of disk space.
Instead of that, saving a projection of the wave-functions can be triggered.\footnote{%
See flag \texttt{DUMPSNAPS} in subsection~\ref{flag:DUMPSNAPS} on page~\pageref{flag:DUMPSNAPS}.}
Note that, at some point $e^{-E_{i} \tau}$ will become smaller than the machine precision.
If we consider the decomposition of a wave-function in an eigen-energy basis, the evolution of each component $\Psi_{i}(\vec{r},\tau)$ with $\tau$ is given by Eq.~\eqref{eq:iter_decay}.
Hence, it decays as $\sim e^{-E_i\tau}$, where $E_i$ is the eigen-energy associated with the respective eigen-state.
The iterative procedure keeps $\sum_{\vec{r} \in V} \lvert\Psi^2(\vec{r},\tau)\rvert = 1$, for all $\tau$ by renormalizing the wave-function after a certain number of iteration steps.
While the ground state component remains large, the excited state components decay exponentially compared to the ground state.
That is, $\sum_{\vec{r} \in V} \lvert\Psi_i^2(\vec{r},\tau)\rvert/\sum_{\vec{r} \in V} \lvert\Psi_0^2(\vec{r},\tau)\rvert \sim e^{-(E_i-E_0)\tau}$.
Hence, the characteristic time $\tau_{\text{decay},\,i}$ during which an excited state $i$ vanishes during the iterative procedure is
\begin{equation}
\tau_{\text{decay},\,i} \sim \frac{-\log\epsilon}{E_i-E_0} \,,
\end{equation}
where $\epsilon$ is the computer-$\epsilon$.

For the question whether two excited states $\Psi_{i}(\vec{r})$ and $\Psi_{j}(\vec{r})$ can be separated in the iterative (or overlap) procedure, their energy difference $\lvert E_i-E_j \rvert$ has to be sufficiently large, or they have to be distinguishable in terms of another quantum number by which they can be separated.
Hence, if all snapshots are taken at large values of $\tau$, the only wave-function that can be recovered is the ground state and the overlap procedure will not be able to recover any excited states unless snapshots at much smaller times are included as well, as we will see in detail in the examples in the next section.
The logic behind this iterative procedure is the assumption that at the latest available time any contamination from excited states is already negligible, such that the estimate given by this large-time wave-function is a good representation of the ground state.
The next step in this line of reasoning is that the preceding snapshot contains additional non-negligible contributions of only one excited state.
An estimate of this excited state can then be extracted once the projection to the estimate of the ground state has been removed.
Finally, the iterative procedure assumes that the second excited state is non-negligible at the next earlier time snapshot, removes the projections to the estimates of all lower states, and takes the remaining wave-function as the estimate of the second excited state.
In principle this logic could be iterated successively in order to get one excited state after the other, as long as their energies differ by a sufficient amount, such that there is only one additional state at each earlier snapshot.
In practice, isolating more than two excited states is very often infeasible.

Unless the ground state is successfully isolated at the latest time, its estimate will be a linear combination of it with one or a few of the lowest excited states instead.
Thus, the estimate of the excited states are in turn also linear combinations of the ground state and a similar or even larger set of excited states.
This is the reason why none of the states is correctly identified at small simulation times.
Only if the simulation runs long enough to successfully isolate the ground state, reliable results for any of the other states become possible, too.
In order to get accurate results for the excited states, one snapshot has to be available in each time interval between the freezeout times of any two excited states.
Since the code internally uses the two latest snapshots, better projections could be obtained during post-processing if the snapshot times are chosen in a manner particularly suitable for the energies of the system in question.

We improve the previous version of the code by adding the option of using a kinetic term $H_{K}^{(i)}$, $i = 1,2,3$, based on the FFT scheme of subsection~\ref{subsec:kin_terms}, instead of the one based on finite differences, $H_{K}^{(0)}$.
In all cases, the values of $\mathcal{A}$ and $\mathcal{B}$ are taken to be the same as in~\cite{Strickland:2009ft},
\begin{equation}
\mathcal{A} = \frac{1 - \frac{\Delta\tau}{2} V_{b}(\vec{r})}{1 + \frac{\Delta\tau}{2} V_{b}(\vec{r})} \,,\quad \mathcal{B} = \frac{1}{1 + \frac{\Delta\tau}{2} V_{b}(\vec{r})} \,.
\end{equation}
The time step $\Delta\tau$ is an independent parameter that should be chosen as small as necessary to reduce the discretization errors of the time evolution and achieve good convergence, while being as large as permissible to minimize the computational expense.
Both of these limitations depend on the combination of the kinetic term and the potential term.

\subsection{Finite volume normalization}
\label{subsec:vol_norm}

Close to the edges, the wave-functions are always affected by severe finite volume effects, albeit differently for Dirichlet or periodic boundary conditions.
Namely, the same distance $r$ realized by a separation along any one of the axes is much closer to the boundary than any realization in terms of an off-axis separation, e.g., $r = 3n \cdot A$ implemented via $\vec{r} = (3,0,0)n \cdot A$ vs. $\vec{r} = (2,2,1)n \cdot A$.
On the level of post-processing scripts\footnote{%
See subsections~\ref{subsec:post_processing_scripts}~and~\ref{subsec:post_processing_python_module}.} we use a parameter $\texttt{min\_\allowbreak{}sep\_\allowbreak{}edge}$ to remove contributions from sites closest to the boundary from expectation values, and restrict wave-functions to an \emph{effective lattice volume} $V^{\Box} \equiv ((1-\texttt{min\_\allowbreak{}sep\_\allowbreak{}edge})AN)^3$.
This finite volume normalization in the effective lattice volume, $V^{\Box}$, or \emph{box normalization} for short, is different from the full volume normalization used within the \texttt{quantumfdtd~v3} code, cf., Sec.~\ref{subsec:iterative_procedure}.
Instead it is defined as
\begin{equation}
\sum\limits_{\vec{r} \in V^{\Box}} \lvert\Psi(\vec{r})\rvert^2 = 1 \,.
\end{equation}
We provide a script that normalizes the wave-function by enforcing $\sum_{\vec{r} \in V^{\Box}} \lvert\Psi(\vec{r}\rvert^2 = 1$, within the effective lattice volume $V^{\Box}$.
As a very simple and intuitive measure for the sensitivity to finite volume effects we employ the average radii~in this box normalization, $\langle \rho \rangle^{\Box}$, that are computed after normalizing the wave-functions (obtained through the iterative procedure) in the effective lattice volume $V^{\Box}$,
\begin{equation}
\label{eq:box_normalization}
\langle \rho \rangle^{\Box} = \sum\limits_{\vec{r} \in V^{\Box}} \rho(\vec{r}) \cdot \lvert \Psi(\vec{r}) \rvert^{2} \,.
\end{equation}
Unless $\langle \rho \rangle^{\Box} \ll N/2$ we have to expect significant distortions of the wave-function due to finite volume effects.

\subsection{Parity projection}
\label{subsec:parity_projection}

We include scripts for computing the positive, $P^{+}$, negative, $P^{-}$, and negative-around-an-axis, $P^{-}_{p_{k}}$, parity projections of the wave-function,
\begin{align}
\label{eq:parity} & P^{\pm} \Psi(\vec{r}) = \frac{1}{2} \left(\Psi(\vec{r}) \pm \Psi(-\vec{r})\right) \,, \\
\label{eq:parity_axis} &\begin{aligned}
P^{-}_{\vec{p}_{k} = \hat{e}_{3}} \Psi(x_{1},x_{2},x_{3}) & = \frac{1}{2} \left[ \Psi(x_{1},x_{2},x_{3}) + \Psi(-x_{1},-x_{2},x_{3}) \right. \\
& \left. - \Psi(x_{1},x_{2},-x_{3}) - \Psi(-x_{1},-x_{2}-x_{3}) \right] \,.
\end{aligned}
\end{align}
Here, we have chosen $\vec{p}_{k} = \hat{e}_{3} = (0,0,1)$, although the script accepts $\hat{e}_{1}$ and $\hat{e}_{2}$ as well.
The aim is to separate the positive parity $s$-waves from the negative-parity $p$-waves.
For the $p$-wave states, the negative-around-an-axis, $P^-_{\vec{p}_k}$ separate the states with different magnetic quantum number $m \in \{0,\ \pm 1\}$.

The weight of a parity projection $P$ is defined as
\begin{equation}
\langle \Psi \lvert P \rvert \Psi \rangle = \sum\limits_{\vec{r} \in V^{\Box}} \lvert P\Psi(\vec{r}) \rvert^2 \,.
\end{equation}
Note that within the effective lattice volume $V^{\Box}$, $\langle \Psi \lvert P^+ \rvert \Psi \rangle + \langle \Psi \lvert P^- \rvert \Psi \rangle = 1$ because of box normalization (unitarity) and the fact that parity projections are orthonormal.
In this sense, we talk about the \emph{weight} of a parity projection $P$ as being one particular projection.

It is straightforward to extend these scripts for projection to the higher multipoles, if the need should arise.

\section{Examples}
\label{sec:examples}

\subsection{Studied cases}
\label{subsec:cases}

\begin{figure}[ht]
\centering
\ig{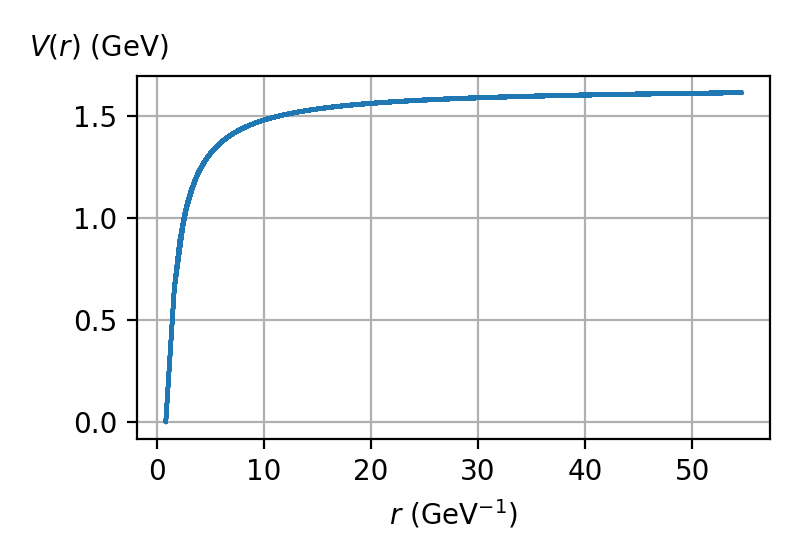}\hfill
\ig{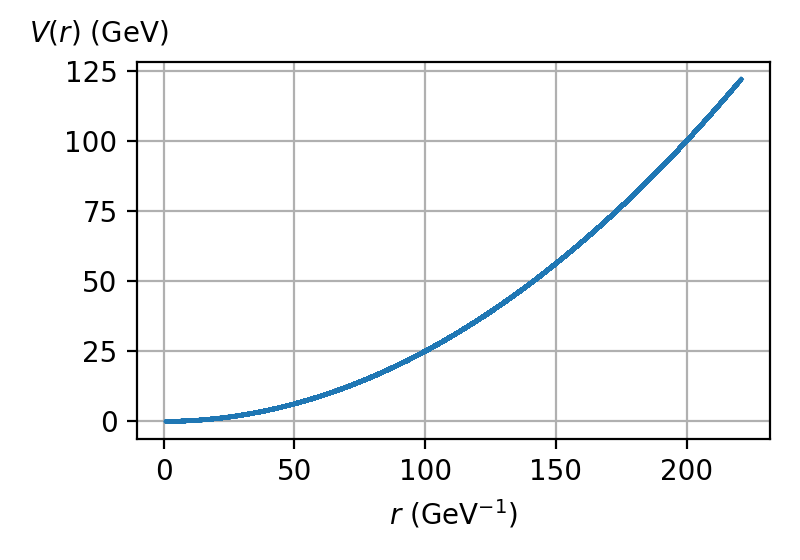}
\caption{\label{fig:pots}
Left: The Coulomb potential used for the following analysis.
The lattice spacing is $A = 0.12$~fm.
We use $N = 64$ points in each spacial direction, and the potential is centered in the lattice volume at the point $(N_H,N_H,N_H)$ with $N_H = (N+1)/2 = 32.5$.\\
Right: The harmonic oscillator potential used for the analysis in subsection~\ref{subsec:rel_kin_term_harmonic}.
We use $N = 256$ points in each spatial direction, and the potential is centered in the lattice volume as well.}
\end{figure}

We discuss the convergence of the iterative procedure (see the previous subsections~\ref{subsec:kin_terms} and~\ref{subsec:iterative_procedure}) in the following.
Furthermore, we study the performance of the overlap method for separating excited states (see subsection~\ref{subsec:iterative_procedure}), as well as the behavior of the parity projections (see subsection~\ref{subsec:parity_projection}).

We first use the Coulomb-like potential (see subsection~\ref{subsec:hard_coded_potentials}) as our benchmark, setting \texttt{POTENTIAL=2}, as it is shown in the left panel of Fig.~\ref{fig:pots}.
This Coulomb potential diverges outside of the grid at $(N_H,N_H,N_H)$ with $N_H = (N+1)/2$, i.e., in the center of the unit cell at $(N/2,N/2,N/2)$.
Alternatively, one could implement a Coulomb potential centered on the grid at $(0,0,0)$ via \texttt{POTENTIAL=102};
the regularization of the divergence at its center gives rise to a discretization artefact resulting in suppressed contributions from distances of the order of the lattice spacing, $r \sim A$.
Different paths bridging the same radial distance may belong to different representations of the cubic group, i.e., $(3,0,0)$ or $(2,2,1)$, which both correspond to $r/A = 3$.
Hence, the derivatives along these different paths still exhibit a substantial roughness as functions or $r/A$, which, however, reduces rapidly for larger $r/A$.
This is the dominant discretization artefact in all of those following examples.
Discretization artefacts can be parameterized through a series in dimensionless products of the lattice spacing and any other parameter of non-trivial mass dimension.
We set the lattice spacing $A = 0.12$~fm and use a reduced mass $M = 0.3$~GeV with no other dimensionful parameters; hence, the discretization artefacts can be understood in terms of factors of a mixed series in $A\cdot M$ or $A/r$, where the latter are relevant only if $r$ is not much larger than $A$.
Because each kinetic term is an even function of $A$, the discretization artefacts are restricted to even terms in $A$, too.
Due to $(A\cdot M)^2 \approx 0.03$, we expect mild finite mass discretization errors; discretization errors of the form $(A/r)^2$ are negligible in comparison for $r/A \gg 6$.
Along these lines we use the Coulomb-like initial conditions, setting \texttt{INITCONDTYPE=2} (see subsection~\ref{subsec:initial_conditions}).

As a second example we use the harmonic oscillator potential together with the relativistic kinetic term, setting \texttt{POTENTIAL=4} as it is shown in the right panel of Fig.~\ref{fig:pots}.
We keep the Coulomb-like initial conditions (\texttt{INITCONDTYPE=2}) but we reduce the lattice spacing to $A = 0.006$~fm while increasing the number of points in each direction to $N = 256$ and we also increase the mass to $M = 30$~GeV which results in $(A \cdot M)^{2} \approx 0.81$ which should result in reasonably small finite mass discretization errors\footnote{%
In order to be able to compare to existing literature results we had to slightly change the potential and use a large mass.
See the discussion in subsection~\ref{subsec:rel_kin_term_harmonic} for more details.}.
Note that in contrast to the Coulomb potential, the harmonic oscillator potential has no divergence at its origin.

In all of the cases, we do not plot points that are closer than $0.15N$ to one of the edges and use the box normalization corresponding to $\texttt{min\_\allowbreak{}sep\_\allowbreak{}edge} = 0.3$\footnote{%
See the definition of the \texttt{load\_wf} function in subsection~\ref{flag:min_sep_edge} on page~\pageref{flag:min_sep_edge})}, in order to avoid zones where finite volume effects are prominent, see subsection~\ref{subsec:vol_norm}.
We compare -- wherever possible -- to the theory expectation obtained from a continuum, infinite volume wave-function integrated in the same effective lattice volume $V^{\Box}$.

We plot the positive parity projection $P^+$, Eq.~\eqref{eq:parity}, and the negative parity around-an-axis $P^-_{\vec{p}_k}$, Eq.~\eqref{eq:parity_axis}.
We do not plot a projected wave-function if its weight is less than $10^{-5}$.
However, we need to stress that the overlapping procedure used for extracting ground and excited states out of the iterative procedure (see subsection~\ref{subsec:iterative_procedure}) can also introduce numerical artefacts affecting excited states.

This section is organized as follows:
We start with the Coulomb potential.
In subsection~\ref{subsec:rel_kin_term}, we study the convergence of the relativistic kinetic term $H_{K}^{(3)}$.
A selection of wave-functions at different $\tau$ is shown, as well as the evolution of the energies and the weights of the parity projections for each state.
In subsection~\ref{subsec:comp_non_rel_kin_terms}, we compare the non-relativistic kinetic terms $H_{K}^{(0)}$ (FDTD based), $H_{K}^{(1)}$, and $H_{K}^{(2)}$ (both FFT-based) to one another.
We additionally compare $H_{K}^{(0)}$ with $H_{K}^{(2)}$ at an increased lattice volume using $N = 128$.
In subsection~\ref{subsec:comp_high_mass}, we increase the reduced mass to $M = 3$~GeV, and compare the relativistic kinetic term $H_{K}^{(3)}$ with the non-relativistic one $H_{K}^{(2)}$.
In subsection~\ref{subsec:parity_fixing}, we study the parity fixing machinery of \texttt{quantumfdtd}, for both symmetrization and anti-symmetrization.
We use the non-relativistic kinetic term $H_{K}^{(0)}$ (FDTD) as an example.
We then switch to the harmonic oscillator potential and study the relativistic kinetic term $H_{K}^{(3)}$ in subsection~\ref{subsec:rel_kin_term_harmonic}.
Finally, in subsection~\ref{subsec:convergence_conditions}, we summarize the convergence conditions and numerical behavior of the program.

\subsection{The relativistic kinetic term \texorpdfstring{$H_{K}^{(3)}$}{H\_K(3)}}
\label{subsec:rel_kin_term}

\begin{figure}[t]
\centering
\ig{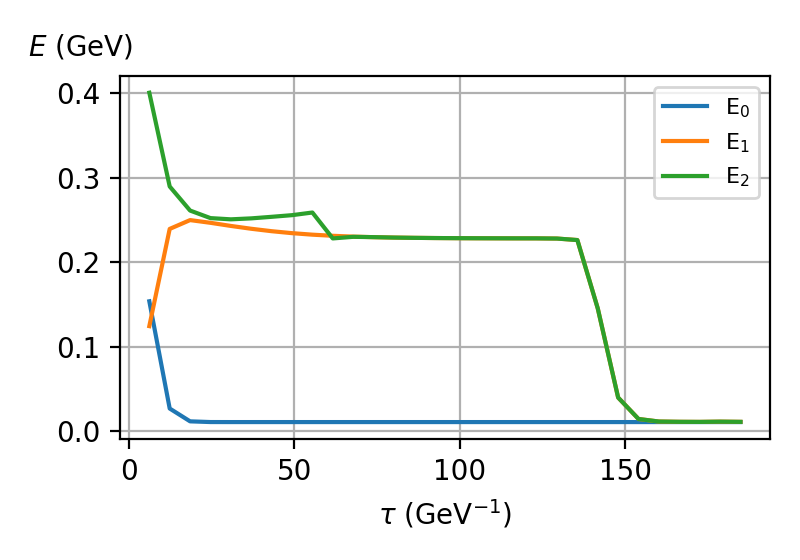}%
\hfill%
\ig{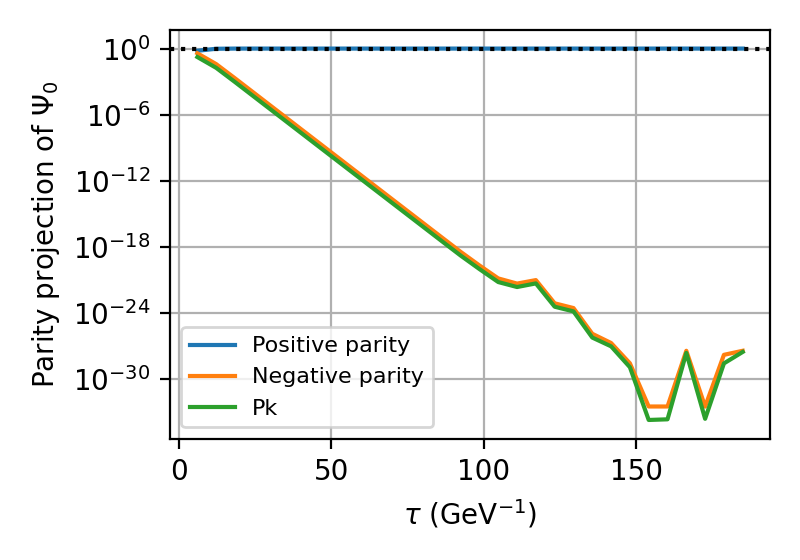}%
\\
\ig{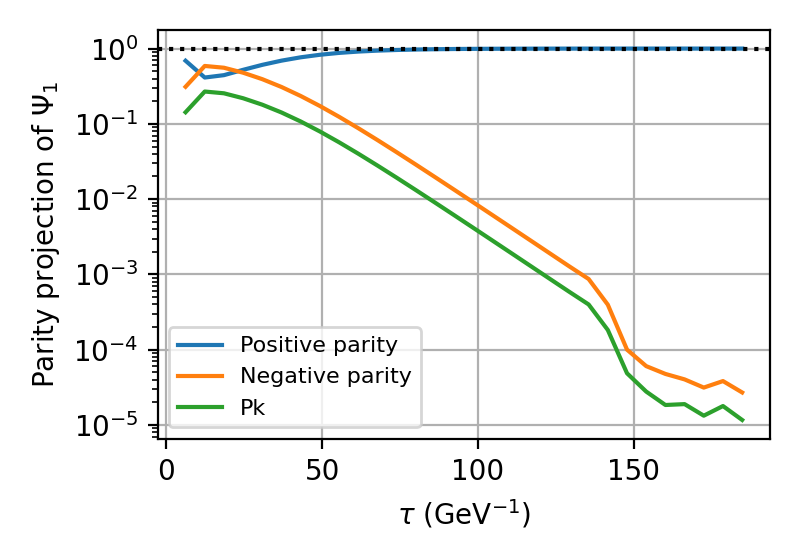}%
\hfill%
\ig{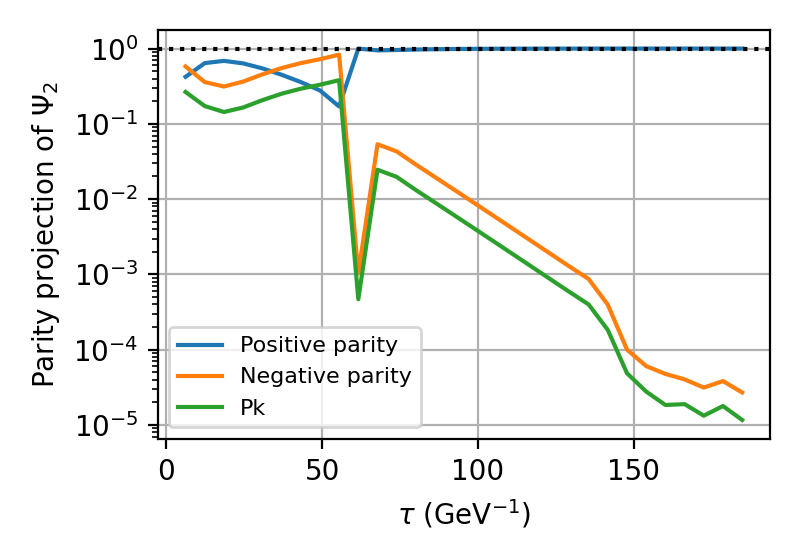}%
\caption{\label{fig:energ_proj_KIN3}
Energies obtained as a function of $\tau$ using the relativistic kinetic term $H_{K}^{(3)}$ and the Coulomb potential (top left).\\
Weights of the parity projection of the corresponding ground state (top right), and of the first (bottom left) and second (bottom right) excited state as a function of $\tau$.}
\end{figure}

\begin{figure}[t]
\centering
\igt{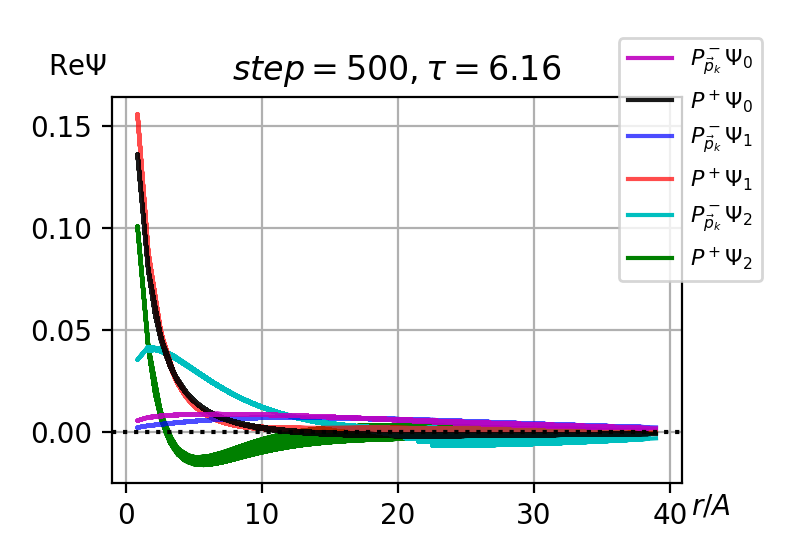}%
\hfill%
\igt{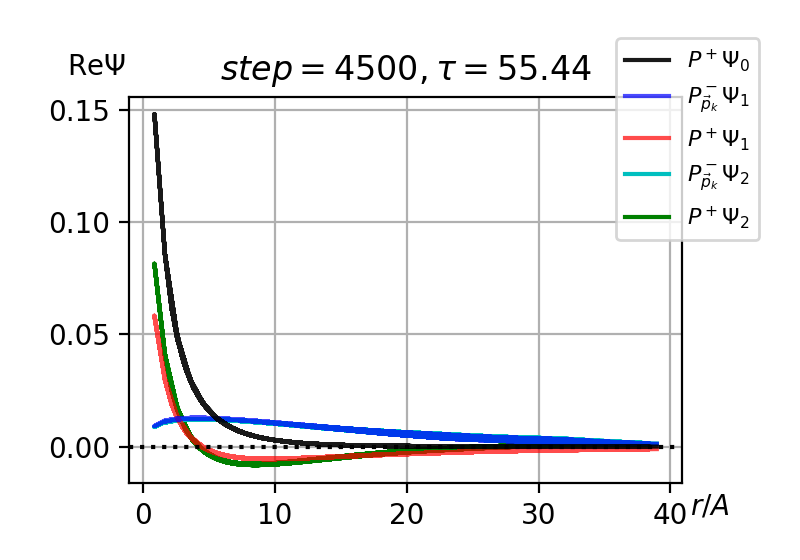}%
\hfil%
\igt{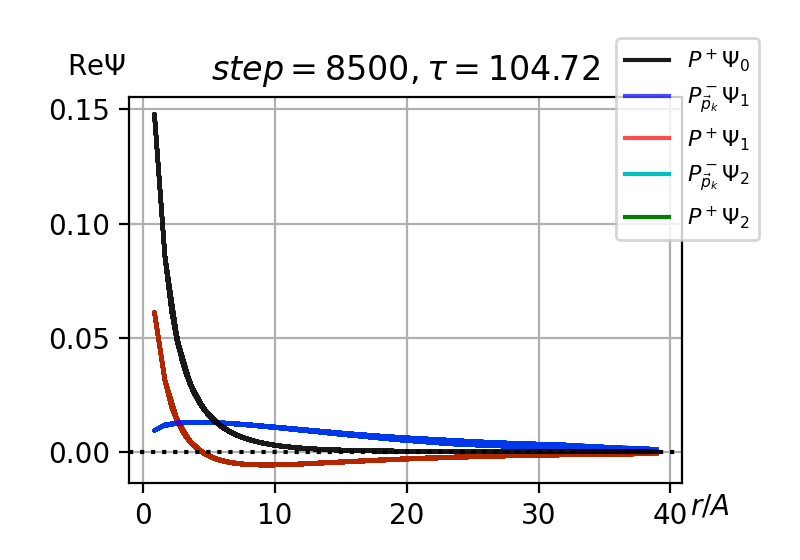}%
\caption{\label{fig:wf_KIN3}
Parity-projected wave-functions as a function of $r/A$ using the relativistic kinetic term $H_{K}^{(3)}$ at three different snapshot time steps.
After the last displayed time step the wave-functions do not change by eye which is supported by the values of the additional average radii at the later time step.
The average radii in units of $A$ are given by}
\small
\begin{tabular}{r|c|c|c|c|c|c}
& $P_{\vec{p}_{k}}^{-} \Psi_{0}$ & $P^{+} \Psi_{0}$ & $P_{\vec{p}_{k}}^{-} \Psi_{1}$ & $P^{+} \Psi_{1}$ & $P_{\vec{p}_{k}}^{-} \Psi_{2}$ & $P^{+} \Psi_{2}$ \\
\hline
$\langle \rho \rangle^{\Box}_{\text{left}}$ & $18.17$ & $6.11$ & $19.58$ & $4.97$ & $10.26$ & $8.78$ \\
\hline
$\langle \rho \rangle^{\Box}_{\text{middle}}$ & $-$ & $3.36$ & $15.95$ & $16.49$ & $16.28$ & $11.83$ \\
\hline
$\langle \rho \rangle^{\Box}_{\text{right}}$ & $-$ & $3.36$ & $15.79$ & $15.86$ & $15.79$ & $15.86$ \\
\hline
$\langle \rho \rangle^{\Box}_{\text{step $=10000$}}$ & $-$ & $3.36$ & $15.78$ & $15.87$ & $15.78$ & $15.87$
\end{tabular}
\end{figure}

In this subsection, the relativistic FFT-based kinetic term $H_{K}^{(3)}$, Eq.~\eqref{eq:HK3}, is studied together with a Coulomb potential.

We use $N = 64$ points in each direction and plot the convergence of the energies as a function of $\tau$ (top left plot in Fig.~\ref{fig:energ_proj_KIN3}) and the relative weights of the parity projections for the 3 lowest states versus $\tau$ (top right and bottom plots in Fig.~\ref{fig:energ_proj_KIN3}).
Finally we plot the parity projections of the wave-functions for three selected values of $\tau$ (Fig.~\ref{fig:wf_KIN3}).

Note that, for $\tau > 125$~GeV$^{-1}$, the weights of the excited states fall below the computational accuracy, $\epsilon$, of the computer.
For a 64 bit architecture, $\epsilon \sim 10^{-16}$.
Below this threshold we expect numerical artefacts which can be seen in the top right plot of Fig.~\ref{fig:energ_proj_KIN3}, where the exponential decay breaks for a weight of the minor component of the parity projection $\sim 10^{-21} \sim 10^{-16}/N^3$ ($N = 64$).
See also, on the top left side of Fig.~\ref{fig:energ_proj_KIN3}, that the energy of the first excited state decays to that of the ground state because the overlap method is no longer able to separate the first excited state from the ground state once numerical artefacts show up.

For large values of $\tau$, the energy of the \emph{excited states} decay to that of the ground state, whose energy eigen-value remains stable throughout the evolution after its correct value has been reached.

Note also, on the top left plot of Fig.~\ref{fig:energ_proj_KIN3}, a first crucial point at $\tau \sim (60-65)$~GeV$^{-1}$, where the energy of the second excited state, $E_2$, degenerates with the one of the first excited state, $E_1$.
This can also be seen in the relative weights of the parity projections of the second excited state shown in the bottom right plot of Fig.~\ref{fig:energ_proj_KIN3}.
Actually, we are seeing that the second excited state has fallen below the machine precision.
Hence, the only remaining states are the ground state and the first excited state.

For high values of $\tau$, but not high enough to hit the aforementioned cancellation issues, the parity projected wave-functions are stable as a function of $\tau$.
The negative parity projections around an axis, $P^-_{\vec{p}_k}$, of different separated states are compatible (they actually come from the lowest $P^-_{\vec{p}_k}$ odd state).
Additionally, there are 2 positive parity projections, $P^+$, the ground state and the first excited state in the $s$-wave.
This can be seen in Fig.~\ref{fig:wf_KIN3} where we plot the parity projected wave-functions for three different snapshot time steps.

We add an additional row corresponding to 10000 steps, i.e., $\tau \sim 123$~GeV$^{-1}$.
The wave-functions do not change by eye which is also supported by the values of the average radii at that later time step.
Their values lie well within the considered lattice volume such that we do not expect considerable finite volume effects.
Note also that, for $\tau > 100$~GeV$^{-1}$, these average radii become stable as a function of $\tau$~(see the table in the caption of Fig.~\ref{fig:wf_KIN3}).
The average radii of the parity projections, $P^{+}\Psi_1$ and $P^{+}\Psi_2$, are compatible.
Similarly for the parity projections, $P^{-}_{\vec{p}_k}\Psi_1$ and $P^{-}_{\vec{p}_k}\Psi_2$.
That is, the overlapping procedure of subsection~\ref{subsec:iterative_procedure} can separate the first excited states of each parity, but cannot identify the higher states.

\subsection{Comparison between the different non-relativistic kinetic terms}
\label{subsec:comp_non_rel_kin_terms}

\begin{SCfigure}[\sidecaptionrelwidth][ht]
\centering
\ig{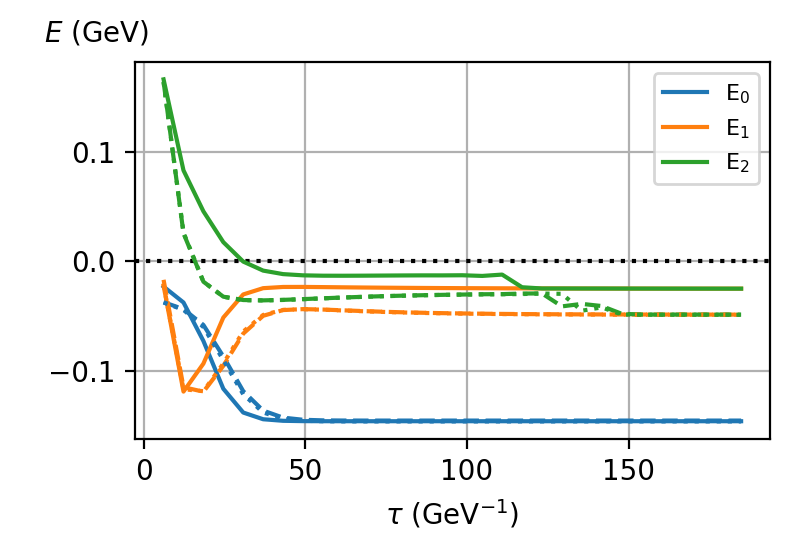}
\caption{\label{fig:comp_KIN_0_KIN_1_KIN_2}
Energies obtained as a function of $\tau$ using the non-relativistic kinetic terms $H_{K}^{(0)}$ (solid), $H_{K}^{(1)}$ (dashed), and $H_{K}^{(2)}$ (dotted) and the Coulomb potential.
The energies for the two operators with periodic boundary conditions ($H_{K}^{(1)}$ and $H_{K}^{(2)}$) are almost indistinguishable with the exception of $\tau \approx 125$~GeV$^{-1}$, where the degeneration of the second excited state with the first one due to reaching machine precision follows a slightly different pattern.}
\end{SCfigure}

We now compare the three non-relativistic kinetic terms introduced in subsection~\ref{subsec:kin_terms}:
$H_{K}^{(0)}$, Eq.~\eqref{eq:HK0}, based on the FDTD method with Dirichlet boundary conditions; $H_{K}^{(1)}$, Eq.~\eqref{eq:HK1}, based on a discretized FFT implementation of the continuum kinetic term $H_{K}^{nr}$, Eq.~\eqref{eq:HKnr}; and $H_{K}^{(2)}$, Eq.~\eqref{eq:HK2}, based on the discretized FFT implementation of the FDTD kinetic term.
Both $H_{K}^{(1)}$ and $H_{K}^{(2)}$ use periodic boundary conditions.

The eigen-energies obtained using the different kinetic terms are shown as a function of $\tau$ in Fig.~\ref{fig:comp_KIN_0_KIN_1_KIN_2}.
The lines that correspond to $H_K^{(1)}$ and $H_K^{(2)}$, respectively, are pretty much coincident, since they are both non-relativistic and FFT-based implementations of the kinetic term.
The only difference is a small lattice correction in the kinetic term (see Eqs.~\eqref{eq:HK1} and~\eqref{eq:HK2}).
The $H_K^{(0)}$ line, on the other hand, is based on the FDTD method for computing the (non-relativistic) kinetic term.

\label{discussion:boundary_conditions}
This method assumes Dirichlet boundary conditions (vanishing wave-function on the boundary) instead of periodic ones, so that disagreement with FFT-based methods is expected when finite volume effects are prominent, i.e., in states similarly large as the box.
This qualitative difference shows up as a significant deviation of the $H_K^{(0)}$ line from the $H_K^{(1,2)}$ ones.
The difference is small and basically negligible for the ground state, that stabilizes at $\tau \sim 50$~GeV$^{-1}$ and whose eigen-energy value, $E_{0}$, shows no discrepancy between $H_K^{(0)}$ and $H_K^{(1,2)}$ after that.
However, when considering the first excited state, the corresponding energy, $E_{1}$, stabilizes at different values for $H_K^{(0)}$ or $H_K^{(1,2)}$, respectively.
The second excited state's energy, $E_{2}$, decays to $E_{1}$ in all the cases, and it also shows the discrepancy between $H_K^{(0)}$ and $H_K^{(1,2)}$.
Note that the $1s$ ground state of the Coulomb potential is the one most concentrated around $r = 0$, and, hence, the one that should show less finite volume effects.
This implies that the differences between $H_K^{(0)}$ and $H_K^{(1,2)}$ due to different boundary conditions (finite volume effects) should be less prominent in the $1s$ state.
This is consistent with the fact that, in Fig.~\ref{fig:comp_KIN_0_KIN_1_KIN_2}, the $E_0$ lines corresponding to all different non-relativistic kinetic terms ($H_K^{(0,1,2)}$) are very close to one another.

In the wave-function plots shown below in Figs.~\ref{fig:wf_comp_KIN_0} to~\ref{fig:wf_comp_KIN_2}, we show with dashed lines the analytic solutions using the \emph{non-relativistic} kinetic term with a Coulomb potential, whereas the solid lines are those computed via the \texttt{quantumfdtd} program.
We use $N = 64$ points in each direction.
The analytic solutions are the well known eigen-states of the Hydrogen atom, $\Psi(\vec{r}) = R(r) Y(\Omega)$, with
\begin{equation}
\label{eq:non-rel-theo}
R_{1s}(r) = N_{1s} \, e^{-Mr} \,, \quad R_{2s}(r) = N_{2s} \, (2 - Mr) \, e^{-Mr/2} \,, \quad R_{2p}(r) = N_{2p} \, Mr \, e^{-Mr/2} \,,
\end{equation}
where $M$ is the reduced mass assuming unit charge and the $Y(\Omega)$ are the spherical harmonics.
The continuum normalization is given by
\begin{equation}
N_{1s}^{\infty} = 2M^{3/2} \,, \quad N_{2s}^{\infty} = \frac{M^{3/2}}{2\sqrt{2}} \,, \quad N_{2p}^{\infty} = \frac{M^{3/2}}{2\sqrt{6}}\,.
\end{equation}
The corresponding average radii in the continuum are defined through integrals as
\begin{equation}
\langle r \rangle^{\infty} = \int_{V} dV r \cdot \lvert R(r) Y(\Omega) \rvert^{2} \,, \quad \text{with } \int_{V} dV \lvert R(r) Y(\Omega) \rvert^{2} \overset{!}{=} 1 \,.
\end{equation}
Explicitly, they are given by
\begin{equation}
\label{eq:analytic_avg_radii}
\langle r \rangle^{\infty}_{1s} = \frac{3}{2M} \,, \quad \langle r \rangle^{\infty}_{2s} = \frac{6}{M} \,, \quad \langle r \rangle^{\infty}_{2p} = \frac{5}{M} \,.
\end{equation}
However, for the sake of comparability, we use the box normalization as defined by Eq.~\eqref{eq:box_normalization}.

Although the $1s$, $2p$, and $2s$ states can be recognized, and the wave-functions that can be extracted via parity projection from different \texttt{quantumfdtd} states (see subsection~\ref{subsec:parity_projection}) are compatible, there are some differences between the 3 kinetic terms, due to finite volume effects or due to discretization errors.
A study of the former is performed at the end of the section, where we increase the lattice volume to $N = 128$.

For the FDTD kinetic term, $H_{K}^{(0)}$, the most prominent finite volume effect is a faster than expected decay of the $2p$ and $2s$ wave-functions close to the borders of the lattice volume which can be seen in Fig.~\ref{fig:wf_comp_KIN_0}.
This is expected because of the Dirichlet boundary conditions.
The $1s$ wave-function decays exponentially, so that its theoretical value at the borders of the finite volume is also low and therefore, this wave-function is not as severely affected by the Dirichlet boundary conditions.

\begin{figure}[t]
\centering
\ig{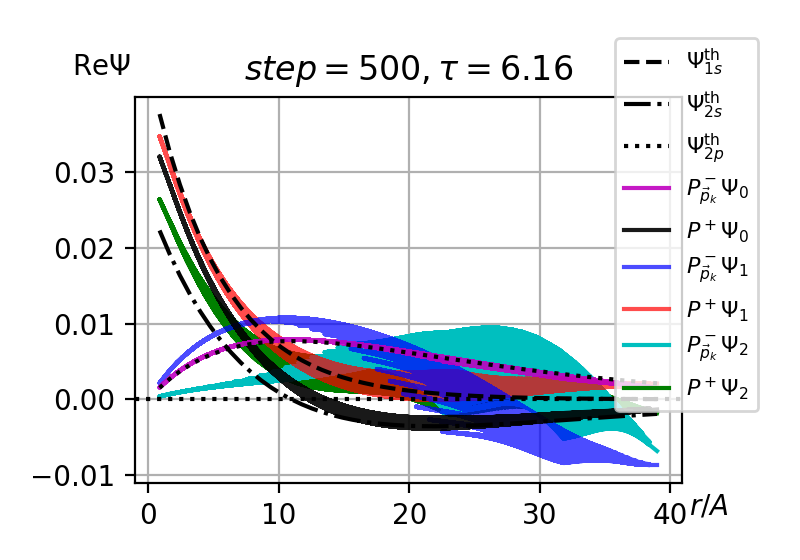} \hfill
\ig{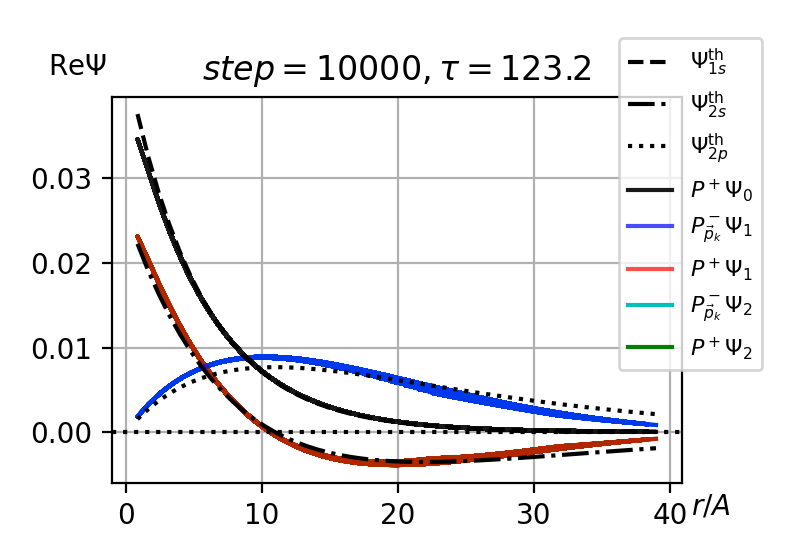}
\caption{\label{fig:wf_comp_KIN_0}
Parity-projected wave-functions as a function of $r/A$ using the non-relativistic kinetic term $H_{K}^{(0)}$.
The broad line of the $P^-_{\vec{p}_k}\Psi_2$ wave-function in the left plot is due to higher excited states that have not yet decayed (see subsection~\ref{subsec:iterative_procedure}).
The average radii in units of $A$ are given by}
\small
\begin{tabular}{r|c|c|c|c|c|c|c|c|c}
& $P_{\vec{p}_{k}}^{-} \Psi_{0}$ & $P^{+} \Psi_{0}$ & $P_{\vec{p}_{k}}^{-} \Psi_{1}$ & $P^{+} \Psi_{1}$ & $P_{\vec{p}_{k}}^{-} \Psi_{2}$ & $P^{+} \Psi_{2}$ & $\Psi_{1s}^{\text{th}}$ & $\Psi_{2s}^{\text{th}}$ & $\Psi_{2p}^{\text{th}}$ \\
\hline
$\langle \rho \rangle^{\Box}_{\text{left}}$ & $18.74$ & $15.58$ & $17.25$ & $9.60$ & $21.41$ & $13.94$ & \multirow{2}{*}{$8.14$} & \multirow{2}{*}{$20.30$} & \multirow{2}{*}{$19.06$} \\
\cline{1-7}
$\langle \rho \rangle^{\Box}_{\text{right}}$ & $-$ & $8.38$ & $17.63$ & $18.90$ & $17.63$ & $18.90$ & & &
\end{tabular}
\end{figure}

\begin{figure}[t]
\centering
\ig{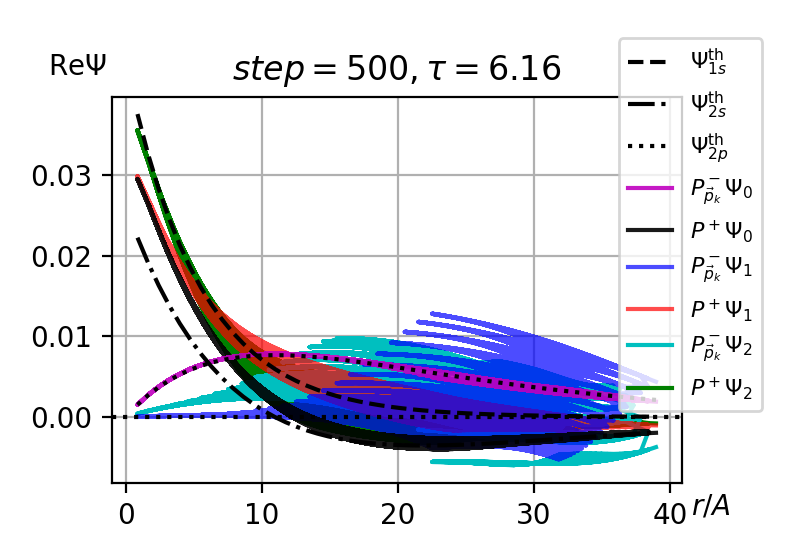} \hfill
\ig{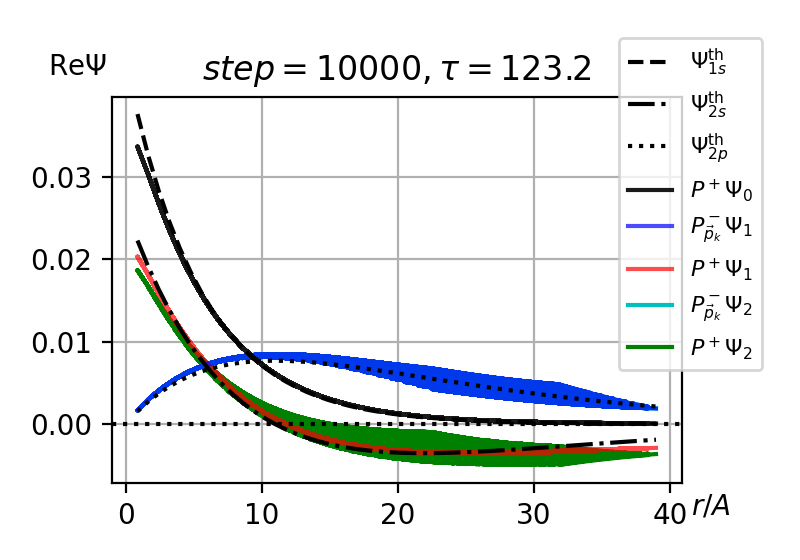}
\caption{\label{fig:wf_comp_KIN_1}
Parity projected wave-functions as in Fig.~\ref{fig:wf_comp_KIN_0} using the non-relativistic kinetic term $H_{K}^{(1)}$.
The average radii in units of $A$ are given by}
\small
\begin{tabular}{r|c|c|c|c|c|c|c|c|c}
& $P_{\vec{p}_{k}}^{-} \Psi_{0}$ & $P^{+} \Psi_{0}$ & $P_{\vec{p}_{k}}^{-} \Psi_{1}$ & $P^{+} \Psi_{1}$ & $P_{\vec{p}_{k}}^{-} \Psi_{2}$ & $P^{+} \Psi_{2}$ & $\Psi_{1s}^{\text{th}}$ & $\Psi_{2s}^{\text{th}}$ & $\Psi_{2p}^{\text{th}}$ \\
\hline
$\langle \rho \rangle^{\Box}_{\text{left}}$ & $18.85$ & $16.66$ & $24.50$ & $9.60$ & $20.11$ & $9.26$ & \multirow{2}{*}{$8.14$} & \multirow{2}{*}{$20.30$} & \multirow{2}{*}{$19.06$} \\
\cline{1-7}
$\langle \rho \rangle^{\Box}_{\text{right}}$ & $-$ & $8.49$ & $18.47$ & $20.93$ & $18.47$ & $21.70$ & & &
\end{tabular}
\end{figure}

\begin{figure}[t]
\centering
\ig{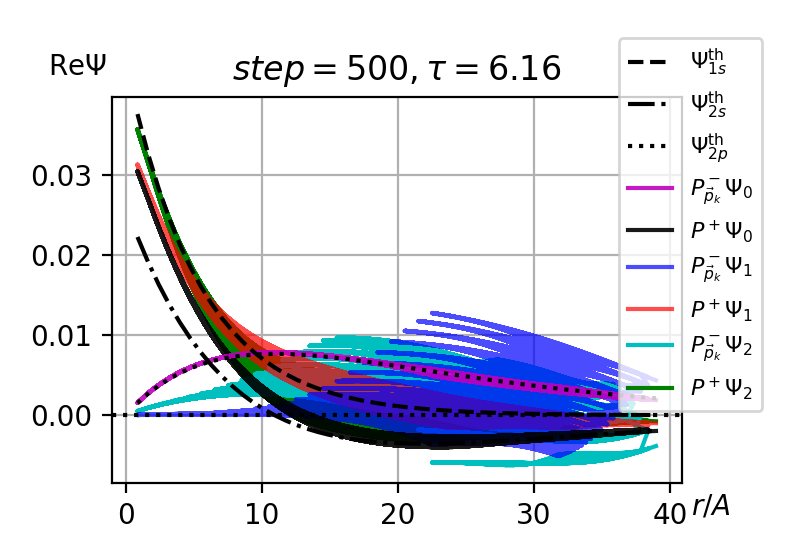} \hfill
\ig{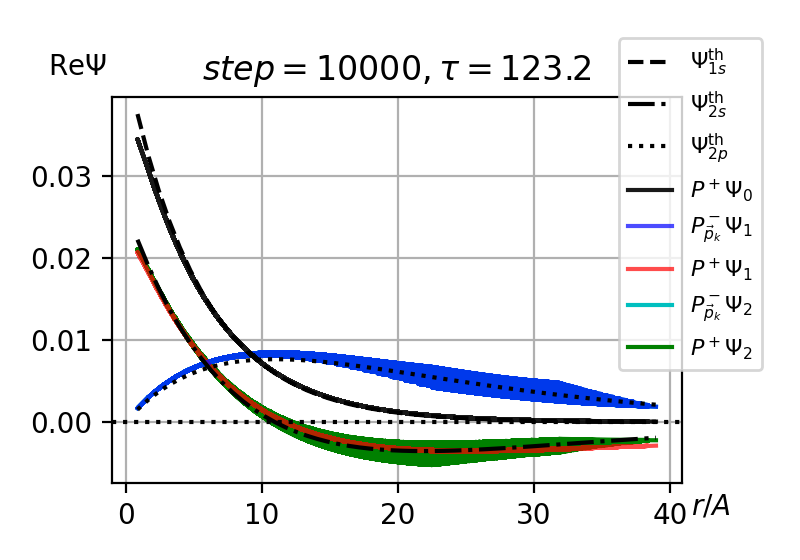}
\caption{\label{fig:wf_comp_KIN_2}
Parity projected wave-functions as in Fig.~\ref{fig:wf_comp_KIN_0} using the non-relativistic kinetic term $H_{K}^{(2)}$.
The average radii in units of $A$ are given by}
\small
\begin{tabular}{r|c|c|c|c|c|c|c|c|c}
& $P_{\vec{p}_{k}}^{-} \Psi_{0}$ & $P^{+} \Psi_{0}$ & $P_{\vec{p}_{k}}^{-} \Psi_{1}$ & $P^{+} \Psi_{1}$ & $P_{\vec{p}_{k}}^{-} \Psi_{2}$ & $P^{+} \Psi_{2}$ & $\Psi_{1s}^{\text{th}}$ & $\Psi_{2s}^{\text{th}}$ & $\Psi_{2p}^{\text{th}}$ \\
\hline
$\langle \rho \rangle^{\Box}_{\text{left}}$ & $18.84$ & $16.55$ & $24.47$ & $9.40$ & $20.07$ & $9.29$ &\multirow{2}{*}{$8.14$} & \multirow{2}{*}{$20.30$} & \multirow{2}{*}{$19.06$} \\
\cline{1-7}
$\langle \rho \rangle^{\Box}_{\text{right}}$ & $-$ & $8.43$ & $18.45$ & $20.93$ & $18.45$ & $20.46$ &  &  &
\end{tabular}
\end{figure}

For the FFT-based kinetic terms, $H_{K}^{(1)}$ and $H_{K}^{(2)}$, the most prominent discretization artefacts can be seen in Figs.~ \ref{fig:wf_comp_KIN_1}, and~\ref{fig:wf_comp_KIN_2} and are located at values of $r$ close to the lattice spacing $A$.
It is very pronounced due to the hard UV cutoff of the high momentum modes in the FFT.
This effect is not present for the $2p$ wave-function, that vanishes at the origin and effectively suppresses contributions from the hardest modes.
But it is especially prominent for the $2s$ wave-function.

For low values of $\tau$, higher states with quantum number $l>0$ show up as broad lines on the plots.
Such states decay exponentially as a function of $\tau$, according to Eq.~\eqref{eq:iter_decay}, so that they do not show up at larger $\tau$.
In particular, the negative parity-projected ground state, $P^-_{\vec{p}_k}\Psi_0$, is dominated by the $\Psi_{2p}$ contribution and the positive parity-projected first excited state, $P^+\Psi_1$, is actually dominated by the true ground state $\Psi_{1s}$.
Note that the ground state, $\Psi_{1s}^{\text{th}}$, and the first excited state, $\Psi_{2p}^{\text{th}}$, have $P^+$ and $P^-$ parities, respectively.
This is why $P^-_{\vec{p}_k}\Psi_0$ and $P^+\Psi_1$ are actually dominated by the first excited state and by the ground state, respectively.
According to subsection~\ref{subsec:iterative_procedure}, the overlap procedure allows for an extraction of the ground state and of the excited states.
However, due to numerical inaccuracy, this separation is not perfect.
Otherwise, we would have $P^-_{\vec{p}_k}\Psi_0 = 0$.
For all of the higher states contributions with both parities persist, because both states are degenerate for a non-relativistic Coulomb problem (infinite volume and continuum).

The continuum average radii in units of $A$ are given by $\langle \rho \rangle^{\infty}_{1s} = 8.21$, $\langle \rho \rangle^{\infty}_{2s} = 32.83$, and $\langle \rho \rangle^{\infty}_{2p} = 27.36$.
However, in the considered effective lattice volumes, the average radii of the continuum solutions of Eq.~\eqref{eq:non-rel-theo} in units of $A$ are $\langle \rho \rangle^{\Box}_{1s} = 8.14$, $\langle \rho \rangle^{\Box}_{2s} = 20.30$, and $\langle \rho \rangle^{\Box}_{2p} = 19.06$.
The lower values are linked to finite volume effects, that make larger values of $r$ being discarded.
Since the radial probability distribution of the $2s$ has its maximum at larger radii than the radial probability distribution of the $2p$, it is expected that the even parity excited states are a bit more sensitive to a periodic boundary condition than their odd parity counterparts.
In fact, radii of large even parity states approach the continuum value from below for Dirichlet boundary condition and from above for periodic boundary condition.
These box normalized theoretical average radii should be compared with those listed in the captions of Figs.~\ref{fig:wf_comp_KIN_0} to~\ref{fig:wf_comp_KIN_2}.
The wave-functions shown in those figures, with $\tau = 6.16$~GeV$^{-1}$, have still not stabilized.
Hence, there are big differences between theoretical and simulated values.
This is especially prominent for the $1s$ state, that still has non-vanishing components of higher excited positive parity states with larger radii like the $2s$.

The evaluation has stabilized in the right panels of Figs.~\ref{fig:wf_comp_KIN_0} to~\ref{fig:wf_comp_KIN_2}, where $\tau = 123.2$~GeV$^{-1}$.
The values of $\langle \rho \rangle^{\Box}_{P^+\Psi_0}$ get closer to the theory value $\langle \rho \rangle^{\Box}_{1s}$, Eq.~\eqref{eq:non-rel-theo}, the ones of $\langle \rho \rangle^{\Box}_{P^-_{\vec{p}_k}\Psi_{1,2}}$ get closer to $\langle \rho \rangle^{\Box}_{2p}$, and the ones of $\langle \rho \rangle^{\Box}_{P^+\Psi_{1,2}}$ get closer to $\langle \rho \rangle^{\Box}_{2s}$.
However, we note that $\langle \rho \rangle^{\Box}_{2p} \approx 20.30$ is pretty close to $\langle \rho \rangle^{\Box}_{2s} \approx 19.06$.
Indeed, the only difference between these states is the angular quantum number $l$, while $n = 1$ is equal.
This fact makes the comparisons harder.
We find that the average radii of the wave-functions of the FFT-based kinetic terms $H_{K}^{(1)}$ (Fig.~\ref{fig:wf_comp_KIN_1}) and $H_{K}^{(2)}$ (Fig.~\ref{fig:wf_comp_KIN_2}) are similar.
On the other hand, the FDTD kinetic term $H_{K}^{(0)}$ (Fig.~\ref{fig:wf_comp_KIN_0}) shows -- with the notable exception of the ground state -- after stabilization, systematically smaller average radii than the FFT-based or continuum results due to the Dirichlet boundary conditions.
This is especially prominent for the $2s$ and $2p$ states.

Looking at the tail of the wave-functions, the decay of the excited state wave-functions is faster in the FDTD case.
This fast decay is explicitly enforced by the Dirichlet boundary conditions.
For the FFT-based kinetic terms, periodic boundary conditions are used which do not enforce $\Psi(\text{boundary}) \equiv 0$, and thus the tails of the corresponding wave-functions are closer to the continuum ones as can be seen in Figs.~\ref{fig:wf_comp_KIN_1} and~\ref{fig:wf_comp_KIN_2}.
However, this more realistic, yet slower decay implies an increased sensitivity to the boundary.
The broad bands are due to the distinction of on-axis and off-axis separations, which effectively feel different sizes of the box, see Figs.~\ref{fig:wf_comp_KIN_1}, and~\ref{fig:wf_comp_KIN_2}.
The $1s$ state is less affected by finite volume effects because its theoretical average radius $\langle \rho \rangle^{\Box}_{1s} \approx 8.14$ is well within the finite lattice volume.

We now compare the FDTD kinetic term, $H_{K}^{(0)}$, (solid lines in all of the following figures) with the FFT-based kinetic term, $H_{K}^{(2)}$, (dashed lines in all of the following figures) for a increased lattice volume (original $N = 64$ on the left and increased $N = 128$ on the right in all of the following Figs.~\ref{fig:energ_comp_KIN_128} through~\ref{fig:wf_comp_KIN_128_2}).
The other parameters are unchanged.

\begin{figure}[t]
\centering
\ig{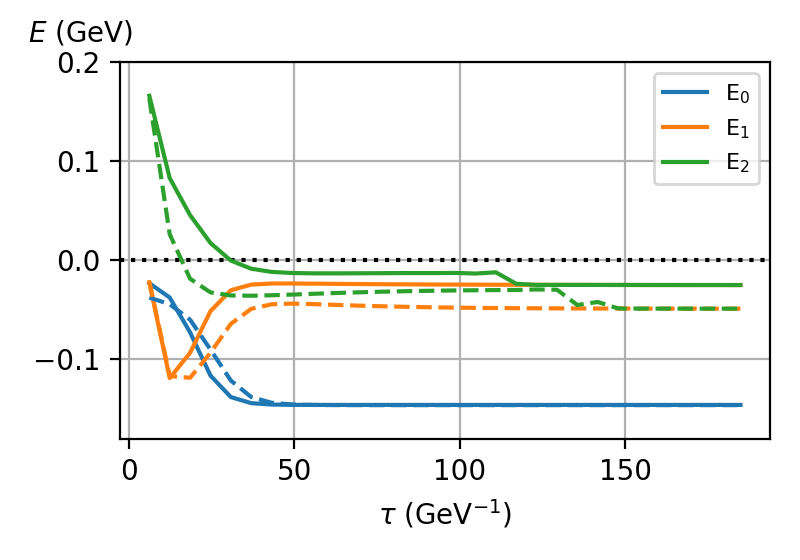} \hfill
\ig{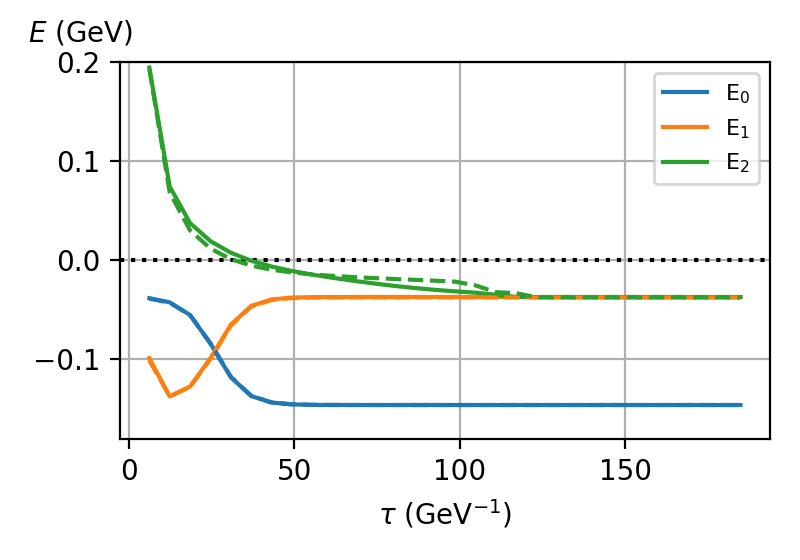}
\caption{\label{fig:energ_comp_KIN_128}
Energies obtained as a function of $\tau$ using the Coulomb potential and the non-relativistic kinetic terms $H_{K}^{(0)}$ (FDTD based, solid lines) and the non-relativistic kinetic term $H_{K}^{(2)}$ (FFT-based, dashed lines) and the Coulomb potential.
Left: $N = 64$; right: $N = 128$.}
\end{figure}

In Fig.~\ref{fig:energ_comp_KIN_128} we plot the energy eigen-values for the $N = 64$ and $N = 128$ simulations.
The relative weights of the parity projections are shown in Fig.~\ref{fig:proj_comp_KIN_128}, and finally in Fig.~\ref{fig:wf_comp_KIN_128_2} we plot the wave-functions for two different values of $\tau$.

For $N = 128$ the differences between $H_K^{(0)}$ and $H_K^{(2)}$ shown in the right plots of Figs.~\ref{fig:energ_comp_KIN_128} and~\ref{fig:proj_comp_KIN_128} are small, whereas for $N = 64$ (left plots, respectively) the differences are quite noticeable.
Comparing the energies in Fig.~\ref{fig:energ_comp_KIN_128}, the evolution of the $N = 128$ green ($E_{2}$) lines are very close to one another.
The ones of the lower states are basically indistinguishable.
However, this is not the case for the $N = 64$ simulations.
Beyond the ground state, the results are not even quantitatively consistent.

This trend continues when looking at the behavior of the relative weights of the parity projections in Fig.~\ref{fig:proj_comp_KIN_128}.
In the right panels we see little to no difference between the two kinetic terms.
The same holds true for the ground state even in the smaller volume (top left panel of Fig.~\ref{fig:proj_comp_KIN_128}).
The parity projections of the first excited state (middle panels of Fig.~\ref{fig:proj_comp_KIN_128}) are similar between $H_K^{(0,2)}$ for $N = 128$ with weights $p_{lm} \approx 1/4$ for each state befitting an approximately degenerate singlet and triplet.
No such degeneracy is reflected in the weights for $N = 64$.
Instead, in the smaller volume one parity survives as the dominant contribution in the first excited state and the other in the second (not shown); this pattern is reversed between both $H_K^{(0)}$ and $H_K^{(2)}$.
This behavior is not totally unexpected given that $H_K^{(0)}$ assumes Dirichlet boundary conditions, whereas $H_K^{(2)}$ uses periodic boundary conditions.
The implications of the different boundary conditions have been discussed above on page~\pageref{discussion:boundary_conditions}.
For the second excited state (not shown) not even the larger volume is sufficient, i.e., we see behavior somewhat similar to $N = 64$.
The $d$-wave is not distinguished from the $s$-wave by the parity projection.
Hence, with $H_K^{(0)}$ their sum ends up with a weight of $p_{\text{even}} \approx p_{00}+5p_{2m} \approx 1/6$, while the sum of the $p$-waves ends up with a weight of $p_{\text{odd}} \approx 3p_{1m} \approx 1/3$.
With $H_K^{(2)}$, the odd parity states are much more suppressed, and even parity dominates completely.
Eventually, the second excited state freezes out, and the same pattern as in the first excited state emerges for $\tau > 125$~GeV$^{-1}$ in both volumes.

The continuum average radii in units of $A$ using the box normalization are given by $\langle \rho \rangle^{\Box}_{1s} = 8.22$, $\langle \rho \rangle^{\Box}_{2s} = 30.77$, and $\langle \rho \rangle^{\Box}_{2p} = 26.18$, respectively.
As it happened for $N = 64$ above, the infinite volume continuum solutions of Eq.~\eqref{eq:non-rel-theo} in units of $A$ have larger average radii: $\langle \rho \rangle^{\infty}_{1s} = 8.21$, $\langle \rho \rangle^{\infty}_{2s} = 32.83$, and $\langle \rho \rangle^{\infty}_{2p} = 27.36$, respectively.
However, note that the difference is not as prominent as in the $N = 64$ case.
Actually, $\rho = 31$ falls well inside the lattice volume of $N^3 = 128^3$, so that finite volume effects are not so drastic.

\begin{figure}[t]
\centering
\ig{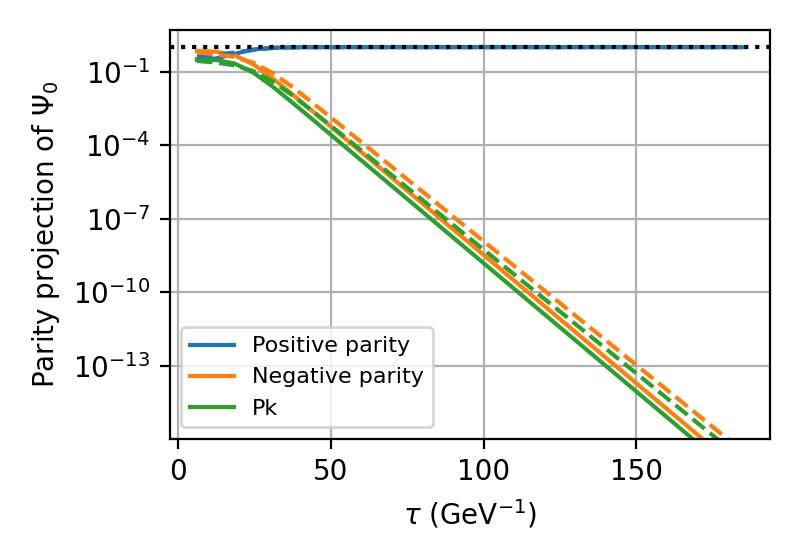}%
\hfill%
\ig{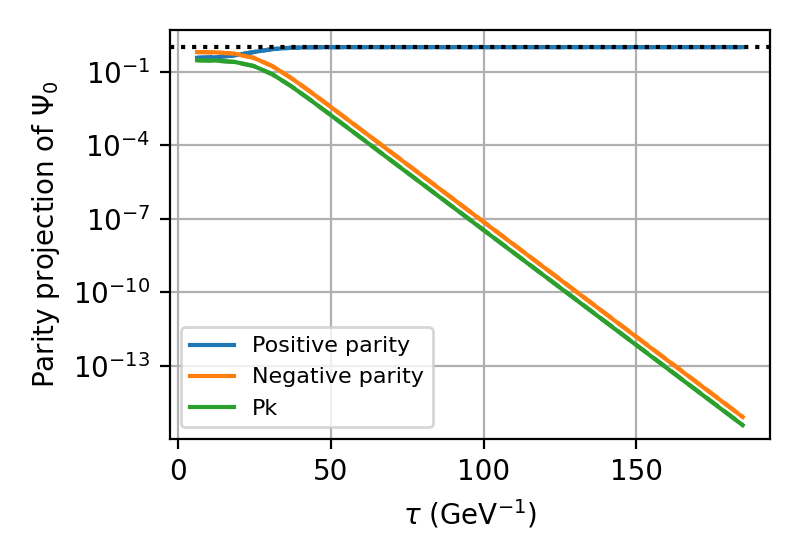}%
\\
\ig{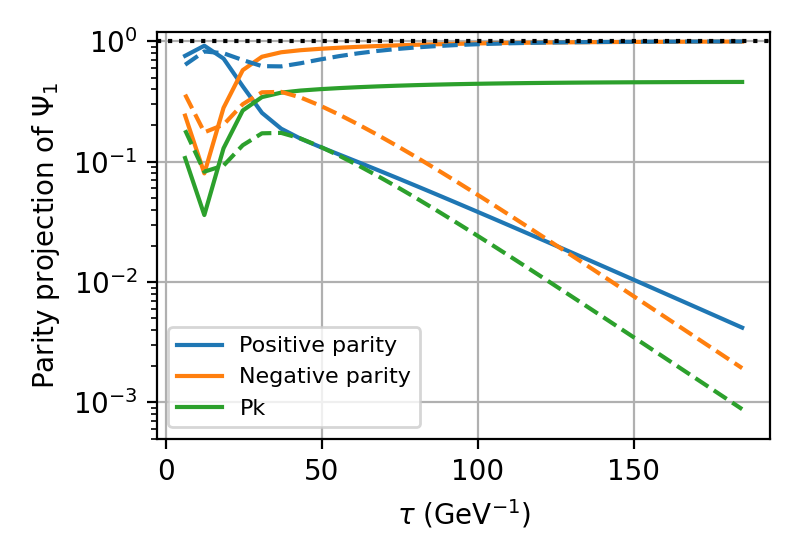}%
\hfill%
\ig{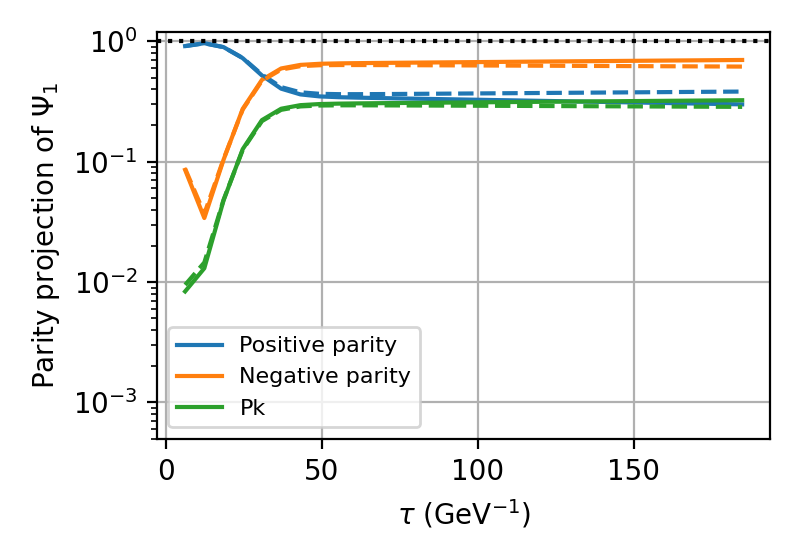}%
\caption{\label{fig:proj_comp_KIN_128}
Weights of the parity projection of the ground state (top row), and first excited state (bottom row) as a function of $\tau$ using the non-relativistic kinetic terms $H_{K}^{(0)}$ (FDTD based, solid lines) and the non-relativistic kinetic term $H_{K}^{(2)}$ (FFT-based, dashed lines).
Left plots: $N = 64$; right plots: $N = 128$.}
\end{figure}

\FloatBarrier

Indeed, when comparing the results for the wave-functions in Fig.~\ref{fig:wf_comp_KIN_128_2} with the ones from above and from the previous sections, we see that the finite volume effects are greatly reduced.
Even if we are comparing Dirichlet boundary conditions (FDTD kinetic term) versus periodic boundary conditions (FFT-based kinetic term).
Additionally, the FDTD ($H_{K}^{(0)}$) and FFT-based ($H_{K}^{(2)}$) solutions agree with one another and with the theoretical predictions for the wavefunctions and average radii (see dashed lines in Fig.~\ref{fig:wf_comp_KIN_128_2} and
the associated table in the caption.
Compared with the previous Figs.~\ref{fig:wf_comp_KIN_0} (FDTD, Dirichlet boundary conditions) and~\ref{fig:wf_comp_KIN_2} (FFT, periodic boundary conditions) each using $N = 64$, we can see that, in particular, the problem with the behavior of the tail in Fig.~\ref{fig:wf_comp_KIN_0}, that was attributed to finite volume effects and the Dirichlet boundary conditions, disappears when the volume is large enough to accommodate most of the wave-functions.

\begin{figure}[t]
\centering
\ig{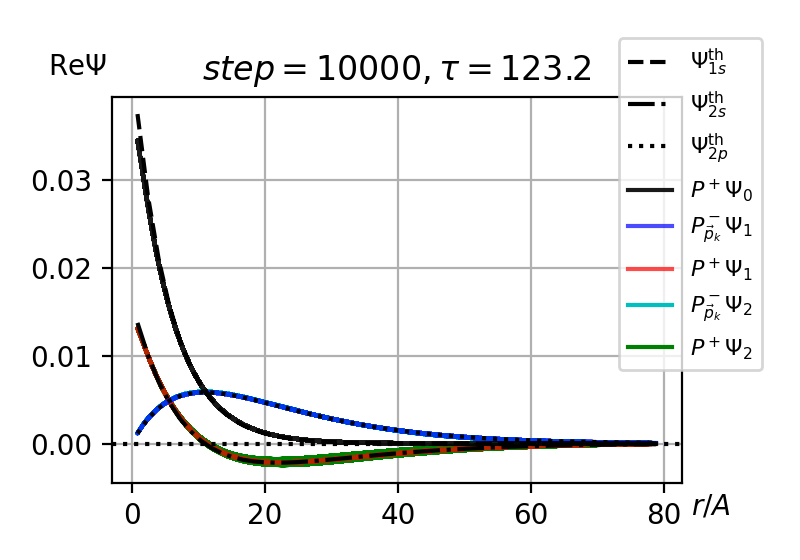} \hfill
\ig{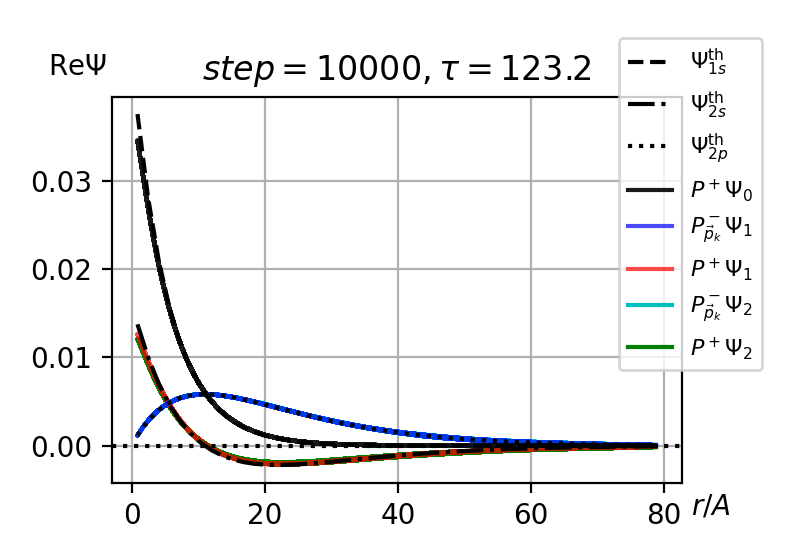}
\caption{\label{fig:wf_comp_KIN_128_2}
Parity-projected wave-functions as a function of $r/A$ using the non-relativistic kinetic term $H_{K}^{(0)}$ (FDTD based, left) and the non-relativistic kinetic term $H_{K}^{(2)}$ (FFT-based, right) using the increased lattice volume ($N = 128$).
The average radii in units of $A$ are given by}
\small
\begin{tabular}{r|c|c|c|c|c|c|c|c|c}
& $P_{\vec{p}_{k}}^{-} \Psi_{0}$ & $P^{+} \Psi_{0}$ & $P_{\vec{p}_{k}}^{-} \Psi_{1}$ & $P^{+} \Psi_{1}$ & $P_{\vec{p}_{k}}^{-} \Psi_{2}$ & $P^{+} \Psi_{2}$ & $\Psi_{1s}^{\text{th}}$ & $\Psi_{2s}^{\text{th}}$ & $\Psi_{2p}^{\text{th}}$ \\
\hline
$\langle \rho \rangle^{\Box}_{\text{left}}$ & $-$ & $8.50$ & $26.05$ & $30.77$ & $25.85$ & $30.71$ & \multirow{2}{*}{$8.22$} & \multirow{2}{*}{$30.77$} & \multirow{2}{*}{$26.18$} \\
\cline{1-7}
$\langle \rho \rangle^{\Box}_{\text{right}}$ & $-$ & $8.50$ & $26.16$ & $31.69$ & $26.14$ & $32.66$ &  &  &
\end{tabular}
\end{figure}

The ground state radius $\langle \rho \rangle^{\Box}_{P^+\Psi_{0}}$ is minimally larger in the bigger box, and insensitive to the boundary condition as in the smaller box.
On the one hand, the even parity excited states computed with the FDTD (Dirichlet boundary conditions) have average radii in the box normalization, $\langle \rho \rangle^{\Box}_{P^+\Psi_{1,2}}$, that are still systematically smaller than the corresponding theoretical value $\langle \rho \rangle^{\Box}_{2s}$; on the contrary, these state have larger radii when computed through the FFT (periodic boundary conditions).
On the other hand, the odd parity excited states have average radii $\langle \rho \rangle^{\Box}_{P^-_{\vec{p}_k}\Psi_{1,2}}$ that are marginally smaller than the theoretical value $\langle \rho \rangle^{\Box}_{2p}$.

It can finally be seen in Fig.~\ref{fig:wf_comp_KIN_128_2} that, especially for the FFT-based kinetic term ($H_K^{(2)}$, right plot), the behavior of the $2s$ and $2p$ states deviates from the respective continuum result close to the origin.
The continuum result curves lie slightly higher than the numerically computed ones in that region.
This points out the issue with the lattice implementation of the Coulomb potential close to its center as already mentioned and is expected due to the pole at the origin.
Hence, we are facing discretization artifacts originating from the discretization of the potential.
The FFT approach, seems slightly worse than the FDTD approach ($H_K^{(0)}$, left plot of Fig.~\ref{fig:wf_comp_KIN_128_2}) in handling this kind of effect.
The singularity of the Coulomb potential at $r = 0$ makes the derivatives of the potential diverge close to $r = 0$, whereas the FFT sets a cut on their values due to the finite volume in momentum space.
The FDTD is based on finite differences, so that such a drastic cut in momentum space is not present in contrast to the former.

\subsection{Comparison of the relativistic \texorpdfstring{$H_{K}^{(3)}$}{H\_K(3)} and the non-relativistic \texorpdfstring{$H_{K}^{(2)}$}{H\_K(2)} kinetic terms at high mass}
\label{subsec:comp_high_mass}

\begin{figure}[ht]
\centering
\ig{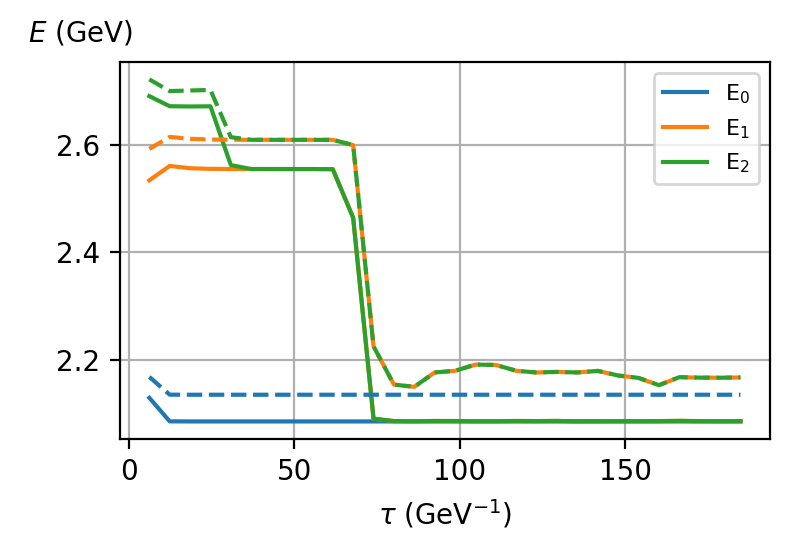}%
\hfill%
\ig{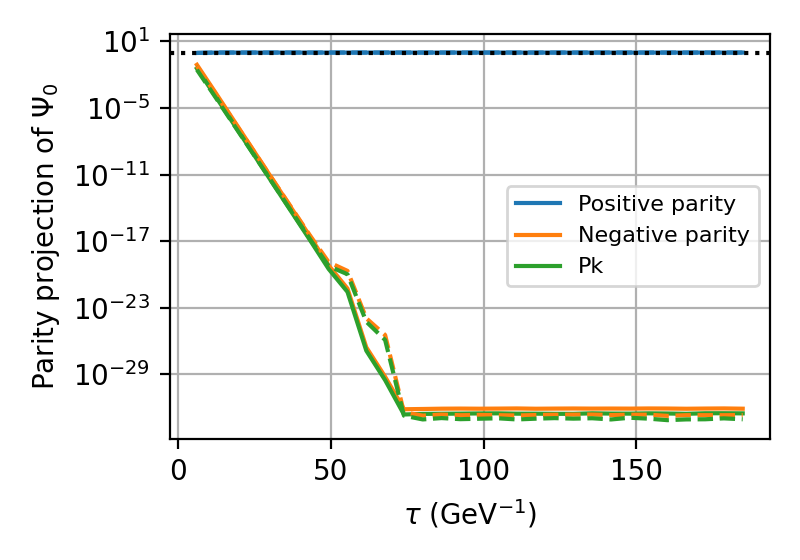}%
\\%
\ig{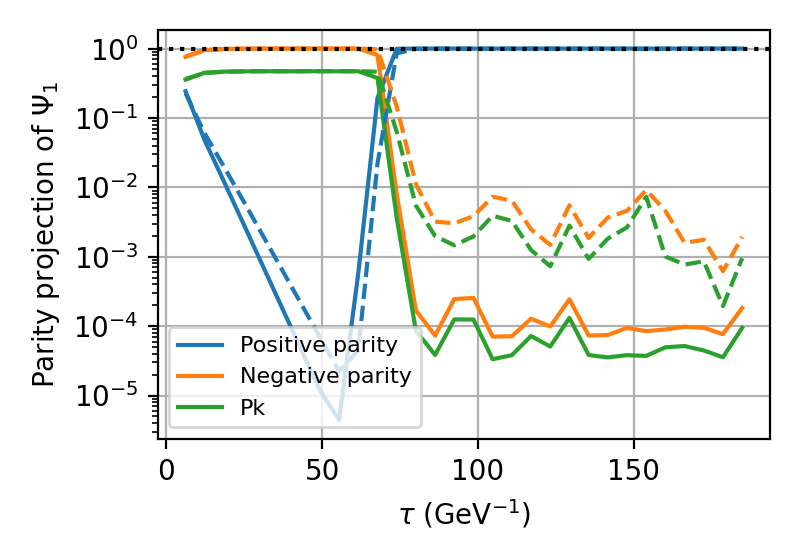}%
\hfill%
\ig{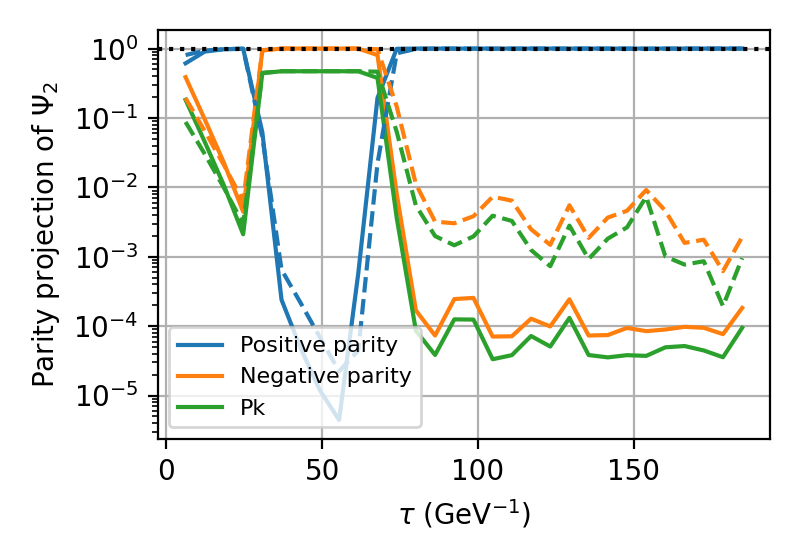}
\caption{\label{fig:energ_HEAVYMASS_proj_HEAVYMASS}
Energies obtained as a function of $\tau$ using the relativistic kinetic term $H_{K}^{(3)}$ (solid) and $H_{K}^{(2)}$ (dashed; shifted by $3$~GeV) using the increased reduced mass $M = 3$~GeV and the Coulomb potential (top left).\\
Weights of the parity projection of the corresponding ground state (top right), and of the first (bottom left) and second (bottom right) excited state as a function of $\tau$ using the relativistic kinetic term $H_{K}^{(3)}$ (solid) and $H_{K}^{(2)}$ (dashed) using the increased reduced mass $M = 3$~GeV.}
\end{figure}

In this subsection, we increase the reduced mass up to $M = 3$~GeV in order to simulate a heavy, non-relativistic system.
In all of the plots in Fig.~\ref{fig:energ_HEAVYMASS_proj_HEAVYMASS}, the results of the relativistic kinetic term, $H_{K}^{(3)}$, Eq.~\eqref{eq:HK3}, are displayed with solid lines, whereas the results of the non-relativistic one, $H_{K}^{(2)}$, Eq.~\eqref{eq:HK2}, are displayed with dashed lines, respectively.

The goal is to compare the numerical behavior of the relativistic and the non-relativistic kinetic terms in the regime of high mass, where they should become more similar.
The non-relativistic energy solutions have to be shifted by $M$ in order to allow for a comparison with the relativistic energy solutions.
Due to $(A \cdot M)^2 \approx 3.3$ we expect finite mass discretization errors.
For this reason, this setup is not appropriate for production purposes, where one should reduce the lattice spacing by (approximately) the inverse factor from $A = 0.12$~fm to $A \lesssim 0.02$~fm.
The energies and parity projections of the wave-functions evolve faster with $\tau$ (see Fig.~\ref{fig:energ_HEAVYMASS_proj_HEAVYMASS}), such that the energy of the second excited state, $E_2$, degenerates with $E_{1}$ at $\tau \sim 25$~GeV$^{-1}$ and $E_1$ degenerates with $E_{0}$ at $\tau \sim 70$~GeV$^{-1}$.
For $\tau > 60-65$~GeV$^{-1}$, numerical cancellation effects start showing up (see parity projections in Fig.~\ref{fig:energ_HEAVYMASS_proj_HEAVYMASS}).
The similarities between the results in the heavy mass limit, however, can be also seen in the top left panel of Fig.~\ref{fig:energ_HEAVYMASS_proj_HEAVYMASS}, where the decay curves are quite similar after taking into account an additive factor of $3$~GeV on the energies due to the rest mass on the relativistic approach.

\begin{figure}[ht]
\centering
\ig{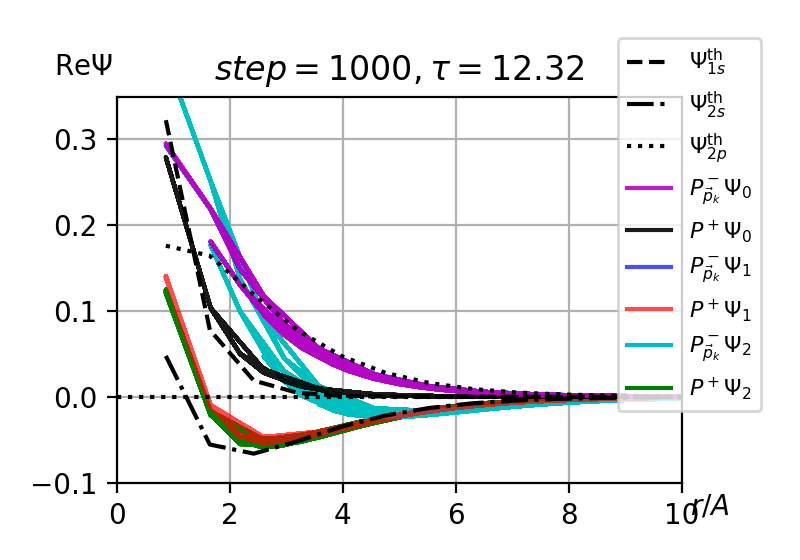}%
\hfill%
\ig{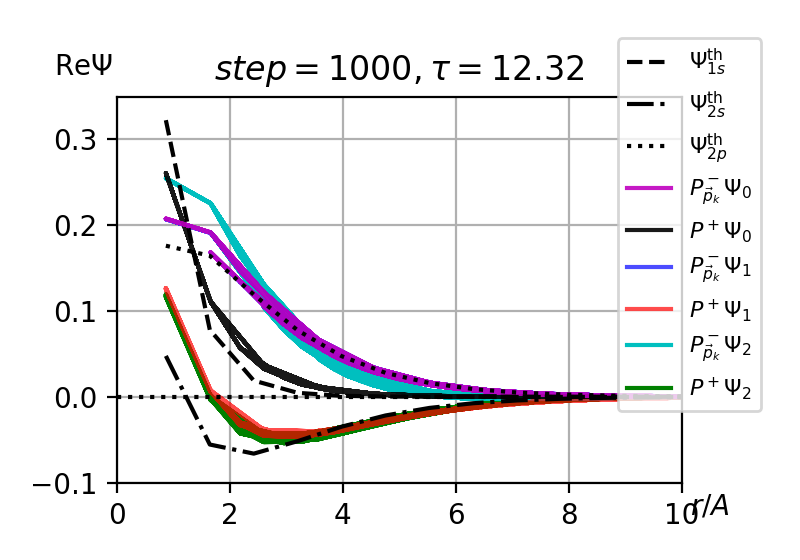}%
\\%
\ig{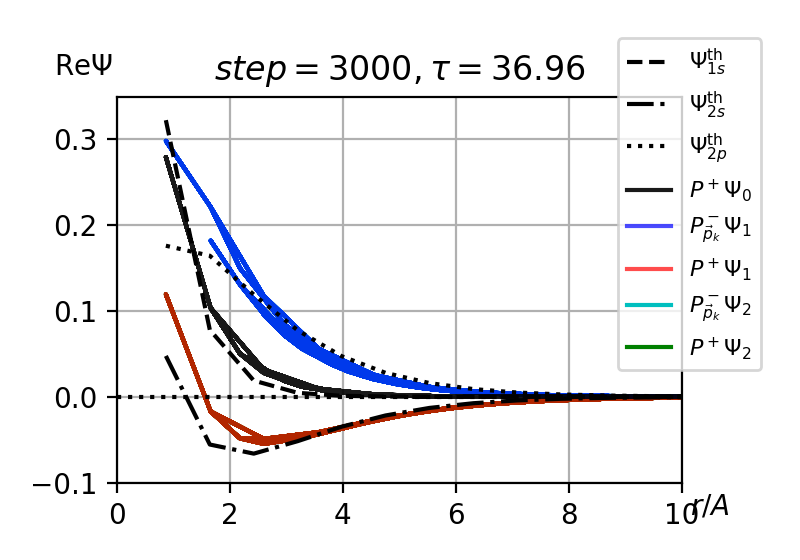}%
\hfill%
\ig{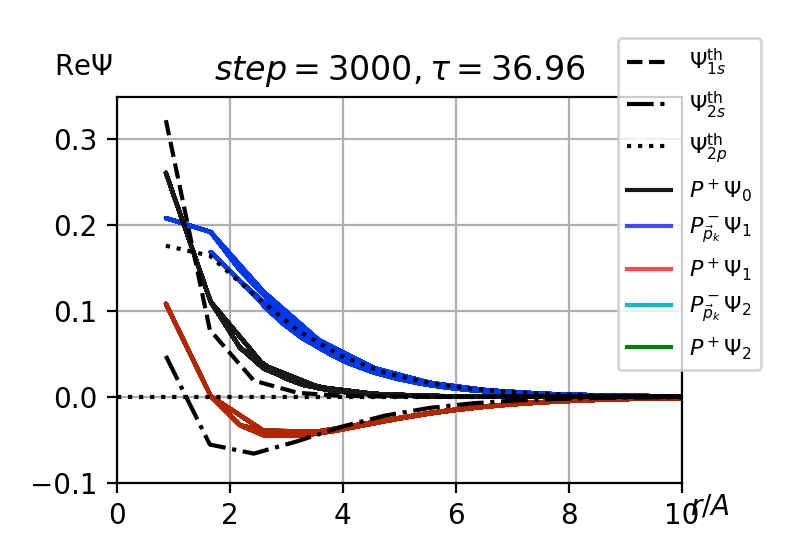}
\caption{\label{fig:wf_HEAVYMASS}
Parity-projected wave-functions as a function of $r/A$ using the relativistic kinetic term $H_{K}^{(3)}$ (left) and the non-relativistic kinetic term $H_{K}^{(2)}$ (right) using the increased reduced mass $M = 3$~GeV at different snapshot time steps.
For comparison we also show the theoretical non-relativistic curves in the left plots that should behave non-relativistically.
The average radii in units of $A$ are given by}
\small
\begin{tabular}{r|c|c|c|c|c|c|c|c|c}
& $P_{\vec{p}_{k}}^{-} \Psi_{0}$ & $P^{+} \Psi_{0}$ & $P_{\vec{p}_{k}}^{-} \Psi_{1}$ & $P^{+} \Psi_{1}$ & $P_{\vec{p}_{k}}^{-} \Psi_{2}$ & $P^{+} \Psi_{2}$ & $\Psi_{1s}^{\text{th}}$ & $\Psi_{2s}^{\text{th}}$ & $\Psi_{2p}^{\text{th}}$ \\
\hline
$\langle \rho \rangle^{\Box}_{\text{top left}}$ & $2.13$ & $1.27$ & $2.15$ & $3.52$ & $2.11$ & $3.52$ & \multirow{4}{*}{1.02} & \multirow{4}{*}{3.39} & \multirow{4}{*}{2.74} \\
\cline{1-7}
$\langle \rho \rangle^{\Box}_{\text{top right}}$ & $2.57$ & $1.37$ & $2.57$ & $3.90$ & $2.11$ & $3.75$ & & & \\
\cline{1-7}
$\langle \rho \rangle^{\Box}_{\text{bottom left}}$ & $-$ & $1.27$ & $2.11$ & $3.65$ & $2.11$ & $3.65$ & & & \\
\cline{1-7}
$\langle \rho \rangle^{\Box}_{\text{bottom right}}$ & $-$ & $1.37$ & $2.56$ & $4.07$ & $2.56$ & $4.07$ & & &
\end{tabular}
\end{figure}

As expected, the wave-functions (see Fig.~\ref{fig:wf_HEAVYMASS}) of the relativistic and the non-relativistic cases are closer to one another due to the increased reduced mass.
For comparison reasons we therefore show the non-relativistic theory curves also for the relativistic kinetic term (left plots).
However, the results are visibly non-smooth due to discretization artifacts.
Comparing these results with the results from the previous subsections~\ref{subsec:rel_kin_term} and~\ref{subsec:comp_non_rel_kin_terms} shows that the increase of the reduced mass from $M = 0.3$~GeV to $M = 3$~GeV by an order of magnitude leads to the expected decrease of the characteristic average radius of the wave-functions.
For the non-relativistic case, $\langle \rho \rangle^{\infty}_{1s} = 0.82$ in units of $A$, which is below a single lattice spacing.
For the $2s$ and $2p$ excited states, the continuum average radii are $\langle \rho \rangle^{\infty}_{2s} = 3.28$ and $\langle \rho \rangle^{\infty}_{2p} = 2.74$, respectively.
The fact that the continuum average radii in units of $A$ are close to $1$ makes a precise analysis of this case more complicated because higher terms in the series in $(A/r)^2$ are substantial contributions to the discretization artefacts (see subsection~\ref{subsec:cases}).

In the right panels of Fig.~\ref{fig:wf_HEAVYMASS}, it can be seen that, although the wave-function computed via the non-relativistic FFT-based kinetic term $H_{K}^{(2)}$ are close to the continuum result, there are clear discrepancies that partially originate from the momentum discretization.
Since the large mass compresses the states to smaller radii, i.e., $\langle r \rangle \propto 1/M$, even the discretization artefacts from the series in $(A/r)^2$ are enhanced by the large mass.
This is very prominent for the $1s$ and $2s$ states, that have non-vanishing values close to the origin.
Anyhow, in Fig.~\ref{fig:wf_HEAVYMASS} it can be seen that the wave-functions computed via the relativistic kinetic term, $H_{K}^{(3)}$, and the non-relativistic one, $H_{K}^{(2)}$, (both based on a FFT with periodic boundary conditions) are pretty close to one another.
This is to be compared against the cases shown in the right panel of Fig.~\ref{fig:wf_comp_KIN_2} (non-relativistic kinetic term $H_{H}^{(2)}$) and bottom right panel of Fig.~\ref{fig:wf_KIN3} (relativistic kinetic term $H_{K}^{(3)}$), where the differences between relativistic and non-relativistic are quite prominent.

\clearpage

\subsection{The \texorpdfstring{\texttt{quantumfdtd}}{quantumfdtd} internal parity fixing machinery}
\label{subsec:parity_fixing}

\begin{figure}[ht]
\centering
\ig{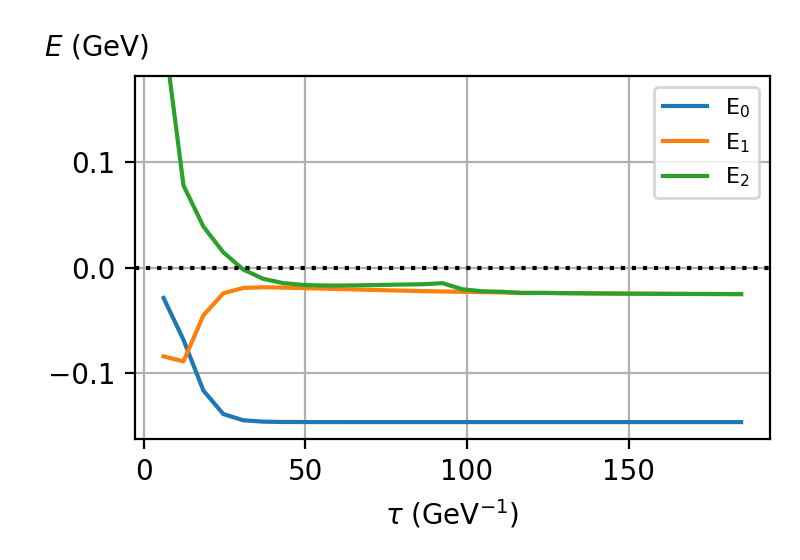} \hfill
\ig{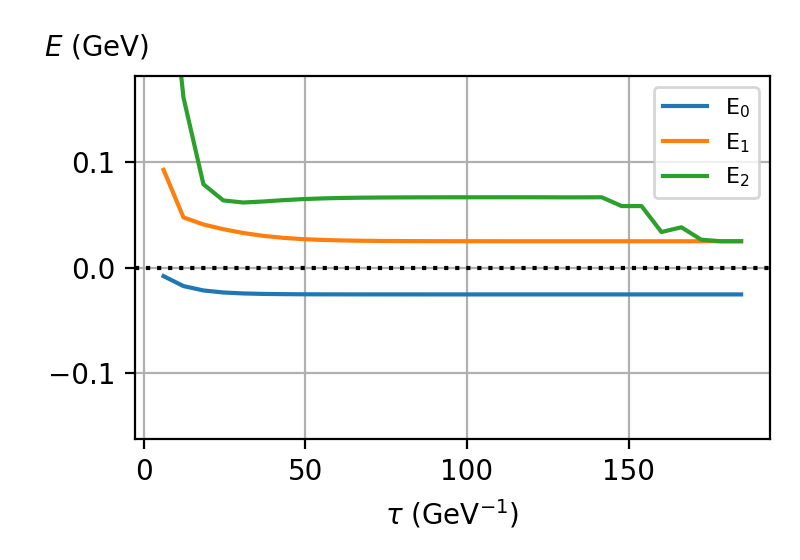}
\caption{\label{fig:energ_comp_parity}
Energies obtained as a function of $\tau$ using the non-relativistic kinetic terms $H_{K}^{(0)}$ with \texttt{INITSYMMETRY=1} (left, symmetric along the $z$ axis) and \texttt{INITSYMMETRY=2} (right, antisymmetric along the $z$ axis) and the Coulomb potential.}
\end{figure}

In this subsection, we investigate the ability to specify the initial spatial symmetry of the wave-function using the \texttt{INITSYMMETRY} parameter (see the definition in subsection~\ref{flag:INITSYMMETRY} on page~\pageref{flag:INITSYMMETRY}).
We set \texttt{INITSYMMETRY=1} for the left plots in all figures, which corresponds to symmetrization along the $z$ axis, and we set \texttt{INITSYMMETRY=2} for the right plots in all figures, which corresponds to anti-symmetrization.
We again use $N = 64$ and $M = 0.3$~GeV and the non-relativist kinetic term $H_{K}^{(0)}$.

\begin{figure}[!ht]
\centering
\ig{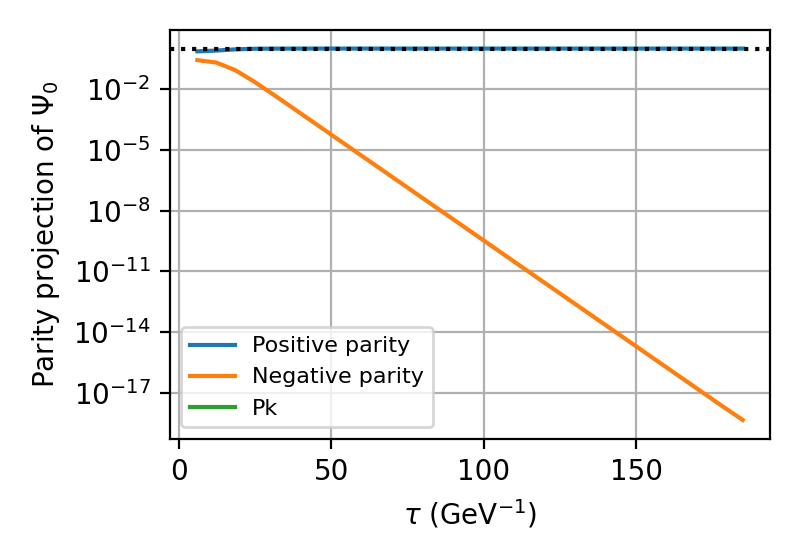} \hfill
\ig{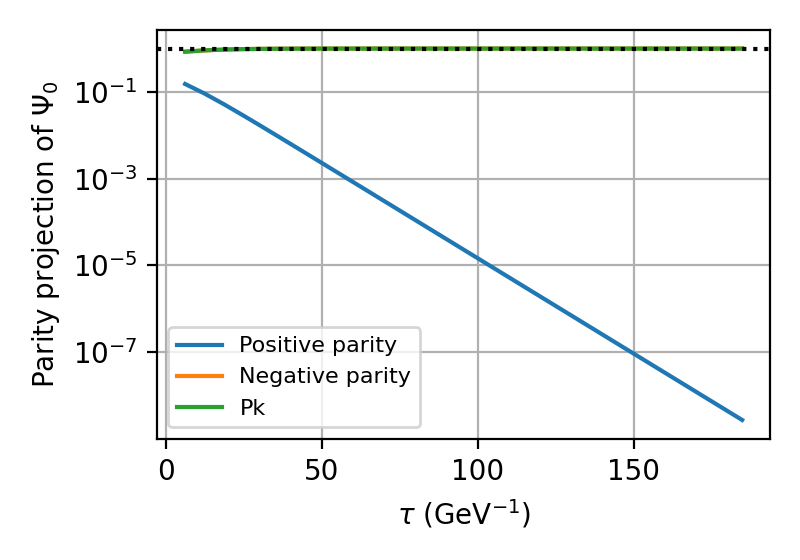}%
\\%
\ig{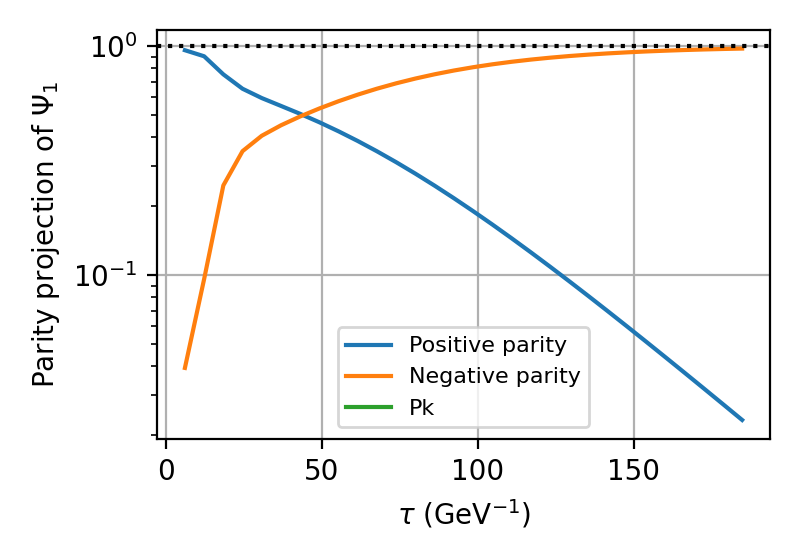} \hfill
\ig{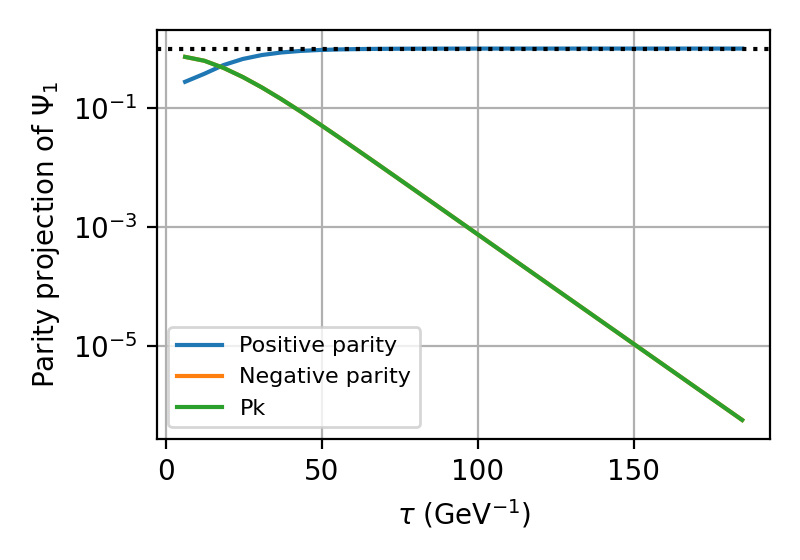}%
\\%
\ig{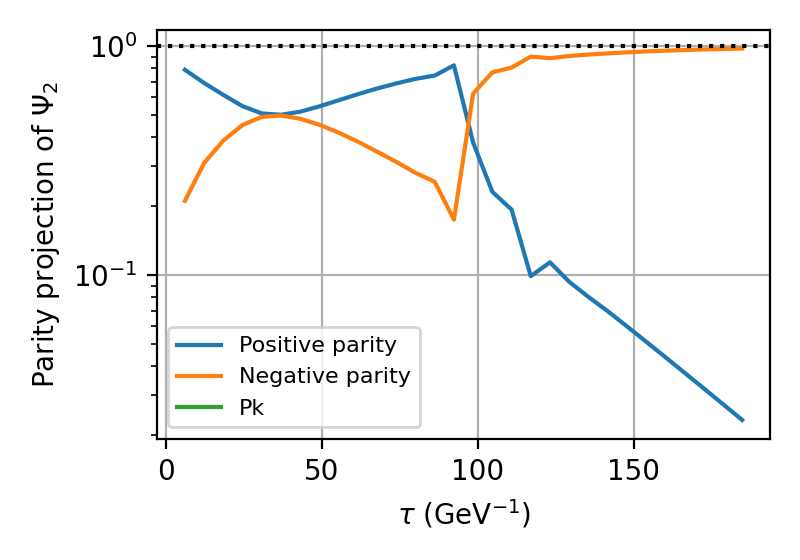} \hfill
\ig{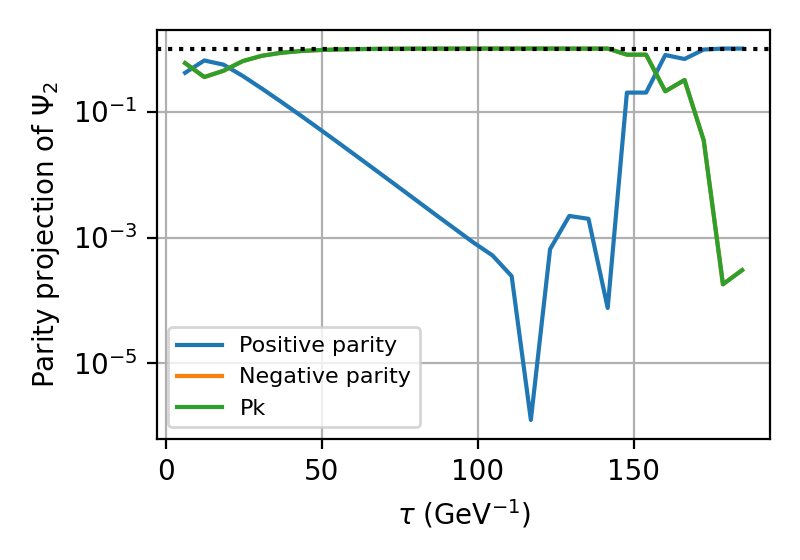}
\caption{\label{fig:proj_comp_parity}
Weights of the parity projection of the ground state (top row), and first (middle row) and second (bottom row) excited state as a function of $\tau$ using the non-relativistic kinetic terms $H_{K}^{(0)}$ with \texttt{INITSYMMETRY=1} (left, symmetric along the $z$ axis) and \texttt{INITSYMMETRY=2} (right, anti-symmetric along the $z$ axis).}
\end{figure}

\begin{figure}[!ht]
\centering
\ig{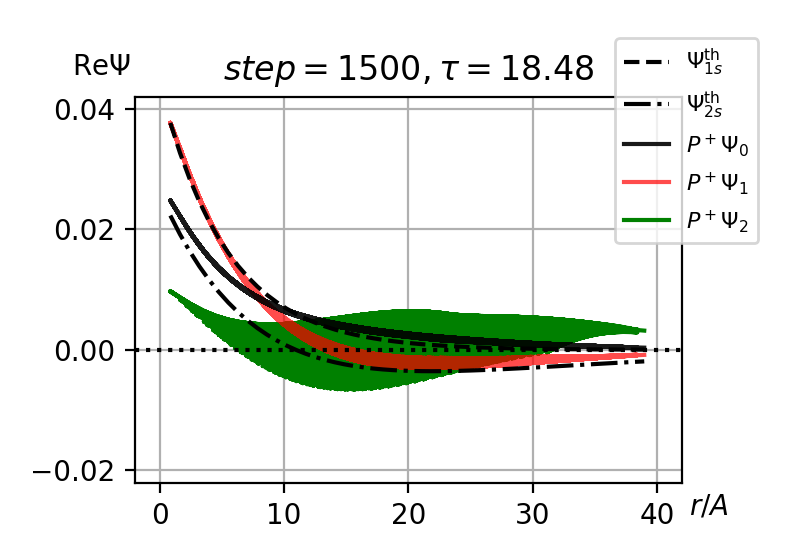}%
\hfill%
\ig{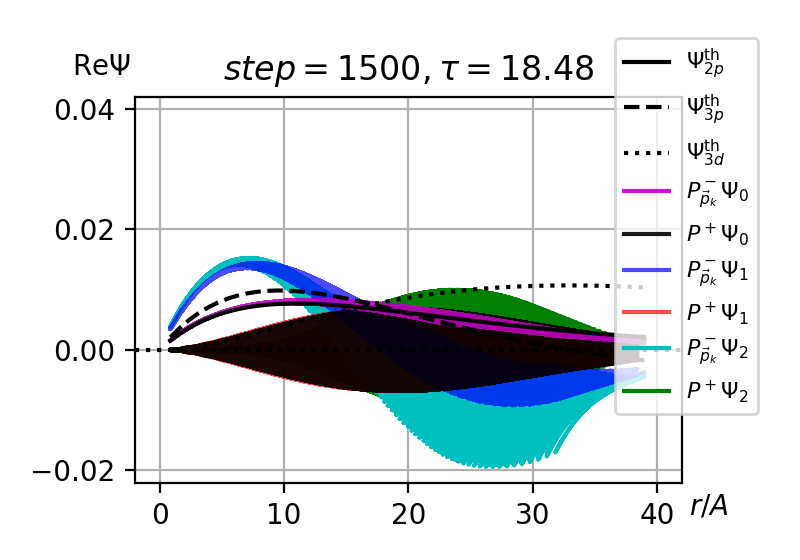}%
\\%
\ig{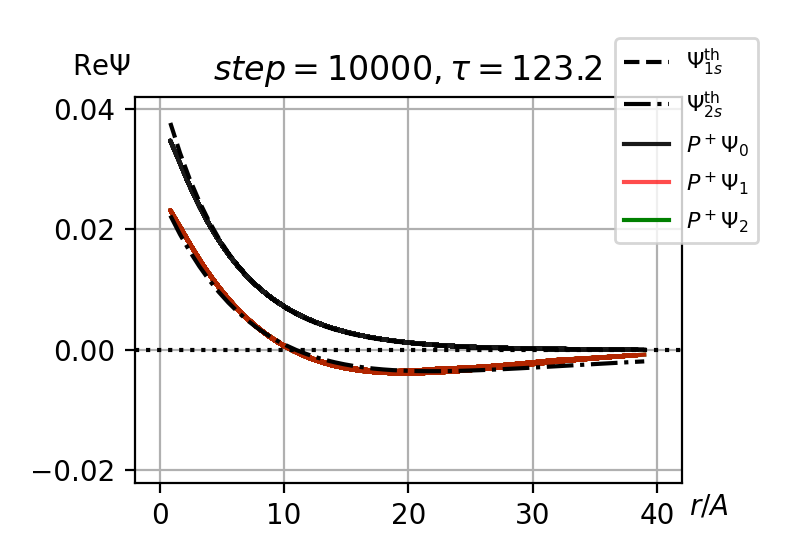}%
\hfill%
\ig{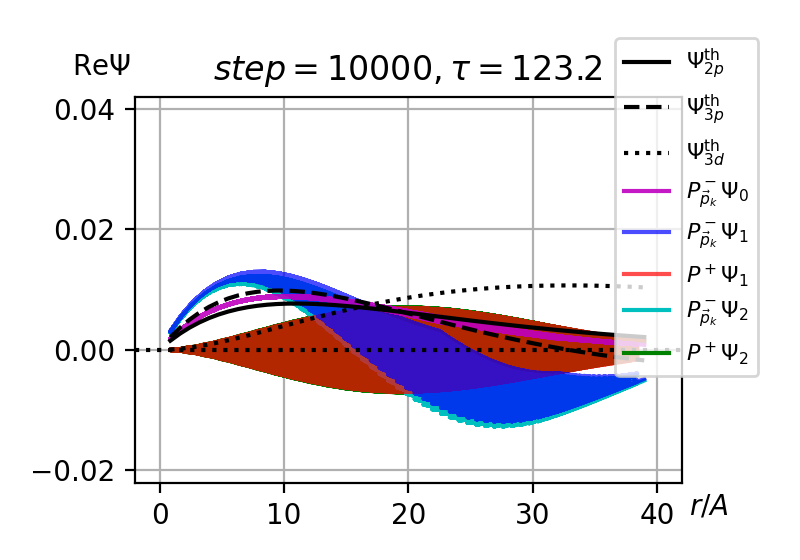}
\caption{\label{fig:wf_comp_parity}
Parity-projected wave-functions as a function of $r/A$ using the non-relativistic kinetic term $H_{K}^{(0)}$ with \texttt{INITSYMMETRY=1} (left, symmetric along the $z$ axis) and \texttt{INITSYMMETRY=2} (right, anti-symmetric along the $z$ axis).
The average radii in units of $A$ are given by}
\small
\begin{tabular}{r|c|c|c|c|c|c|c|c|c|c|c}
& $P_{\vec{p}_{k}}^{-} \Psi_{0}$ & $P^{+} \Psi_{0}$ & $P_{\vec{p}_{k}}^{-} \Psi_{1}$ & $P^{+} \Psi_{1}$ & $P_{\vec{p}_{k}}^{-} \Psi_{2}$ & $P^{+} \Psi_{2}$ & $\Psi_{1s}^{\text{th}}$ & $\Psi_{2s}^{\text{th}}$ & $\Psi_{2p}^{\text{th}}$ & $\Psi_{3p}^{\text{th}}$ & $\Psi_{3d}^{\text{th}}$ \\
\hline
$\langle \rho \rangle^{\Box}_{\text{top left}}$ & $-$ & $12.44$ & $-$ & $11.66$ & $-$ & $21.71$ & \multirow{2}{*}{$8.14$} & \multirow{2}{*}{$20.30$} & $-$ & $-$ & $-$ \\
\cline{1-7}\cline{10-12}
$\langle \rho \rangle^{\Box}_{\text{top right}}$ & $18.64$ & $22.09$ & $15.09$ & $21.92$ & $17.48$ & $23.76$ &  &  & 19.06 & 16.37 & 21.98 \\
\hline
$\langle \rho \rangle^{\Box}_{\text{bottom left}}$ & $-$ & $8.38$ & $-$ & $18.90$ & $-$ & $18.90$ & \multirow{2}{*}{$8.14$} & \multirow{2}{*}{$20.30$} & $-$ & $-$ & $-$ \\
\cline{1-7}\cline{10-12}
$\langle \rho \rangle^{\Box}_{\text{bottom right}}$ & $17.63$ & $-$& $18.28$ & $22.13$ & $19.64$ & $22.01$ &  &  & 19.06 & 16.37 & 21.98
\end{tabular}
\end{figure}

The corresponding energy solutions are shown in Fig.~\ref{fig:energ_comp_parity}.
The relative weights of the parity projections are shown in Fig.~\ref{fig:proj_comp_parity}, and, finally, the wave-functions are shown in Fig.~\ref{fig:wf_comp_parity}.
There the dashed lines again represent the analytic solutions for the non-relativistic kinetic term and a Coulomb potential.
The results for the latter should be compared with the figures of subsection~\ref{subsec:comp_non_rel_kin_terms} and with the FDTD based kinetic term, $H_{K}^{(0)}$, therein.
That is, with Fig.~\ref{fig:wf_comp_KIN_0}.

In Fig.~\ref{fig:proj_comp_parity}, it can be seen that the choice \texttt{INITSYMMETRY=1} (left plots) removes the $2p_z$ state completely.
However, there is still a residual negative parity component, most likely due to higher lying states or due to $2p$ states that are symmetric around the $z$ axis.
On the other hand, \texttt{INITSYMMETRY=2} (right plots in Fig.~\ref{fig:proj_comp_parity}) removes all $s$ states, leaving only $2p$ and higher lying states.
In particular, there are $3d$ states, which are odd under $P^-_{\vec{p}_k}$ for two directions while being even under parity.

This manifests in the $P^-_{\vec{p}_k}\Psi_{1,2}$ and the $P^+\Psi_1$ projections in the bottom right plot of Fig.~\ref{fig:wf_comp_parity} remaining broad (the corresponding bottom left plot had this artefact removed that was present for smaller values of $\tau$ in the top plots).
These states are not separated by the post-processing parity projections (see subsection~\ref{subsec:parity_projection}).
Hence, these positive parity wave-functions mix contributions from states with different geometry that show up as the broad bands in the Figure even for late time steps $\tau$.
Note that, on the right energy plot of Fig.~\ref{fig:energ_comp_parity}, the ground state energy no longer corresponds to the ground state energy of the left plot.
It now corresponds to the first excited state energy of the left plot which is the $2p$ state, since it is the lowest lying state compatible with the anti-symmetrization.
The two excited states in the right-hand plot are corresponding to $3d$ and $3p$.
Although these large states would be degenerate in the continuum in an infinite volume, they both show very strong finite size effects that lift their degeneracy.

The continuum average radii in units of $A$ are given by $\langle \rho \rangle^{\infty}_{1s} = 8.21$, $\langle \rho \rangle^{\infty}_{2s} = 32.83$, and $\langle \rho \rangle^{\infty}_{2p} = 27.36$.
As explained previously in subsection~\ref{subsec:comp_non_rel_kin_terms}, the values are lower when computed using the box normalization, i.e., $\langle \rho \rangle^{\Box}_{1s} = 8.14$, $\langle \rho \rangle^{\Box}_{2s} = 20.30$, and $\langle \rho \rangle^{\Box}_{2p} = 19.06$, respectively.
At sufficiently large times $\tau$, the average radii agree reasonably well with the theoretical expectations.
This behavior is in line with what we have already seen in the previous subsections; especially in subsection~\ref{subsec:comp_non_rel_kin_terms}.
The possibility to exclude certain states from the beginning can be an advantage during demanding calculations and is also a non-trivial cross check.

Because of the excited state contamination with $3p$ and $3d$ state in the case of the anti-symmetric initial condition we also provide their average radii\footnote{%
In the continuum, the analytical formulae are given by $\langle r \rangle^{\infty}_{3p} = 25/(2M)$ and $\langle r \rangle^{\infty}_{3d} = 21/(2M)$ using $R_{3p}(r) = 4\sqrt{2}/9 \, (M/3)^{3/2} \, Mr \, (1 - Mr/6) \, e^{-Mr/3}$ and $R_{3d}(r) = 2\sqrt{2}/(27\sqrt{5}) \, (M/3)^{3/2} \, (Mr)^{2} \, e^{-Mr/3}$.}.
In the continuum they are given by $\langle \rho \rangle^{\infty}_{3p} = 68.52$ and $\langle \rho \rangle^{\infty}_{3d} = 57.55$, while in the box normalization their values read $\langle \rho \rangle^{\Box}_{3p} = 16.37$ and $\langle \rho \rangle^{\Box}_{3d} = 21.98$.
Both states with main quantum number $n = 3$ become extremely compressed within this finite box.
In particular, the wave-function at large radii aligned with any axis is forced to be small due to the Dirichlet boundary condition and continuity.
Yet it may be much larger at similar radii realized in off-axis directions that are still far away from the boundary.
The consequence of these boundary effects are the broad, wavy bands which become narrow only close to either the center or to the corners of the box.

Since the $3p$-state is significantly more deformed by cutting away large parts of the contribution between its two nodes at large radii in the finite box, its box-normalized radius is even smaller than for the $2p$-state.
Nevertheless, the corresponding pattern of a function with two nodes at finite radii is clearly recognizable in both right panels of Fig.~\ref{fig:proj_comp_parity}.
The $3d$-state, on the other hand, is rather localized with only a single node.
The average radius of its box-normalized wave-function is surprisingly similar to the one obtained in the simulation.
With regard to Fig.~\ref{fig:proj_comp_parity} this may be a bit surprising, as its box-normalized radius still substantially undershoots the infinite volume radius.

We also note that the wave-function of the $3d$-state is not projected by the post-processing scripts with the correct symmetries, since projection onto $d$-waves has not been implemented\footnote{%
See section~\ref{subsec:parity_projection}.}.
Hence, the remaining angular dependence is averaged over, which leads to a wave-function spreading over both signs in the smeared band of the positive parity state.
Physical information should not be extracted from these states unless the volumes are larger by at least a factor $4^3$.

\subsection{The relativistic kinetic term \texorpdfstring{$H_{K}^{(3)}$}{H\_K(3)} and the harmonic oscillator}
\label{subsec:rel_kin_term_harmonic}

In this subsection, the relativistic FFT-based kinetic term $H_{K}^{(3)}$, Eq.~\eqref{eq:HK3}, is studied together with a harmonic oscillator potential.
Our goal is to compare to the existing analytic literature results of Ref.~\cite{Li:2005bd}.
There, however, the harmonic oscillator potential comes without the factor of $1/2$ which we also take into account for the simulation (compare to the standard form shown in Eq.~\eqref{eq:hardcoded_harmonic_oscillator}).
Additionally, there, the reduced mass is chosen to be $M = 30$~GeV which we use as well.
In order to combat large mass discretization effects we decrease the lattice spacing to $A = 0.006$~fm which results in $(A \cdot M)^{2} \approx 0.81$ which should result in reasonably small mass discretization effects.
We increase the number of points in each direction to be $N = 256$ to sustain a large enough physical volume.
Smaller masses are possible in principle which would remove the need for such small $A$ and large $N$, however, the computational cost for the analytic solution grows very rapidly (see below).
Such choices would, however, very likely increase the numerical behavior of the simulation.

\begin{figure}[t]
\centering
\ig{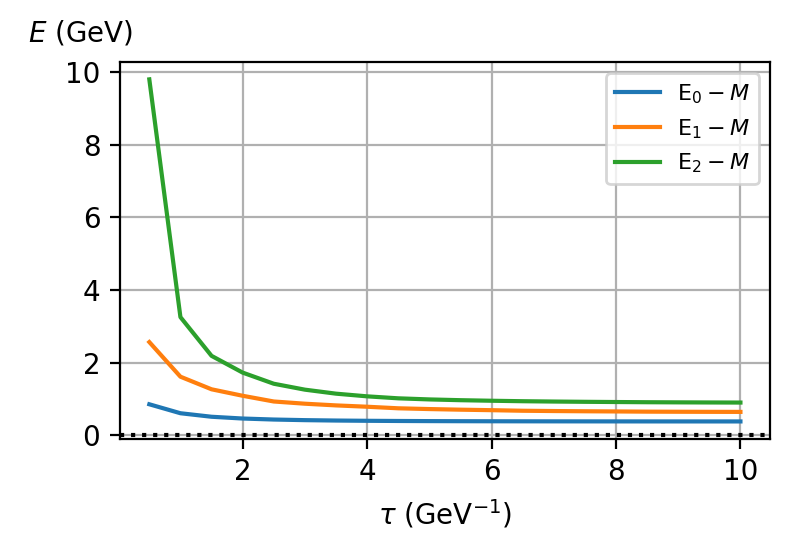}%
\hfill%
\ig{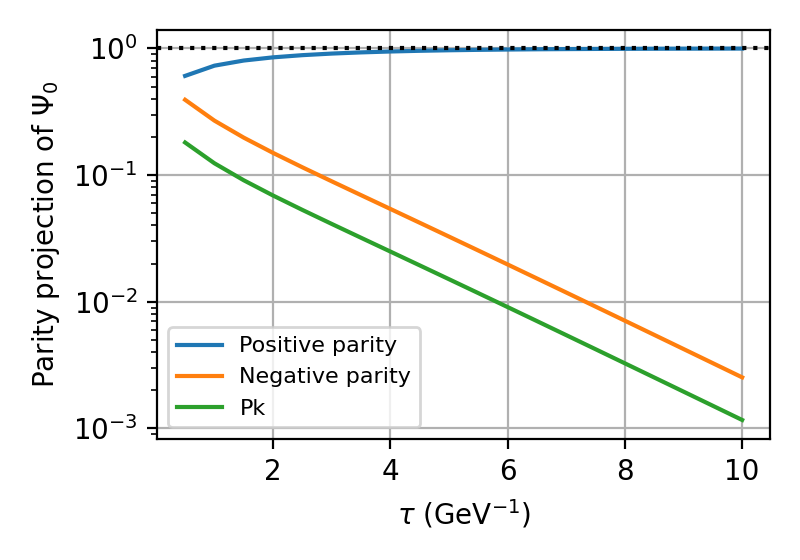}%
\\
\ig{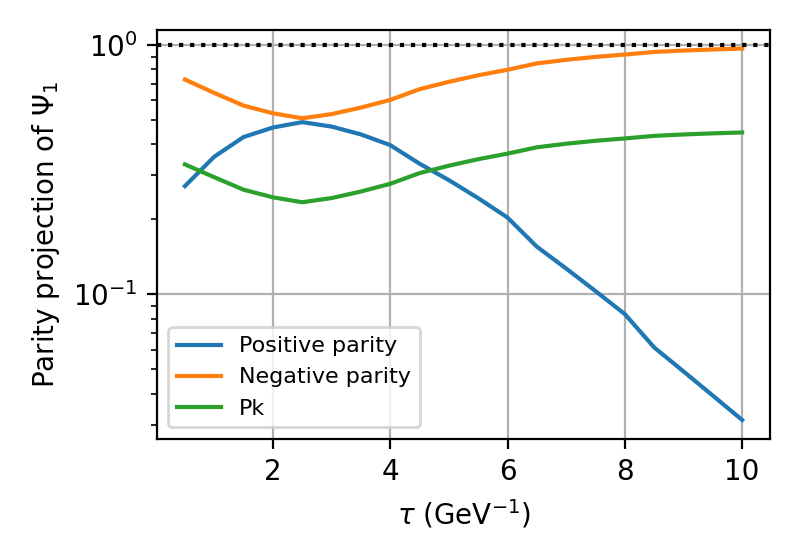}%
\hfill%
\ig{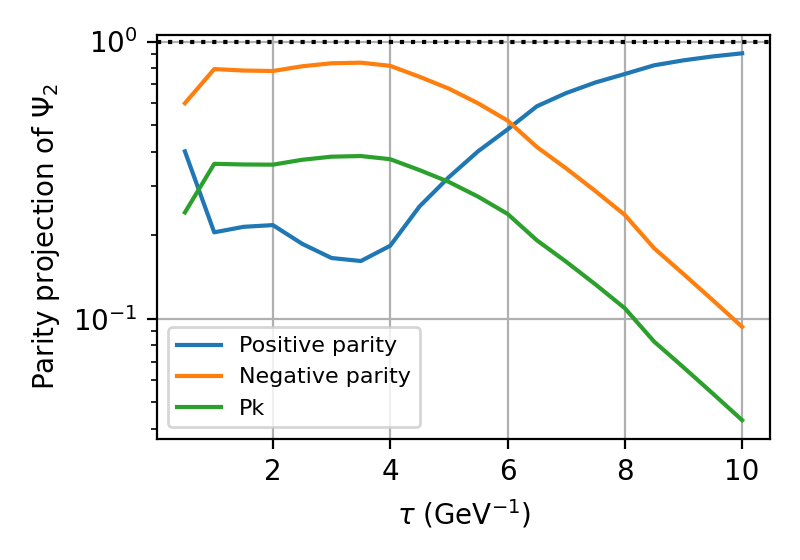}%
\caption{\label{fig:energ_proj_harm}%
Energies obtained as a function of $\tau$ using the relativistic kinetic term $H_{K}^{(3)}$ and the harmonic oscillator potential shifted by the mass $M = 30$~GeV (top left).\\
Weights of the parity projection of the corresponding ground state (top right), and of the first (bottom left) and second (bottom right) excited state as a function of $\tau$.}
\end{figure}

\begin{figure}[ht]
\centering
\ig{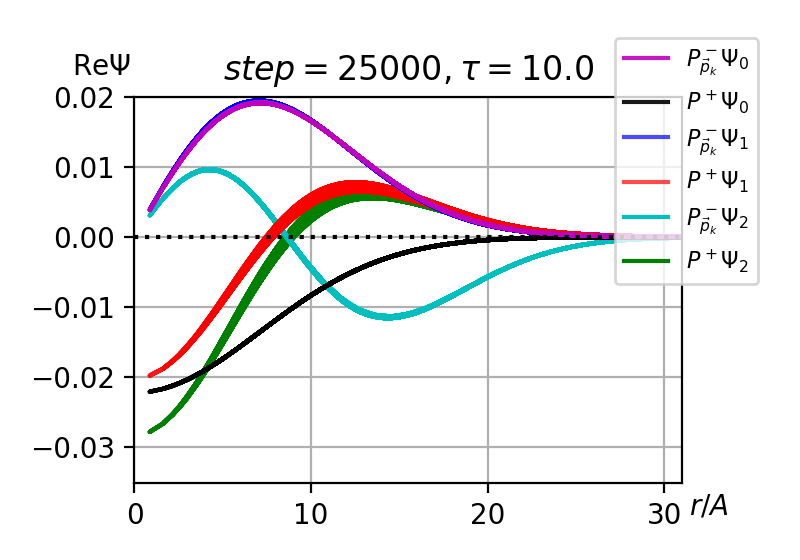}%
\hfill%
\ig{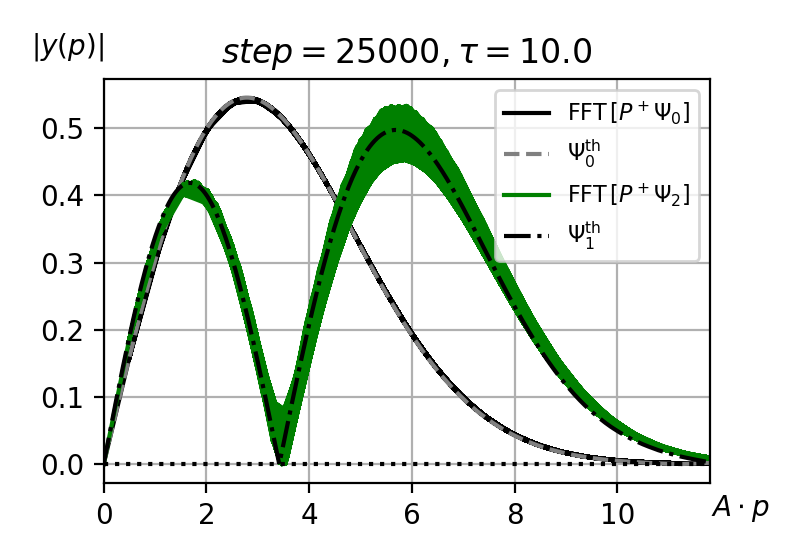}
\caption{\label{fig:wavefunction_harm}%
Parity-projected wave-functions as a function of $r/A$ using the relativistic kinetic term $H_{K}^{(3)}$ with the modified harmonic oscillator potential (left plot).\\
On the right we show the momentum space wave-functions, $\lvert y(p) \rvert$, according to Eq.~\eqref{eq:harmonic_oscillator_momentum_space} as a function of $A \cdot p$ obtained via a FFT, and compare them to the theoretical predictions following Ref.~\cite{Li:2005bd}.}
\end{figure}

We plot the convergence of the energies as a function of $\tau$ (top left plot in Fig.~\ref{fig:energ_proj_harm}) and the relative weights of the parity projections for the 3 lowest states versus $\tau$ (top right and bottom plots in Fig.~\ref{fig:energ_proj_harm}), and finally we plot the wave-functions in Fig.~\ref{fig:wavefunction_harm} where we also compare them to existing theoretical results for $l = 0$.

At small values of $\tau$ we can see some finite mass discretization artefacts, but for values of $\tau > 8$~GeV$^{-1}$ the values of all three states have stabilized and we do not see any decay of excited states.
The ground state is dominated by positive parity as one would expect.
For the first excited state, one would expect negative parity domination of the state, which we can see, but due to the slow convergence caused by the large mass, only at large $\tau$.
Finally, for the second excited state we -- as expected -- see positive parity at large $\tau$.
The above enables us to extract three distinct states with the expected parities and their corresponding energies reliably.

In contrast to the relativistic Coulomb potential, the relativistic harmonic oscillator does not have a divergence at the origin.
Additionally, a closed form analytic solution in momentum space exists~\cite{Li:2005bd} for $l = 0$ states which reads\footnote{%
In the following equations any dimensionful quantities are rescaled in order to be dimensionless; for details see Ref.~\cite{Li:2005bd}.\newline
An implementation of the recursive formula can be found at \url{https://github.com/quantumfdtd/relativistic_harmonic_oscillator}.}
\begin{equation}
\label{eq:harmonic_oscillator_momentum_space}
y(p) \equiv \sqrt{4\pi} p \Psi(\vec{p}) = \sum\limits_{n=0}^{\infty} c_{2n+1} \frac{p^{2n+1}}{(2n+1)!} \,.
\end{equation}
The momentum space wave-function, $\Psi(\vec{p})$, is the Fourier transform of the position space solution\footnote{%
See also the discussion in subsection~\ref{subsec:kin_terms} and the Eqs.~\eqref{eq:eigenmodes} to~\eqref{eq:HK0imp_coordinate_space}.}, $\Psi(\vec{r})$, and the recursive coefficients read
\begin{align}
c_{2n} &= 0 \quad\quad \text{for } n = 0,1,2,\ldots \,, \\
c_{1} &= 1 \,, \quad\quad c_{3} = M-E \,, \\
c_{2n+3} &= (M-E) c_{2n+1} + \sum\limits_{k=1}^{n} \left(\begin{array}{c} 2n+1 \\ 2k \end{array}\right) c_{2n-2k+1} M^{1-2k} (-1)^{k-1} (2k-1) \left[\frac{(2k-2)!}{2^{k-1}(k-1)!}\right]^{2} \,, \\
& \quad\quad \text{for } n = 1,2,3,\ldots \nonumber \,.
\end{align}
Similar to Ref.~\cite{Li:2005bd}, the sum in Eq.~\eqref{eq:harmonic_oscillator_momentum_space} needs to be truncated at some value $\tilde{N}$ for which we take their value of $\tilde{N} = 44$.
The convergence of this momentum space solution gets worse with decreasing $M$, such that very high values of $\tilde{N}$ are required and numerical instabilities due to division of large numbers (due to the factorials) spoil the results.

We show the results in position space as well as the momentum space results obtained via a FFT in Fig.~\ref{fig:wavefunction_harm}.
The latter is compared to the truncated analytic result of Eq.~\eqref{eq:harmonic_oscillator_momentum_space}.

In Fig.~\ref{fig:wavefunction_harm} we show the position space solution for the largest $\tau$ value available in the left panel as a function of $r/A$.
We can separate the states as discussed above but the relatively large contribution of the positive parity state to the first excited state is still visible.
In the right panel we show the momentum space wave-functions, $\lvert y(p) \rvert$, according to Eq.~\ref{eq:harmonic_oscillator_momentum_space} as a function of $A \cdot p$ obtained via a FFT, and compare them to the theoretical predictions following Ref.~\cite{Li:2005bd}.
We only show the $l = 0$ states, i.e., the ground state and the second excited state that we could clearly identify in Fig.~\ref{fig:energ_proj_harm} to be the ones with positive parity, because we can only compare those to the theoretical predictions.

To obtain the right plot of Fig.~\ref{fig:wavefunction_harm}, the FFT over the parity-projected wave-functions (left plot of Fig.~\ref{fig:wavefunction_harm}) is used.
This is similar to the implementation of the kinetic terms $H_K^{(1)}$, $H_K^{(2)}$ and $H_K^{(3)}$ (see subsection~\ref{subsec:kin_terms}).
There, a FFT is used to turn from position to momentum space, where the kinetic term is evaluated, and then an IFFT is used to turn back into position space.
Here, we use a FFT to turn position space into momentum space so that we can compare our results with the ones from Ref.~\cite{Li:2005bd}.

Note that the FFT has to be applied over the whole original lattice of dimension $N^3$, without any notion of effective volume $V^{\Box}$, i.e., without taking into account \texttt{min\_sep\_edge}.
Otherwise, we would be changing the normalization of the momentum coordinates.
After the FFT has been computed, the relation between the lattice indices and the value of the corresponding momentum coordinates is given by the inversion of Eq.~\eqref{eq:eigenmodes} and reads
\begin{equation}
\label{eq:normali_p}
\vec{p} = \frac{2\pi}{NA}(x_1,x_2,x_3) \,,
\end{equation}
where the $x_i$ are in lattice units and satisfy
\begin{equation}
\label{eq:FFT_layout}
x_i = \begin{cases}
x_i \,, & \quad \text{if} \quad x_i \le N/2 \\
N/2 - x_i \,, & \quad \text{otherwise}
\end{cases} \,,
\end{equation}
which is motivated by the usual layout of FFT implementations.

Ref.~\cite{Li:2005bd} plots the radial momentum space wave-function,
\begin{equation}
y(p) = \sqrt{4\pi}p\Psi(p) \,.
\end{equation}
Because only the $l = 0$ states are computed, these states have spherical symmetry.
The normalization of Ref.~\cite{Li:2005bd} is such that $y'(0) = 1$, because their computation amounts to a Taylor expansion around $p = 0$ that solves an ordinary differential equation (ODE) with initial conditions $y(0) = 0$, $y'(0) = 1$.
On the right plot of Fig.~\ref{fig:wavefunction_harm}, we normalize these solutions (theory lines) such that, instead,
\begin{equation}
\label{eq:normali_y}
\int\limits_0^{\max(p)} \lvert y(p) \rvert^2 \, \text{d}p = 1 \,,
\end{equation}
where $\max(p)$ is the maximum $p$ for which $y(p)$ has been evaluated.

For a comparison with the radial wave-functions, $y(p)$, of Ref.~\cite{Li:2005bd} (normalized as in Eq.~\eqref{eq:normali_y}), we have to modify the normalization of our results such that
\begin{equation}
\left(\frac{2\pi}{NA}\right)^3 \cdot \sum\limits_{\vec{p}} \lvert \Psi(\vec{p}) \rvert^2 = 1 \,,
\end{equation}
where $(2\pi/NA)$ is the reciprocal lattice spacing.
Then, these wave-functions, $\Psi(p)$, are multiplied by a factor $\sqrt{4\pi}p$ turning them into spherical coordinates.
Finally, in order to get rid of complex phases due to the FFT, the absolute values of our results and of the theory predictions are plotted for comparison.
With the results presented so far, we reproduce Fig.~1, and the values\footnote{%
The analytic results for $\tilde{N} = 44$ are reported as $(E_0^\mathrm{ana} - M) \approx 0.38627$~GeV and $(E_1^\mathrm{ana} - M) \approx 0.89864$~GeV in Tab.~1 of Ref.~\cite{Li:2005bd}, where the first (radially) excited ($l = 0$) state corresponds to the second excited state in our numerical simulation (since our first excited state has odd parity).
At $\tau = 10$~GeV$^{-1}$ we obtain numerical results for the ground state energy of $(E_0^\mathrm{num} - M) \approx 0.38645$~GeV, 
and first or second excited state energies of $(E_1^\mathrm{num} - M) \approx 0.64889$~GeV 
or $(E_2^\mathrm{num} - M) \approx 0.90567$~GeV 
(corresponding to $E_1^\mathrm{ana}$), respectively, which is agreement at 0.5\textperthousand{} level for $E_0$ or still at sub-percent level for the excited state.} of Tab.~1 of Ref.~\cite{Li:2005bd}.

\subsection{Convergence conditions}
\label{subsec:convergence_conditions}

The convergence of the energies and wave-functions shown in Figs.~\ref{fig:energ_proj_KIN3} to~\ref{fig:wf_comp_parity} is pretty clean.
Discretization errors that are predominantly due to the lattice implementation of the Coulomb potential are visible on the coarser lattices and excited states are affected by finite volume effects in small lattice volumes.
The wave-functions generated by the \texttt{quantumfdtd} can be considered reliable, only after the following conditions are met:
\begin{itemize}
\item Appropriate parity projection and normalization has been performed using the provided post-processing scripts for separating the wave-functions (see subsection~\ref{subsec:parity_projection}).
\item Using the snapshot facility of the code for multiple values of $\tau$ the wave-function has been shown to be stable within small changes of $\tau$ and the energy $E_{s,i}(\tau)$ should be in a plateau, i.e., flat or, at least, in a saddle point.
\item $A < M^{-1}$, such that the typical radius of the ground state is substantially larger than the lattice spacing $A$, $\langle \rho \rangle^{\Box}_{P^+\Psi_0} \gg 1$.
This condition is violated in subsection~\ref{subsec:comp_high_mass} in the example.
\item $2\langle \rho \rangle^{\Box}_{P^{\pm}\Psi_{1,2}} \ll N$, such that typical radii of the excited states are substantially smaller than half of the box size $N$.
\item It is important to choose the temporal grid spacing $\texttt{EPS} \propto \texttt{A}^2$ relative to the lattice spacing \texttt{A} (see the \texttt{EPS} parameter on page~\pageref{flag:EPS}).
\end{itemize}

\clearpage

\section{Compilation and running}
\label{sec:compile_run}

We require $\text{\texttt{\$\{NCORES\}}}$ \texttt{MPI}~processes, with $\text{\texttt{\$\{NCORES\}}}$ being a divisor of $N$.
The following external libraries are used:
\begin{itemize}
\item \texttt{MPI}~library
\item \texttt{FFTW\_MPI}, version~3\footnote{%
\url{http://fftw.org/} and Ref.~\cite{Frigo:2005zln}}
\item GNU Scientific Library (\texttt{GSL})\footnote{%
\url{https://www.gnu.org/software/gsl/} and Ref.~\cite{gsl}}, linked to \texttt{CBLAS}\footnote{%
\url{https://www.netlib.org/blas/} and Refs.~\cite{blas1, blas2, blas3}}
\item For some of the post-processing scripts, \texttt{Python~3} is required\footnote{%
\url{https://www.python.org/downloads/}}
\end{itemize}
There are both, open source (\texttt{openMPI}\footnote{%
\url{https://www.open-mpi.org/}}) and proprietary implementations of the \texttt{MPI}~library.
The open source \texttt{openMPI}~2.1.1 and the proprietary \texttt{IBM~MPI}~1.4 versions of the \texttt{MPI}~library have been used for testing purposes.
Concerning \texttt{FFTW\_MPI} and \texttt{GSL}, both are open source.
Versions \texttt{FFTW\_MPI}~3.5.7 and \texttt{GSL}~2.5 have been used for testing.
Hence, the usage of this code does not require any proprietary software.

The directory structure is as follows:
\begin{description}
\item[\texttt{build/src}:] Source code.
\item[\texttt{build/include}:] Header files.
\item[\texttt{build/obj}:] Folder for intermediate object files.
\item[\texttt{data} and \texttt{data/snapshot}:] Folder for default output files and snapshot files.
The file format is ASCII, and the output consists of (i) the potential (see the \texttt{SAVEPOT} parameter for formating options), (ii) wave-functions (see the \texttt{DATAFOLD} parameter for formating options), and, if the parameter \texttt{DUMPSNAPS} is set, (iii) snapshots (see the parameter \texttt{DUMPSNAPS}) and (iv) eigen-energies.
The snapshots are saved in \texttt{data/snapshot} which must exist if one wants to store them, otherwise, the run will fail.
In the cases (i), (ii), and (iii), the first columns of the ASCII files are the lattice coordinates $x_1$, $x_2$, and $x_3$, being the Cartesian coordinates in lattice units.
The fourth column is the distance to the origin in lattice units which is corrected in case the potential is centered on the lattice center.
The eigen-energies are stored in the files \texttt{decay.dat}, \texttt{ground\_state.out}, \texttt{first\_excited\_state.out}, and \texttt{second\_excited\_state.out}.
The first file is an ASCII list containing:
\begin{equation*}
\text{step} \quad \tau \quad \Real E_{b,0} \quad \Imag E_{b,0} \quad \Real E_0 \quad \Imag E_0 \,.
\end{equation*}
The other 3 files, \texttt{ground\_state.out}, \texttt{first\_excited\_state.out}, and \texttt{second\_excited\_state.out}, contain the eigen-energies of the ground state ($i = 0$), the first ($i = 1$), and the second ($i = 2$) excited state, respectively.
Their layouts are:
\begin{equation*}
\text{step} \quad \tau \quad \Real E_{b,i} \quad \Imag E_{b,i} \quad \Real E_i \quad \Imag E_i \quad \langle x \rangle \quad \langle y \rangle \quad \langle z \rangle \quad N \quad A \quad M \quad \sigma \,.
\end{equation*}
In all the cases, $E_{b,i}$ stands for the \emph{bare} eigen-energy and $E_i$, for the actual eigen-energy (for an explanation see the definition of the parameter \texttt{UPDATE} in subsection~\ref{flag:UPDATE} on page~\pageref{flag:UPDATE}).
$\langle \ldots \rangle$ are the respective expectation values of the coordinates (in lattice units) with respect to the origin of the potential.
\item[\texttt{input}:] Folder for default input files.
The file \texttt{input/params.txt} is necessary for running the program.
\item[\texttt{license}:] GNU Public License.
\item[\texttt{archive}:] Several versions of the \texttt{Makefile}, for different machine configurations (architecture).
\item[\texttt{scripts}:] Examples of scripts for running the program and post-processing of the output data.
This includes the \texttt{Python~3}~\cite{python} module \texttt{quantumfdtd.\allowbreak{}py}, that offers a \texttt{Python} class (\texttt{quantumfdtd}) for the post-analysis inside the Pandas~\cite{reback2020pandas, mckinney-proc-scipy-2010} framework as well as the scripts \texttt{symmetrize.\allowbreak{}sh}, for computing parity projections, and several scripts for extracting plots\footnote{%
\texttt{postprocessing.sh} for the potential, energies and wave-functions and \texttt{postprocessing\_\allowbreak{}sym.sh} for the parity projections.
In both cases, plots of the full 3D wave-functions snapshots are obtained if they have been computed.}.
\end{description}

The root directory contains a \texttt{Makefile} that allows one to build the program provided that all of the external libraries are installed.
If the libraries are installed in non-standard locations, this \texttt{Makefile} should be modified accordingly.

The following commands, executed in the root folder of the provided tarball, are required for building and executing the program:
\begin{verbatim}
$ make
$ mpirun -N ${NCORES} ./mpisolve
\end{verbatim}
Here, \texttt{\$\{NCORES\}} is the number of cores to be used which should be a divisor of the number of the spatial grid points.
The source code is available on GitHub: \url{https://github.com/quantumfdtd/quantumfdtd_v3}.

\section{Parameters of the program}
\label{sec:params}

The parameters listed in the following subsections should be defined in the standard parameter file located at \texttt{input/params.txt}.
The \texttt{input} directory should be located in the directory where the program is also run from.
Note that many of the post-processing scripts require the corresponding input file in the standard location (\texttt{input/params.txt}) to work.
The configuration file accepts full line comments, that are marked with an initial double backslash, \texttt{\textbackslash\textbackslash}, on the commented line.

For convenience, the program accepts command line parameters as well.
They can be parsed as follows:
\begin{verbatim}
$ mpirun -N ${NCORES} ./mpisolve -KEY1 ARG1 -KEY2 ARG2 ...
\end{verbatim}
Unless stated differently, all of the numerical values are separated by tabs in all of the output files.

\subsection{Basic configuration}
\label{subsec:basic_params}

This program is based on an iterative method for finding both the ground state energy and the two first excited state energies.
The corresponding wave-functions are computed as well.
Hence, most of the basic configuration flags that are required for running the program are related to the setup of the lattice (discretization) or to the iterative procedure.
The following parameters are required for running the program:
\begin{description}
\item[\texttt{NUM}:] Number of spatial grid points in each direction.
Note that they should be divisible by the number of computational nodes, i.e., $\text{\texttt{NUM}} \mod N_{\text{\texttt{MPI}~cores}} = 0$.
\item[\texttt{A}:] Spatial grid spacing.
Default units: 1/GeV.
\item[\texttt{MASS}:] Reduced mass of the simulated system.
Default units: GeV.
\item[\texttt{KINTERM}:]\label{flag:KINTERM} Selection of the kinetic term (see subsection~\ref{subsec:kin_terms}) to be used:
\begin{itemize}
\item The non-relativistic kinetic term based on FDTD (\texttt{0}).
\item Two other non-relativistic kinetic terms based on FFT (\texttt{1},\texttt{2}).
\item The relativistic kinetic term also based on FFT (\texttt{3}).
\end{itemize}
\item[\texttt{POTENTIAL}:] Potential to simulate.
There are 6 hard-coded potentials described in subsection~\ref{subsec:hard_coded_potentials} and 4 different options to read the potential from file described in subsection~\ref{subsec:external_potentials}.
\item[\texttt{EPS}:]\label{flag:EPS} Temporal grid spacing.
Should be proportional to $\text{\texttt{A}}^{2}$ with a proportionality constant that can be determined from empirical stability analysis.
\item[\texttt{UPDATE}:]\label{flag:UPDATE} How many steps should be taken before checking the \texttt{SNAP\-UPDATE}, \texttt{SNAP\-DUMP} and \texttt{STEPS} variables, and computing observables\footnote{%
Ground state energy and the expectation values $\braket{r^{2}}$, $\braket{x}$, $\braket{y}$, and $\braket{z}$.
Note that here $\braket{\rho^{2}}$ is with respect to the center of lattice, i.e., with respect to the point $(x_{1},x_{2},x_{3}) = (N_H,N_H,N_H)$ with $N_H = (N+1)/2$.} and printing them to command line.\\
If the \texttt{SAVEDECAY} flag is enabled, observables are also saved to the file \texttt{decay.dat}, as an ASCII table: $T$, $E_{b}$.
Here, $T$ stands for the total time (\texttt{EPS}$\times$\texttt{\#steps}) and $E_{b}$ is the bare ground state energy.\\
Note that some hard-coded potentials split the actual physical potential $V$ in two parts\footnote{%
There is an option for using this splitting for externally provided potentials as well (see subsection~\ref{subsec:external_potentials}).}:
A \emph{bare} ($V_{b}$) and a \emph{subtracted} ($V_{s}$) potential, with the actual physical potential being $V = V_b - V_s$.
This option is inherited from previous versions of the \texttt{quantumfdtd} code~\cite{Strickland:2009ft, Margotta:2011ta}, where it was used to introduce additive scalar constants to the potential, that do not modify the dynamics of the Schrödinger equation and can thus be split off.
$V_{b}$ is the one used inside the iterative procedure, for computing the wave-functions $\Psi_i$ and energies $E_{b}$.
Afterwards, the \emph{subtracted} physical value of the energy, $E_{s}$, is computed from the subtracted potential $V_{s}$ via
\begin{equation}
\label{eq:subtracted_energy}
E_{s,i} = E_{b,i} - \sum\limits_{\vec{r} \in V} V(\vec{r}) \lvert \Psi_{i}(\vec{r}) \rvert^{2} \,,
\end{equation}
where the integral extends over all of the lattice positions.
This is correct for the potentials $V(r)$ in the particular case $V(r) = V_{b}(r) - V_{s}$, where the \emph{subtracted potential} $V_{s}$ is a scalar constant.
If $V_s(r)$ depends on the position, although the code will run, there is no guarantee that the solution is correct.
In the general case, setting $V_b(r) = V(r)$ and $V_s(r) = 0$ should be the preferred way.
We have maintained this option for backwards compatibility with previous versions of the \texttt{quantumfdtd} code.
\item[\texttt{STEPS}:] Maximum number of steps to take.
This limit is enforced each time a number of steps given by the variable \texttt{UPDATE} is evaluated.
Hence, the actual number of steps to take is exactly \texttt{STEPS} only if \texttt{STEPS} is divisible by \texttt{UPDATE}.
Otherwise, it will be the smallest multiple of \texttt{UPDATE} bigger than \texttt{STEPS}.
This is the actual number of steps to take if the \texttt{TOLERANCE} parameter is set to a negative value.
\item[\texttt{SNAPUPDATE}:]\label{flag:SNAPUPDATE} How many steps should be taken before enforcing both normalization of the wave-functions ($\sum_{\vec{r} \in V} \lvert\Psi_{i}(\vec{r})\rvert^{2} = 1$) and their required symmetries (see the flag \texttt{INITCONDTYPE}).
It triggers the check of the tolerance condition (see \texttt{TOLERANCE} flag).
It also triggers the storage of the 2 snapshots that are required for the extraction of the first and second excited states via the overlapping algorithm (see subsection~\ref{subsec:iterative_procedure}).
The post-analysis scripts based on the \texttt{Python~3} module \texttt{quantumfdtd.py} (see subsections~\ref{subsec:post_processing_scripts} and~\ref{subsec:post_processing_python_module}) require both, saving the full snapshots to disk (setting \texttt{DUMPSNAPS=4}), and, setting the \texttt{STEPS} variable to a multiple of \texttt{SNAPUPDATE}, i.e., ensuring that \texttt{STEPS\%SNAPUPDATE=0}.
\item[\texttt{SNAPDUMP}:] In case \texttt{DUMPSNAPS} (see below) is set to a non-null value, how many steps should be taken before storing a wave-function snapshot to disk.
\item[\texttt{TOLERANCE}:] Convergence tolerance with respect to the ground state energy.
The stop condition being\footnote{%
$E_{b,i} - E_{b,i-1}$ stands for the difference between the current bare ground state energy and the one obtained when \texttt{SNAPUPDATE} was triggered previously.} $E_{b,i} - E_{b,i-1} < \texttt{TOLERANCE}$.
It can be disabled by setting a negative value.
\item[\texttt{SAVEWAVEFNCS}:] Save a full snapshot of the 3d wave-functions at the end of the run.
Set to \texttt{1} for activation.
Note that most of the post-processing scripts require the intermediate snapshots, too.
\item[\texttt{SAVEDECAY}:] Save a table of energy versus time.
This is particularly useful for checking the convergence of the iterative procedure to the ground state, especially, when using the FFT-based kinetic terms.
\item[\texttt{SAVEPOT}:] Name of the external file where the potential that is used will be saved to.
The file format is an ASCII list containing:
\begin{equation*}
x_{1} \quad x_{2} \quad x_{3} \quad \rho^2 \quad \Real V \quad \Imag V \,,
\end{equation*}
where $\rho^2$ is the distance to the center of the potential in lattice units, and $x_1$, $x_2$, and $x_3$ are the lattice coordinates.
These $x_i$ are shifted by $-1/2$ if the potential is centered on the lattice volume.
\item[\texttt{DATAFOLD}:] Name of the folder where the full output, wave-functions, and snapshots will be saved to.
By default, is set to \texttt{data}.
Note that many of the post-processing scripts require this standard directory \texttt{data} to work.
The file format is an ASCII list containing:
\begin{equation*}
x_{1} \quad x_{2} \quad x_{3} \quad \rho^2 \quad \Real\Psi_i \quad \Imag\Psi_i \,,
\end{equation*}
where $\rho^2$ is the distance to the center of the potential in lattice units, and $x_1$, $x_2$, and $x_3$ are the lattice coordinates.
For the indices we have: $i = 0$ (ground state), $1$ (first excited state), or $2$ (second excited state).
The coordinates $x_i$ are shifted by $-1/2$ if the potential is centered on the lattice volume.
\item[\texttt{DUMPSNAPS}:]\label{flag:DUMPSNAPS} Dump debugging files containing either "snapshot slices" or full 3D wave-functions:
\begin{itemize}
\item Disables snapshot recording (\texttt{0}).
\item Store "snapshot slices" of the ground state (\texttt{1}).
Only valid for potentials centered on $(N_H,N_H,N_H)$ with $N_H = (N+1)/2$.
Do not use when centering the potential on $(0,0,0)$.
\item Store full 3D wave-functions of the ground state (\texttt{2}).
\item Store "snapshot slices" of the ground state, and of the first and second excited states (\texttt{3}).
Only valid for potentials centered $(N_H,N_H,N_H)$ with $N_H = (N+1)/2$.
Do not use when centering the potential on $(0,0,0)$.
\item Stores full 3D wave-functions of the groundstate, and of the first and second excited states (\texttt{4}).
The post-analysis scripts described in the subsections~\ref{subsec:post_processing_scripts} and~\ref{subsec:post_processing_python_module} require this value for \texttt{DUMPSNAPS}.
\end{itemize}
The full 3D wave-functions are stored in the same way as the ones triggered by \texttt{DATAFOLD}.\\
The file format for the "slices" consists of three concatenated ASCII lists, separated by a row containing "\&\&", respectively.
These lists are, respectively:
\begin{align*}
x_{1} && \Real V[x_{1}, N_H, N_H] && \Imag V[x_{1}, N_H, N_H] \\
x_{2} && \Real V[N_H, x_{2}, N_H] && \Imag V[N_H, x_{2}, N_H] \\
x_{3} && \Real V[N_H, N_H, x_{3}] && \Imag V[N_H, N_H, x_{3}]
\end{align*}
The coordinates of $V[x_1,x_2,x_3]$ are in lattice units, and are shifted by $-1/2$ if the potential is centered on $(N_H,N_H,N_H)$ with $N_H = (N+1)/2$.
These "slices" are a legacy function of \texttt{quantumfdtd~v2} that should not be used if the potential is centered in $(0,0,0)$ as mentioned above because it has not been adapted.
In both cases, the output is stored in the \texttt{data/snapshot} directory.
This directory must exist prior to running the program.
The file names are \texttt{data/\allowbreak{}snapshot/\allowbreak{}wavefunction\_\allowbreak{}step\_\allowbreak{}i.dat}, where \texttt{step} is the step associated to the snapshot ($\tau = \text{EPS} \cdot \text{step}$) and \texttt{i} is an index running over the computational nodes.
\end{description}

\subsection{Initial conditions}
\label{subsec:initial_conditions}

The following parameters allow changing the initial state condition.
The Coulomb-type initial conditions are intended to be a sensible initial guess for spherically symmetric confining central potentials.

\begin{description}
\item[\texttt{INITCONDTYPE}:] Initial condition to use.
Valid values:
\begin{description}
\item[\texttt{0}:] Read initial condition from \texttt{wave-function\_\allowbreak{}0\_\allowbreak{}\#.dat} files in the \texttt{data} directory.
\item[\texttt{1}:] Random Gaussian noise with standard deviation given by the \texttt{SIG} parameter.
\item[\texttt{2}:] Non-relativistic Coulomb-type initial condition, centered at the same point where the potential is centered.
I.e., either in the center of the lattice volume at the point $(N_H,N_H,N_H)$ with $N_H = (N+1)/2$ (this applies to all of the hard-coded potentials with code <\texttt{100}), or in the origin of the lattice at the point $(x_{1},x_{2},x_{3}) = (0,0,0)$, otherwise.
By default, we introduce a linear combination of the states $1s$, $2s$, $2p$ ($m = 0$), and the real part of the $2p$ ($m = \pm 1$).
Note that the selection of the $z$ axis of the $2p$ states is chosen through the flag \texttt{INITCONDAXIS}.
\item[\texttt{3}:] Constant wave-function with value $0.1$.
\item[\texttt{4}:] Boolean test grid, $\Psi(\vec{r}) = (x_{1} \mod 2) \cdot (x_{2} \mod 2) \cdot (x_{3} \mod 2)$.
\item[\texttt{5}:] This was an intermediate step during the development of the new version of \texttt{quantumfdtd}.
The idea was to define an initial wave-function containing a sensible selection of Fourier components.
We ended up with the function shown in Listing~\ref{listing} on page~\pageref{listing}.
If the need arises to modify this function in order to include own guesses for particular potentials, we advise looking into the function \texttt{void setInitialConditions(int seedMult)}, contained in the file \texttt{build/\allowbreak{}src/\allowbreak{}initialconditions.cpp}.
\end{description}
\item[\texttt{INITCONDAXIS}:] For \texttt{INITCONDTYPE} set to \texttt{2}, this flag sets the $z$ axis of the $2p_z$ components of the wave-function.
For the choices \texttt{0}, \texttt{1}, or \texttt{2} the axis is set to $x_{1}$, $x_{2}$, or $x_{3}$, respectively.
\item[\texttt{INITSYMMETRY}:]\label{flag:INITSYMMETRY} Symmetrize the wave-function.
Valid values are:
\begin{description}
\item[\texttt{0}:] Disable option.
\item[\texttt{1}:] Symmetric with respect to the $x_{3}$ direction.
\item[\texttt{2}:] Anti-symmetric with respect to the $x_{3}$ direction.
\item[\texttt{3}:] Symmetric with respect to the $x_{2}$ direction.
\item[\texttt{4}:] Anti-symmetric with respect to the $x_{2}$ direction.
\end{description}
Note that because of the internal memory layout, it is not possible to offer options in the $x_{1}$ direction without a high \texttt{MPI}~overhead.
\item[\texttt{SIG}:] Standard deviation - used when \texttt{INITCONDTYPE} is set to \texttt{1}.
\end{description}

{
\captionsetup[table]{name=Listing}
\begin{table}[ht]
\centering
\begin{minipage}[t]{\textwidth}
\centering
\begin{lstlisting}[frame=single]
for (int skx=1; skx<=NUM; skx++)
for (int sky=1; sky<=NUM; sky++)
for (int skz=1; skz<=NUM; skz++){

  int kx = skx-1;
  int ky = sky-1;
  int kz = skz-1;

  if (2*kx>NUM) kx -= NUM;
  if (2*ky>NUM) ky -= NUM;
  if (2*kz>NUM) kz -= NUM;

  int k2 = kx*kx+ky*ky+kz*kz;

  double fact=((double)(pow(2,k2)*NUM*NUM*NUM));
  fact = 1./fact;

  for (sx=1;sx<=NUMX;sx++)
  for (sy=1;sy<=NUM;sy++)
  for (sz=1; sz<=NUM;sz++){

    int x = sx+NUMX*(nodeID-1)-1;
    int y = sy-1;
    int z = sz-1;

    w[sx][sy][sz] = fact*
      exp(dcomp(0.,M_PI*((double)kx*x)/((double)NUM)))*
      exp(dcomp(0.,M_PI*((double)ky*y)/((double)NUM)))*
      exp(dcomp(0.,M_PI*((double)kz*z)/((double)NUM)));

    }
  }
\end{lstlisting}
\end{minipage}
\caption{\label{listing}
Implementation of \texttt{INITCONDTYPE=5}: a selection of Fourier components as initial guess.}
\end{table}
}

\clearpage

\subsection{Hard-coded potentials}
\label{subsec:hard_coded_potentials}

The program comes bundled with several hard-coded potentials, that were already present in the original code~\cite{Strickland:2009ft, Dumitru:2009ni, Dumitru:2009fy}.
The user should be aware that some of these potentials may be unphysical depending on the boundary conditions.
We stress that all coordinates $(x_{1},x_{2},x_{3})$ and corresponding radii $(r,~\tilde{r})$ are dimensionless numbers.
\begin{description}
\item[\texttt{0}:] No potential ($V(\vec{r}) = 0$).
This is equivalent to an infinitely deep 3d well due to boundary conditions - provided that $\text{\texttt{KINTERM}} = 0$ is used.
\item[\texttt{1}:] 3d square well in the center of the lattice volume:
\begin{equation}
V(x_{1},x_{2},x_{3}) = -10~\text{GeV} \,\cdot \Theta(N/4 - \lvert x_{1} \rvert) \cdot \Theta(N/4 - \lvert x_{2} \rvert) \cdot \Theta(N/4 - \lvert x_{3} \rvert) \,,
\end{equation}
where $\Theta(x)$ is the Heaviside Theta function.
\item[\texttt{2}:] Radial Coulomb potential:
\begin{equation}
V(x_{1},x_{2},x_{3}) =
\begin{cases}
-\frac{1}{A\rho} \,, & \text{for } \rho > 1 \,, \\
-\frac{1}{\texttt{A}} \,, & \text{for } \rho \leq 1 \,,
\end{cases}
\end{equation}
where $\rho = \sqrt{x_{1}^2 + x_{2}^2 + x_{3}^2}$.
\item[\texttt{3}:] Elliptical Coulomb potential:
\begin{equation}
V(x_{1},x_{2},x_{3}) =
\begin{cases}
-\frac{1}{A\tilde{\rho}} \,, & \text{for } \tilde{\rho} > 1 \,, \\
-\frac{1}{\texttt{A}} \,, & \text{for } \tilde{\rho} \leq 1 \,,
\end{cases}
\end{equation}
where $\tilde{\rho} = \sqrt{x_{1}^2 + x_{2}^2 + 4 x_{3}^2}$.
\item[\texttt{4}:] 3d harmonic oscillator:
\begin{equation}
\label{eq:hardcoded_harmonic_oscillator}
V(x_{1},x_{2},x_{3}) = -\frac{A^2}{2} (x_{1}^2 + x_{2}^2 + x_{3}^2)~\text{GeV}^{3} \,.
\end{equation}
\item[\texttt{5}:] Complex 3d harmonic oscillator:
\begin{equation}
V(x_{1},x_{2},x_{3}) = -\frac{A^2}{2} (1 + \text{i}) (x_{1}^2 + x_{2}^2 + x_{3}^2)~\text{GeV}^{3} \,.
\end{equation}
\item[\texttt{6}:] Cornell potential with string breaking:\\
We take the string-breaking scale to be $r_{\text{sb}} = 5.5745$~GeV$^{-1}$ and define
\begin{equation}
\hat{r} =
\begin{cases}
A\rho \,, & \text{for } A\rho < r_{\text{sb}} \,, \\
r_{\text{sb}} \,, & \text{for } A\rho \geq r_{\text{sb}} \,,
\end{cases}
\end{equation}
where $\rho = \sqrt{x_{1}^2 + x_{2}^2 + x_{3}^2}$, and
\begin{equation}
V(x_{1},x_{2},x_{3}) = -\frac{0.385}{\hat{r}} + \texttt{SIGMA} \, \hat{r} + 4 \, \texttt{MASS} \,,
\end{equation}
where \texttt{MASS} is the \texttt{quantumfdtd} reduced mass (see subsection~\ref{subsec:basic_params}) and \texttt{SIGMA} is an independent parameter specified in the input file.
\item[>\texttt{99}:] If $100$ is added to any of the previous potential codes, the corresponding potential will be loaded with its origin centered at the point $(x_{1},x_{2},x_{3}) = (0,0,0)$ instead of at the point $(x_{1},x_{2},x_{3}) = (N_H,N_H,N_H)$ with $N_H = (N+1)/2$.
This is especially useful for periodic boundary conditions.
\end{description}

\subsection{External potentials}
\label{subsec:external_potentials}
The following values of \texttt{POTENTIAL} offer different options to read external potentials from files:
\begin{description}
\item[\texttt{90}:] Read external potential as a table containing
\begin{equation*}
x_{1} \text{\textvisiblespace} x_{2} \text{\textvisiblespace} x_{3} \text{\textvisiblespace} \Re(V) \text{\textvisiblespace} \Im(V) \,,
\end{equation*}
separated by spaces.
\item[\texttt{91}:] Read external potential as a table containing
\begin{equation*}
\rho^{2} \text{\textvisiblespace} \Re(V) \text{\textvisiblespace} \Im(V) \,,
\end{equation*}
separated by spaces.
Pay attention to the fact that the variable is $\rho^2$ and not $\rho$.
\item[>\texttt{189}:] As in the previous subsection~\ref{subsec:hard_coded_potentials}, \texttt{190} and \texttt{191} are, respectively, the same as \texttt{90} and \texttt{91} but centered on $(0,0,0)$.
\end{description}

The following configuration flags are only valid when loading potentials from external files\footnote{%
Values of the \texttt{POTENTIAL} parameter: \texttt{90}, \texttt{91}, \texttt{190} and \texttt{191}}:
\begin{description}
\item[\texttt{EXPOT}:] Name of the file containing the external potential.
The file format depends on the selected kind of external potential as described above.
\item[\texttt{POTCRITR}:] Defines a point from which on (in units of the lattice spacing), the potential will be a $\chi^{2}$ adjustment to the following function:
\begin{equation}
V_{A,B,\sigma}(\rho) = A + B/\rho + \sigma \cdot \rho \,.
\end{equation}
The parameters $A$, $B$, and $\sigma$ are written to console.
Only valid for values of the \texttt{POTENTIAL} parameter: \texttt{90} and \texttt{190}.
\item[\texttt{POTFLATR}:]\label{flag:POTFLATR} Defines a point from which on (in units of the lattice spacing), the potential is flat taking the value of the point previous to it.
Only valid for \texttt{POTENTIAL} values \texttt{90} and \texttt{190}.
\end{description}

The potentials are implemented in the function \texttt{potential} in the file \texttt{build/src/potentials.cpp}.
The potential number is a global variable called \texttt{POTENTIAL}.
There already is a loop in which new potentials could be easily implemented.
The function \texttt{potentialSub} in the same source file implements the \emph{subtracted}\footnote{%
See the definition of the \texttt{UPDATE} flag in subsection~\ref{flag:UPDATE} on page~\pageref{flag:UPDATE}.} potential.

\subsection{Post-processing scripts}
\label{subsec:post_processing_scripts}

The following \texttt{Shell} and \texttt{Python~3} scripts have been included for the post-processing of the wave-functions.
They can be found in the \texttt{scripts} folder.
Note that these scripts rely on the full snapshots being written to disk (by setting \texttt{DUMPSNAPS=4}), and, setting the \texttt{STEPS} variable to a multiple of \texttt{SNAPUPDATE}, i.e., ensuring that \texttt{STEPS\%SNAPUPDATE=0}.
The only scripts that do not explicitly require the full snapshots are \texttt{symmetrize\_wf.py} and \texttt{normalize.py}.
The other ones are part of a processing chain that post-process said snapshots explicitly and thus require their existence.

Unless otherwise stated, they should be run from the parent folder where both the \texttt{data} and \texttt{input} folders are located:
\begin{description}
\item[\texttt{symmetrize.sh}:] Computes the projections and normalizations of all of the wave-functions, using the scripts \texttt{symmetrize\_\allowbreak{}wf.py} and \texttt{normalize.py}.
It accepts direct output stored in the \texttt{data} folder or it accepts wave-functions which are stored in \texttt{gz} compressed \texttt{tar} archives named \texttt{wave-\allowbreak{}function.\allowbreak{}tgz} which allows for batch evaluation of results from many runs.
Non-normalized output files named \texttt{wave-function\_\allowbreak{}\%i\_\allowbreak{}\%p.dat} and normalized ones named \texttt{wave-function\_\allowbreak{}\%i\_\allowbreak{}\%p\_\allowbreak{}norm.dat} are generated.
\texttt{\%i}=0,1,2 denotes the ground, first and second excited states, respectively.
\texttt{\%p} stands for the projection of the wave-function, as follows:
\begin{description}
\item[\texttt{wave-function\_\allowbreak{}\%i\_\allowbreak{}all.dat.gz}:]
The wave-function as computed by \texttt{quantumfdtd}, but in a single file instead of one file per computational node.
\item[\texttt{wave-function\_\allowbreak{}\%i\_\allowbreak{}p.dat.gz}:] Positive parity projection $P^+$, Eq.~\eqref{eq:parity}.
\item[\texttt{wave-function\_\allowbreak{}\%i\_\allowbreak{}m.dat.gz}:] Negative parity projection $P^-$, Eq.~\eqref{eq:parity}.
\item[\texttt{wave-function\_\allowbreak{}\%i\_\allowbreak{}pk.dat.gz}:] Negative parity projection around-an-axis $P^-_{\vec{p}_k}$, Eq.~\eqref{eq:parity_axis}.
\item[\texttt{wave-function\_\allowbreak{}\%i\_\allowbreak{}all\_\allowbreak{}norm.dat.gz}:] Full normalized wave-function.
\item[\texttt{wave-function\_\allowbreak{}\%i\_\allowbreak{}p\_\allowbreak{}norm.dat.gz}:] Normalized $P^+$ projection.
\item[\texttt{wave-function\_\allowbreak{}\%i\_\allowbreak{}m\_\allowbreak{}norm.dat.gz}:] Normalized $P^-$ projection.
\item[\texttt{wave-function\_\allowbreak{}\%i\_\allowbreak{}pk\_\allowbreak{}norm.dat.gz}:] Normalized $P^-_{\vec{p}_k}$ projection.
\item[\texttt{wave-function\_\allowbreak{}\%i\_\allowbreak{}proj.dat}:] Weights of the parity projections.
\end{description}
For all of the cases but the normalized projection $P^-_{\vec{p}_k}$, the output is a gzipped ASCII file containing:
\begin{equation*}
x_{1} \quad x_{2} \quad x_{3} \quad \rho^2 \quad \Real\Psi_i \quad \Imag\Psi_i \,,
\end{equation*}
For the normalized projection $P^-_{\vec{p}_k}$, the output follows the format:
\begin{equation*}
x_{1} \quad x_{2} \quad x_{3} \quad \rho \quad \cos\theta \quad \Real\Psi_i \quad \Imag\Psi_i \,,
\end{equation*}
\item[\texttt{postprocess.sh}:] Generates plots of the potential and the wave-functions.
\item[\texttt{postprocess\_\allowbreak{}sym.sh}:] Generates plots of the parity projected wave-functions obtained with \texttt{symmetrize.sh}.
\item[\texttt{allpostprocess.sh}:] Runs \texttt{postprocess.sh}, \texttt{symmetrize.sh} and \texttt{postprocess\_\allowbreak{}sym.sh}, sequentially.
\item[\texttt{cleandatafiles.sh}:] Runs \texttt{rm -rf data/*.* \allowbreak{}data/snapshot/*.* \allowbreak{}debug/*.*} for cleaning up a previous run.
\item[\texttt{symmetrize\_\allowbreak{}wf.py}:] Computes the positive and negative parity projections of the given wave-function.
It is called as:\\
\texttt{../scripts/symmetrize\_\allowbreak{}wf.py [params.txt] wave-function\_\allowbreak{}\%i\_\allowbreak{}\%d.dat}\\
The optional parameter \texttt{[params.txt]} encode the path to the config file, from which the simulation parameters like $N$, $A$, or the centering of the lattice are taken.
If not provided, the script expects to be called from the \texttt{data} folder and tries to load the config file from the relative path \texttt{../input/params.txt}.\\
\texttt{\%i} should be substituted by the numbers 0, 1 and 2 for the ground, first and second excited states, respectively.\\
\texttt{\%d} should be left on the actual call.
It means that the script will look for files where \texttt{\%d} is a numerical index starting on 0.
This procedure has been implemented because \texttt{quantumfdtd} splits the output data between as many files as computational nodes are used.
This \texttt{\%d} is the numerical index of these files.
In practice, for the ground state wave-function, the call (from the \texttt{data} folder) is \texttt{../scripts/symmetrize\_\allowbreak{}wf.py wave-function\_\allowbreak{}0\_\allowbreak{}\%d.dat}
\item[\texttt{normalize.py}:] Normalizes the wave-functions to unity over the lattice volume.
It is expected to be run from inside the \texttt{data} folder, but it can be run from any folder where wave-functions are stored.
It is called \texttt{../scripts/symmetrize\_\allowbreak{}wf.py wave-function.dat out.dat}.
\end{description}

\subsection{Post-processing \texorpdfstring{\texttt{Python}}{Python} module}
\label{subsec:post_processing_python_module}

We provide a post-processing \texttt{Python~3} module in \texttt{scripts/\allowbreak{}quantumfdtd.\allowbreak{}py}, whose class \texttt{quantumfdtd} aims at helping with the analysis via \texttt{Python~3}~\cite{python}, Pandas~\cite{reback2020pandas, mckinney-proc-scipy-2010}, NumPy~\cite{oliphant2006guide, 5725236, harris2020array}, and Matplotlib~\cite{Hunter:2007}.
It accepts as input the output of either \texttt{symmetrize.sh} or \texttt{allpostprocess.sh} from the previous subsection~\ref{subsec:post_processing_scripts}.
Note that every script relying on the class functions \texttt{load\_wf}, \texttt{load\_proj\_components}, or \texttt{get\_snaps} requires the full snapshots to be present.

The \texttt{quantumfdtd} class functions are:
\begin{description}
\item[\texttt{\_\_init\_\_(self, case, config=None)}:] Constructor.
It requires a \texttt{case} parameter that is a parent folder where \texttt{input} and \texttt{data} folders of an individual run of \texttt{quantumfdtd} and \texttt{allpostprocess.sh} are stored.
There is an optional parameter, \texttt{config}, to provide an absolute or relative path to a \texttt{params.txt} (config file).
The function \texttt{load\_\allowbreak{}energies}, that loads the energies of the final wave-function (and the intermediate snapshots, if they are taken), is also called by the constructor.
\item[\texttt{load\_case(self, case, config=None)}:] In case the default constructor is called, this function accepts the same arguments as the constructor in order to perform a proper initialization of the class.
It also allows to load a different \emph{case}\footnote{%
Path where a run of \texttt{quantumfdtd} is stored.} without destroying and creating a new class.
\item[\texttt{load\_energies(self)}:] This function is implicitly called by the constructor and by the \texttt{load\_\allowbreak{}case} function, so that it does not need to be called explicitly.
It loads the eigen-energies as a function of $\tau$ into the \texttt{self.e\_wf} variable, as a Pandas \texttt{DataFrame}.
It also loads the \texttt{decay.dat} table, that encodes the eigen-energy of the ground state with a higher frequency on $\tau$, and returns it as a \texttt{DataFrame} via the \texttt{self.e\_dec} variable.
\item[\texttt{load\_potential(self)}:] This function loads the potential and returns it as a \texttt{DataFrame} with the columns $x$, $y$, $z$ (integers, coordinates in lattice space in units of $A$), $r$ (in units of $A$), \texttt{re\_v} and \texttt{im\_v} (real and imaginary part of the potential, in GeV).
\item[\texttt{load\_wf(self, snap, state, figure, min\_sep\_edge=-1., normalization=True)}:]\label{flag:min_sep_edge}
This function loads a wave-function and returns it as a \texttt{DataFrame}.
If \texttt{snap=-1}, the final wave-function is loaded.
Otherwise, an intermediate 3D snapshot is loaded, whose \texttt{STEP} value is given by the \texttt{snap} parameter.
Remember that $\tau = \text{\texttt{STEP}}\cdot\text{\texttt{EPS}}$.
The argument \texttt{state} is an integer value from 0 to 2, corresponding to the ground state, or to the first or second excited state.
The \texttt{figure} variable is \texttt{m}, \texttt{p}, \texttt{pk} or \texttt{all}, corresponding to the projection that is going to be loaded: negative parity projection $P^{-}$, positive parity projection $P^{+}$, negative-around-an-axis $P^{-}_{p_{k}}$ and non-projected wave-function, respectively.
The \texttt{min\_sep\_edge} variable keeps points that are closer to the edge than $(\text{min\_sep\_edge} \times N/2)$ out of the loaded wave-function.
Disabled if set to negative value (default).
Finally, if \texttt{Normalization} is set to true, the loaded wave-function is normalized.
This happens after the potential dropping of points due to the \texttt{min\_sep\_edge} flag.
\item[\texttt{fix\_positive\_wf(self, wf, lim=0.05)}:] This function returns the wave-function, inverting its sign if the integration of its real values for the points $r < A \cdot N \cdot \text{\texttt{lim}}$ is negative.
\item[\texttt{get\_avg\_r(self, wf)}:] This function returns the average radius, $\langle r \rangle$, of the wave-function passed by the \texttt{wf} parameter.
\item[\texttt{load\_proj\_components(self)}:] This function adds additional rows to \texttt{self.e\_wf}.
In these rows, the weights of the parity projections of each excited state (see subsection~\ref{subsec:parity_projection}) are stored.
If snapshots are taken, this option is very useful for measuring the decay of the different excited states through the iterative procedure explained in subsection~\ref{subsec:iterative_procedure}.
\item[\texttt{get\_snaps(self)}:] This function returns an \texttt{array} of snapshots.
For each snapshot, the value of \texttt{STEP} is stored on the array, where $\tau = \text{\texttt{STEP}}\cdot\text{\texttt{EPS}}$.
\end{description}

\section{Conclusions}
\label{sec:conclusion}

In this work, the \texttt{quantumfdtd} code version~2.1~\cite{Strickland:2009ft, Margotta:2011ta} has been extended to include the case of a relativistic Schrödinger equation.
This is accomplished via a FFT implementation for the RQM inspired kinetic energy operator and working with the iterative solver in imaginary time.
Additionally, we have extended the program to have support for arbitrary external potentials, introduced as ASCII tables, and included several analysis scripts.
One of them allows to extract parity projections of the extracted wave-functions.
This greatly improves the flexibility and usability of the original program and makes this version~3 valuable even if the original FDTD solver (for the non-relativistic Schrödinger equation) is being used.

We have seen in section~\ref{sec:examples} that the ground-state wave-function is stable for high values of the time evolution parameter $\tau$.
The output wave-functions are linear combinations of the first few excited states.
The overlap method is an approximation to project out the first and second excited state from the full wave-function.
The evolution Eq.~\eqref{eq:iter_decay} (together with a periodic internal normalization of the evolved wave-function), makes the weight of the excited states in the wave-function decay exponentially.
These weights will eventually decay below the machine precision, eliminating the information about excited states from the runs with the highest values of $\tau$.
The parity projection scripts are a useful tool for separating excited states when they are nearly degenerated in energy which is especially noticeable for the $2s$ and the $2p$-states.
For various center-symmetric Hamiltonians we have demonstrated that this degeneracy is quantitatively restored in large enough volumes.
Wherever analytic theory predictions were available, they are reproduced quite well by the simulations.
The possibility to exclude certain states from the beginning by means of choosing the initial symmetry can be an advantage during demanding calculations and is also a non-trivial cross check.

To sum up: the ground-state wave-function should be easy to recover in all of the situations, provided that the corresponding energy is in a plateau and the wave-function is stable with respect to $\tau$.
For increased precision, the positive parity projection can be used.
The first opposite parity state, i.e., typically the $2p$-state, can be separated by means of the negative-around-an-axis parity projection operator.
The $2s$-state can be also separated, but is more challenging due to the behavior of the tail after the first node on a periodic potential.
A large lattice volume can help here.
In the examples presented in this paper, we mainly focused on relatively small lattices ($N \leq 256$) {that are most relevant for interfacing with lattice gauge theory simulations} in order to assess the numerical accuracy of the newly introduced relativistic and non-relativistic kinetic energy terms.
In practice, due to the parallelized implementation of both the FDTD and FFT-based algorithms it is possible to use \texttt{quantumfdtd} on quite large lattices.
For example, there have been past studies with lattices as large as $N = 1024$~\cite{Strickland:2009ft}.

Possible future improvements include:
\begin{itemize}
\item Implementing an automatized algorithm in order to look for saddle points of $E_{s,i}(\tau)$ for $i = 0,1,2$.
\item Implementing new projection operators on the post-processing stage (see subsection~\ref{subsec:parity_projection}).
For instance, separation of $d$-wave states.
For larger overlaps this may involve including contributions from higher states in the initial wave-function (see subsection~\ref{subsec:initial_conditions}).
\item Implementing a suitable binary format for the wave-functions.
\item Allowing for variations of the coupling strength of each potential.
This could be realized by making the coupling a parameter that can be defined like, e.g., the mass or the lattice spacing.
A more involved way would be to, e.g., allow for a file of pairs of scale and coupling to be read in and thereby allow to implement a rudimentary running of couplings with, e.g., the scale $\mu \sim 1/r$.
Finally, linking the program to existing software like \texttt{RunDec}~\cite{Chetyrkin:2000yt, Schmidt:2012az, Herren:2017osy}, would allow for an exact running of the strong coupling in applications that would benefit from that.
Anyhow, note that new analytic potentials can be easily implemented (see the end of subsection~\ref{subsec:hard_coded_potentials}) and arbitrary external potentials can be loaded from files (see subsection~\ref{subsec:external_potentials}).
\item Implementing an implicit Crank-Nicolson method.
This exists for diffusion equations and even for the real-time non-relativistic Schrödinger equation, and makes the evolution unconditionally stable regardless of the choice of time step.
The main issue with this concerning the relativistic Schrödinger equation, however, is that the relativistic Schördinger equation is a non-linear partial differential equation, due to the non-linear kinetic term $\sqrt{M^2 + \sum_{i=1,2,3} p_i^2}$.
This spoils the discretization of the FDTD method, requiring solving of non-linear equations inside the finite-difference time discretization.
This numerical solution of a system of non-linear algebraic equations inside each step of the FDTD method is also likely to introduce further numerical instabilities.
\item Permitting the (anti-)-symmetric solution for a two-particle system in a background potential with an interaction potential between the two particles would be of interest for tetraquark phenomenology.
Evolving the wave-functions of two non-interacting particles in a background potential would be a straightforward, and a factor $2$ more expensive.
Implementing a nontrivial interaction between the two particles would require implementation of new MPI communication procedures in a rather straightforward manner, but would increase the cost of evaluating the potential by a factor \texttt{2+\$\{NCORES\}} instead of $2$ and create some additional communication overhead.
In both cases two-particle (anti-) symmetrization could be easily realized during the postprocessing stage, or during the \texttt{SNAPUPDATE} stage, where geometric (anti-) symmetrization is performed if requested.
\end{itemize}
Such improvements are beyond the scope of the present work.
However, it would be useful to consider them as a guidance for further improvements of the \texttt{quantumfdtd} program.

\section*{Acknowledgments}
The \texttt{quantumfdtd~v3} code was extensively used and tested in the \texttt{Computational Physics II} online course~\cite{patella_weber_2020} at the Humboldt-Universität Berlin in winter semester 2020/21 (course by A.~Patella and J.~H.~Weber).
We thank the student participants for their extensive testing, and the identification of various issues in the code and scripts, or inaccuracies in the description.

R.~L.~Delgado was financially supported by the Ram{\'o}n Areces Foundation post-doctoral fellowship, the Istituto Nazionale di Fisica Nucleare (INFN) post-doctoral fellowship AAOODGF-2019-0000329 and the Spanish grant MICINN: PID2019-108655GB-I00.

M.~Strickland was supported by the U.S.~Department of Energy, Office of Science, Office of Nuclear Physics Award No.~DE-SC0013470.

J.~H.~Weber was supported by the U.S.~Department of Energy, Office of Science, Office of Nuclear Physics and Office of Advanced Scientific Computing Research within the framework of Scientific Discovery through Advanced Computing (SciDAC) award Computing the Properties of Matter with Leadership Computing Resources.
His research was funded by the Deutsche Forschungsgemeinschaft (DFG, German Research Foundation) - Projektnummer 417533893/GRK2575 "Rethinking Quantum Field Theory".

We acknowledge the support by the Deutsche Forschungsgemeinschaft (DFG, German Research Foundation) under Germany's Excellence Strategy -- EXC-2094 -- 390783311 via the Excellence Cluster "ORIGINS".

The simulations have been carried out on the computing facilities of the Computational Center for Particle and Astrophysics (C2PAP) and the Leibniz Supercomputing Center (SuperMUC), on the local theory cluster (T30 cluster) of the Physics Department of the Technische Universität München (TUM), and on local computing facilities at the INFN-Firenze.

\bibliographystyle{elsarticle-num}
\bibliography{\jobname}

\end{document}